\documentclass[times,11pt]{article}
\pdfoutput=1

\usepackage{color}
\usepackage{diagrams}

\usepackage{times}
\usepackage{epsfig}
\usepackage{latexsym}
\usepackage{graphicx}
\usepackage{verbatim}
\usepackage{amssymb}
\usepackage{amsthm}
\usepackage{amsmath}

\newtheorem{theorem}{Theorem}[section]
\newtheorem{proposition}[theorem]{Proposition}
\newtheorem{lemma}[theorem]{Lemma}

\theoremstyle{definition}\newtheorem{example}[theorem]{Example}
\theoremstyle{definition}
\theoremstyle{definition}\newtheorem{definition}[theorem]{Definition}
\theoremstyle{definition}
\theoremstyle{definition}\newtheorem{remark}[theorem]{Remark}
\theoremstyle{definition}
\theoremstyle{definition}
\theoremstyle{definition}
\theoremstyle{definition}
\theoremstyle{definition}\newtheorem{notation}[theorem]{Notation}

\newcommand{\bit}{\begin{itemize}}
\newcommand{\eit}{\end{itemize}\par\noindent}
\newcommand{\ben}{\begin{enumerate}}
\newcommand{\een}{\end{enumerate}\par\noindent}
\newcommand{\beq}{\begin{equation}}
\newcommand{\eeq}{\end{equation}\par\noindent}
\newcommand{\beqa}{\begin{eqnarray*}}
\newcommand{\eeqa}{\end{eqnarray*}\par\noindent}
\newcommand{\beqn}{\begin{eqnarray}}
\newcommand{\eeqn}{\end{eqnarray}\par\noindent}

\def\bR{\begin{color}{red}}
\def\bB{\begin{color}{blue}}
\def\bM{\begin{color}{magenta}}
\def\bC{\begin{color}{cyan}}
\def\bW{\begin{color}{white}}
\def\bBl{\begin{color}{black}}
\def\bG{\begin{color}{green}}
\def\bY{\begin{color}{yellow}}
\def\e{\end{color}}

\newcommand{\bpf}{\noindent {\it Proof:} }
\def\endproof{\hfill$\Box$}

\def\II{{\rm I}}

\title{Picturing classical and quantum Bayesian inference}
\author{Bob Coecke and Robert W. Spekkens\\ \ \\ \em Oxford University Computing Laboratory and Perimeter Institute for Theoretical Physics\em}
\date{}
\begin{document}
\maketitle


\begin{abstract}
We introduce a graphical framework for Bayesian inference that is sufficiently general to accommodate not just the standard case but also recent proposals for a theory of quantum Bayesian inference wherein one considers density operators rather than probability distributions as representative of degrees of belief.  The diagrammatic framework is stated in the graphical language of symmetric monoidal categories and of compact structures and  Frobenius structures therein, in which Bayesian inversion boils down to transposition with respect to an appropriate compact structure.
We characterize classical Bayesian inference in terms of a graphical property and demonstrate that our approach eliminates some purely conventional elements that appear in common representations thereof, such as whether degrees of belief are represented by probabilities or entropic quantities.  We also introduce a quantum-like calculus wherein the Frobenius structure is noncommutative and show that it can accommodate Leifer's calculus of `conditional density operators'.  The notion of conditional independence is also generalized to our graphical setting and we make some preliminary connections to the theory of Bayesian networks.  Finally, we demonstrate how to construct a graphical Bayesian calculus within any dagger compact category.

\end{abstract}

\tableofcontents




\section{Introduction}

In this paper we introduce a graphical calculus and corresponding axiomatics in terms of monoidal categories for a very general notion of Bayesian inference.
It enables one to reason at a highly abstract level about theories more general than 'classical' Bayesian inference, including a proposal for a theory of quantum Bayesian inference introduced by Leifer \cite{Leifer1} and subsequently developed by Leifer and Poulin \cite{Leifer2}, and Leifer and Spekkens \cite{LeiferSpekkens}. This theory is a noncommutative generalization of the theory of classical Bayesian inference wherein one replaces probability distributions over a set of random variables by density operators over a set of systems, marginals by reduced density operators, and conditionals by positive operators for which the partial trace over the conditioned system yields the identity operator on the conditionning system.
The graphical language exploits the two-dimensional diagrammatic representation to distinguish givens and conclusions.  Bayesian inversion is diagrammatic transposition in terms of the \em compact structures \em  \cite{Kelly,KellyLaplaza}.  \em Frobenius structures \em  \cite{CarboniWalters}   will be our vehicle for expressing notions such as \em conditionalization \em and relations of \em conditional independence\em.
`Classical' Bayesian inference is characterized in terms of a condition of commutativity for the Frobenius structure and therefore this structure is key to expressing \em Bayesian updating \em in the specific case of classical Bayesian inference.

There already exists a graphical language for expressing general quantum physical processes on quantum states over a finite-dimensional Hilbert space, namely Selinger's  $CPM(\textbf{FdHilb})$ \cite{Selinger}. As shown by Coecke, Paquette and Pavlovic,  this together with certain kinds of Frobenius structures defines a language for classical stochastic processes as a special case \cite{CPaqPav}.  However, a Bayesian inference need not correspond to a physically-realizable process taking initial quantum states to final states. Rather, it is a \emph{computational} process, taking givens to conclusions. Therefore, such inferences cannot in general be expressed within $CPM(\textbf{FdHilb})$ and the present work provides the extension of the graphical language that is required to accommodate Bayesian inference.

An abstract representation of Bayesian inference allows one to identify which aspects of the standard probability calculus are merely conventional.  For instance, in the context of R.~T.~Cox's derivation of the rules of classical Bayesian inference \cite{Cox}, the standard assumption that one's degree of belief about a proposition $a$ ought to be represented by a number $p(a)$ between 0 and 1 and that one \em multiplies \em a conditional probability with a marginal to get the joint probability, i.e. $p(a,b)=p(a|b)p(b)$  is seen to be a consequence of a choice of convention.  One could equally well represent this degree of belief by any bijective function of $p(a)$ such as $s(a)=-\log p(a)$, in which case $s(a,b)=s(a|b)+s(b)$ and one replaces the standard form of Bayes' rule, $p(a|b)=p(b|a)p(a)/p(b)$, with its ``entropic'' form $s(a|b)=s(b|a)+s(a)-s(b)$.  The abstract approach taken in this work finds a similar result and thereby contributes to the project of extracting the elements of Bayesian inference that are independent of convention.

Our graphical representation of Bayesian inference is also likely to have a close connection with the theory of Bayesian networks, and therefore may shed light on quantum analogues of these \cite{Leifer2}.  This has practical interest in the field of quantum information theory as quantum analogues of belief propagation algorithms are a natural avenue to quantum error correction schemes.  As an example of this connection, the quantum analogue of Bayes' rule has the same form as the approximate reversal channel of Barnum and Knill \cite{BarnumKnill}.  Furthermore, given that Bayesian networks provide a powerful tool for inferring something about the causal relations that hold among propositions from the relations of conditional independence that exist in their correlations \cite{Pearl}, we also hope that our graphical calculus might ultimately help to infer causal relations from quantum correlations and shed light on the quantum violation of Bell's notion of local causality.

Finally, there has been a great deal of interest recently about general probabilistic theories that are distinct from both classical probability theory and quantum theory, e.g.~\cite{Barrett, Convex, BarnumWilceCats}.  By considering a broad landscape of theories, one can hope to identify which aspects of quantum theory are shared with all operational probabilistic theories and which are unique to it.
The framework we develop here provides a novel way of attacking this problem.  By considering quantum theory as a theory of Bayesian inference, one is led to question which aspects of the theory are shared by all theories of Bayesian inference (insofar as one can define such a set) and which are unique to it.


\paragraph{The logic of categorical graphical languages.}
A pedestrian introduction to the graphical calculi for symmetric monoidal categories is in \cite{CPaqII} and a comprehensive survey on these kinds of results is in \cite{Selinger2}.   These graphical languages trace back to Penrose's work in the early 70's.

Compact categories show up in a range of areas of mathematical physics including knot theory and the Temperley-Lieb algebra \cite{Turaev, YetterBook} and the theory of quantum groups \cite{StreetBook}. \em Dagger compact categories \em have recently been exploited by Abramsky and Coecke in quantum information theory \cite{AC} and in proposals for quantum gravity \cite{Baez}.   Frobenius structures trace back to Ferdinand Georg Frobenius' work on the representation theory of finite groups.  They   provide a very concise presentation of topological quantum field theories \cite{BaezDolan, Kock}, they provide a bridge between classical and linear logic \cite{Melies}, and allow diagrammatic axiomatization of quantum observables and C*-algebras \cite{CPV, Jamie}.  Similarly, they allow to distinguish between classical and quantum states \cite{CPaqPav}.

To know how much one can actually prove in a diagrammatic language one relies on the correspondence between graphical languages and certain kinds of monoidal categories, for example:

\begin{theorem}[Joyal-Street 1991 \cite{JS}]
An equation follows from the axioms of symmetric monoidal categories if and only if it can be derived in the corresponding graphical language.
\end{theorem}

\begin{theorem}[Kelly-Laplaza 1980, Selinger 2007 \cite{KellyLaplaza, Selinger}]\label{thm:coherdagcomp}
An equation follows from the axioms of (dagger) compact categories if and only if it can be derived in the corresponding graphical language.
\end{theorem}

If one knows to which categorical structure a certain graphical calculus corresponds, one may ask the question wether there exist complete models of these.\footnote{That is, which enable to embed the corresponding free such categories, and hence, which are such that an equational statement holds in all models if and only if it is a consequence of the axioms of the categorical structure.}  We are aware of two  results of this nature:

\begin{theorem}[Hasegawa-Hofmann-Plotkin 2008 \cite{Plotkin}]
An equation follows from the axioms of traced monoidal categories if and only if it holds in
finite dimensional vector spaces.
\end{theorem}

\begin{theorem}[Selinger 2010 \cite{SelingerCompleteness}]\label{thm:SelingerCompleteness}
An equation follows from the axioms of dagger compact categories if and only if it holds in
finite dimensional Hilbert spaces.
\end{theorem}

Theorem \ref{thm:SelingerCompleteness} is a highly surprising and powerful theorem.   When paired with Theorem \ref{thm:coherdagcomp} it implies 
that an important set of equational statements in quantum theory holds if and only if it can be derived in the graphical calculus.  This result is moreover not only  relevant for quantum mechanics related theories, but also classical probabilistic ones, since the latter can be represented in the category of Hilbert spaces, linear maps and the tensor product by means of  Frobenius structures \cite{CPaqPav}. Unfortunately, there are no completeness results yet of the above kind directly involving  Frobenius structures.

Results like Theorem \ref{thm:SelingerCompleteness} are obviously also important in the context of   automated reasoning, and important steps towards automated reasoning with compact structures and Frobenius structures have already been made \cite{DixonDuncan,DixonDuncanKissinger}.  The developments in this paper make these tools available to the study of (generalized) Bayesian inference.

Given the importance of dagger compact categories in the light of Theorem \ref{thm:SelingerCompleteness}, we construct a class of theories  that generalize   quantum Bayesian inference to any dagger compact category, and for which the concrete non-commutative Frobenius structures arise from the underlying commutative compact structures.  

\paragraph{Structure of the paper.}
In Section \ref{Section:background} we review compact structures, compact categories and dagger compact categories, dagger Frobenius structures therein, the interaction of the latter with compact structures, and the graphical calculus of all of these. In Section \ref{sec:generalBayescal} we define general Bayesian graphical calculi, and also define  the restricted case of classical Bayesian graphical calculi.  We provide an example of a classical Bayesian graphical calculus and show that it is in fact canonical.   We show how entropies provide a model of  classical Bayesian graphical calculus.  We also provide an example of a non-classical one, namely the one introduced by Leifer \cite{Leifer1, Leifer2}, to illustrate the generality of the framework. Typically, while for classical Bayesian graphical calculi the Frobenius multiplications will act commutatively, for non-classical ones it will act non-commutatively.  We also observe that the key structural component of Bayesian graphical calculi, the Frobenius comultiplication, is in fact a logical broadcasting operation  (a map from one object to a pair of these such that the final state has both marginals equal to the initial state). Section \ref{sec:causal} relies on compact structures to obtain a graphical presentation where givens are inputs and conclusions are outputs.  In this setting, Bayesian inversion is nothing but transposition,  and looks as follows:
\[
\raisebox{-0.60cm}{\epsfig{figure=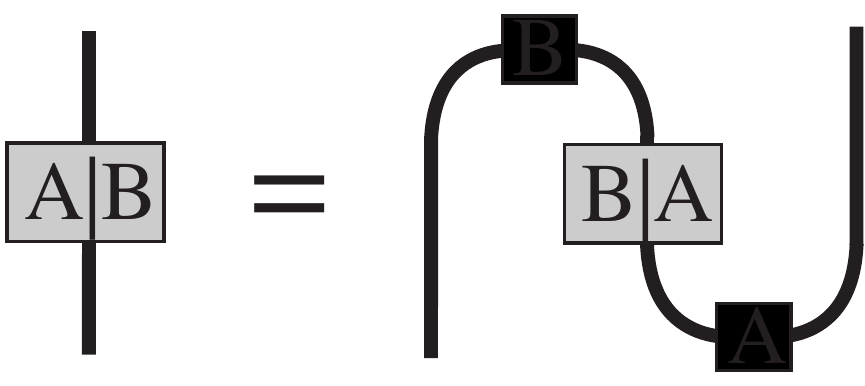,width=88pt}}\ .
\]
 In Section \ref{ConditInd} we define and study the important concept of conditional independence.  We provide a simple example of how an assumption of conditional independence leads to a generalized formula for pooling multiple states of belief, and briefly discuss the semi-graphoid axioms and the connection between our work and Bayesian networks.   Finally, in Section \ref{sec:Qcalc_in_CPM},
  we provide an explicit construction for any dagger compact category of a non-commutative Frobenius multiplication, of which composition of density operators is a special case.  This also gives an explicit presentation of the non-completely positive Frobenius comultiplication which plays the role of a logical broadcasting operation.

\section{Background: dagger Frobenius structures and compact structures}\label{Section:background}

In this paper we work within the graphical language of symmetric monoidal categories (SMCs).
In such a graphical calculus  associativity and unit natural isomorphisms are always strict, that is:
 \beq
( A\otimes B)\otimes C=A\otimes (B\otimes C)\qquad\mbox{and}\qquad A\otimes \II=A=\II\otimes A\,.
 \eeq
General morphisms (or operations)  $f:A\to B$, which we interpret as `processes', are represented as:
 \beq
\raisebox{-0.20cm}{\epsfig{figure=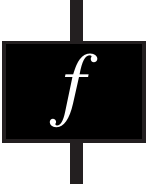,width=20pt}}
 \eeq
while points (or elements) $e:\II\to A$, which we interpret as `states', are represented as:
 \beq
\raisebox{-0.20cm}{\epsfig{figure=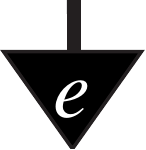,width=20pt}}
 \eeq
 To emphasize that a state $e:\II\to A\otimes B$ is bipartite we represent it as:
  \beq
\raisebox{-0.20cm}{\epsfig{figure=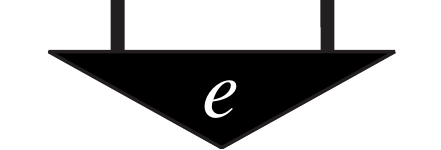,width=60pt}}
 \eeq
Composition and tensoring are respectively represented as:
 \beq
\raisebox{-0.20cm}{\epsfig{figure=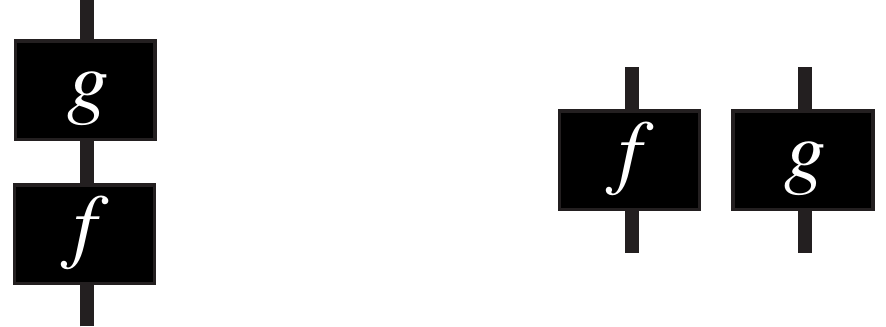,width=125pt}}
 \eeq

A  \em compact structure \em on an object $A$ consists of another object $A^*$ together with a pair of morphisms:
\beq
\eta_A=     \raisebox{-0.20cm}{\epsfig{figure=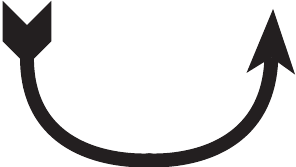,width=30pt}}   :{\rm I}\to A^*\otimes A
\qquad\qquad
\epsilon_A =     \raisebox{-0.20cm}{\epsfig{figure=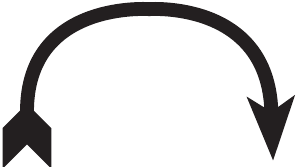,width=30pt}}    :A\otimes A^*\to{\rm I}\,,
\eeq
sometimes referred to as `cups' and `caps',  which satisfy the `yanking' equations:
\beq\label{eq:yanking}
\raisebox{-0.50cm}{\epsfig{figure=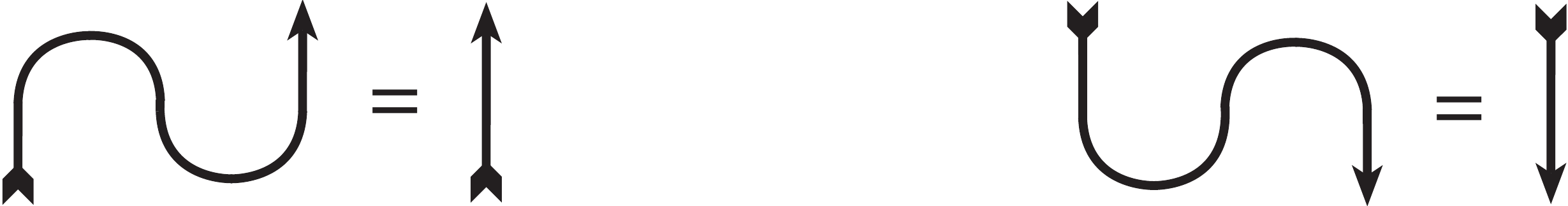,width=250pt}}\ .
\eeq
Hence we depict $A$ by an upward arrow and $A^*$ by a downward one.  We call $A^*$ the \em dual \em of $A$.

A category ${\bf C}$ is a \em compact category \em (CC) \cite{Kelly, KellyLaplaza} if each object comes with a compact structure, which interact in a coherent manner  \cite{Kelly, KellyLaplaza}.  For a number of reasons, including `planarity' of the graphical representation of compact structures on compound objects, one usually adopts the convention that duals are (strictly) contravariant with respect to the tensor, that is,
\beq\label{eq:contravarianttensor}
(A\otimes B)^*=B^*\otimes A^*\qquad\qquad\mbox{and}\qquad\qquad \II^*=\II\,.
\eeq
Cups and caps on a compound object $A\otimes B$ then become:
\beq
\raisebox{-0.40cm}{\epsfig{figure=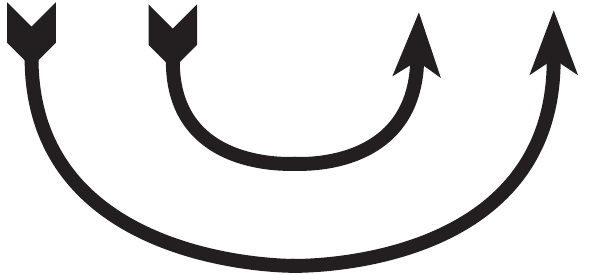,width=60pt}}\qquad\qquad\qquad\qquad
\raisebox{-0.40cm}{\epsfig{figure=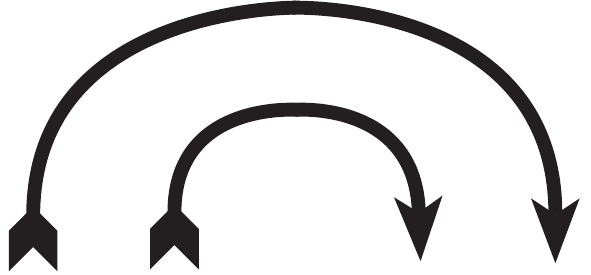,width=60pt}}\,.
\eeq

When we moreover have that $A^{**}=A$  then the direction of arrows clearly distinguishes between `no $*$' and `$*$'.  In this case, coherence\footnote{This means that structural morphisms of the same type are equal. Here that is, if by means of composing and  tensoring  symmetry, (identities), cups and caps one can obtain morphisms
 $f, g: A\to B$ of the same type, then these have to be equal.} requires us to set
\beq\label{eq:compactcoherence}
\eta_{A^*}=\raisebox{-0.36cm}{\epsfig{figure=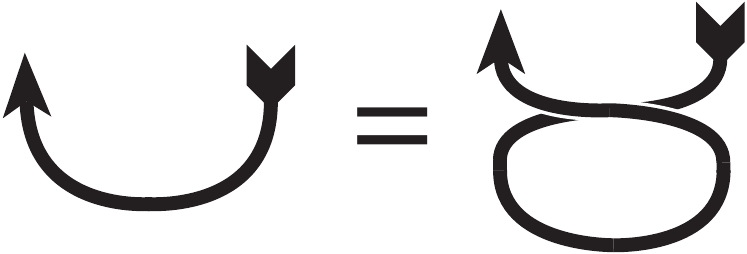,width=76pt}}=\sigma_{A^*\!,A}\circ\eta_A
\qquad
\epsilon_{A^*}= \raisebox{-0.36cm}{\epsfig{figure=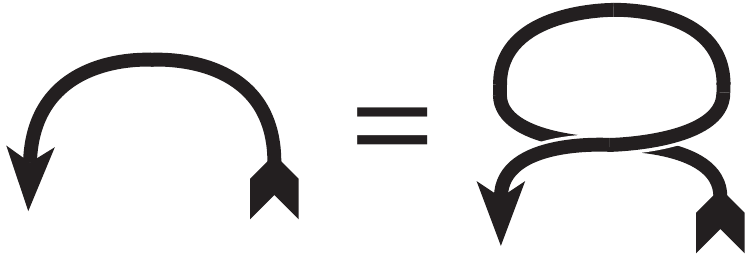,width=76pt}}=\epsilon_A\circ\sigma_{A^*,A}\,,
\eeq
where $\sigma_{A,B}:A\otimes B \to B\otimes A$ is the morphism that simply swaps the objects $A$ and $B$.  We refer to such a CC as \em strict\em.  In this paper all CCs will be strict.

\begin{remark}
The over/under-crossings of wires in the pictures have no formal meaning (cf.~braiding), but only serve to make pictures more readable.
\end{remark}

In any CC each morphism $f:A\to B$ has a \em  transpose \em
\beq\label{eq:transpose}
f^T:=  (1_{A^*}\otimes \epsilon_B)\circ (1_{A^*}\otimes f\otimes 1_{B^*})\circ(\eta_A\otimes 1_{B^*})
=\raisebox{-0.40cm}{\epsfig{figure=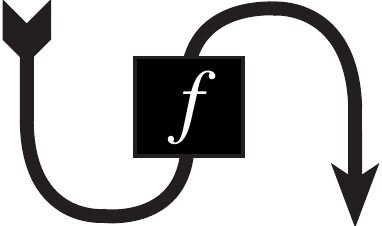,width=48pt}}
: B^*\to A^*\,.
\eeq
Contravariance of $(-)^*$ on objects implies that
\beq
(f\otimes g)^T=g^T\otimes f^T\,.
\eeq

A CC is a \em dagger compact category \em (dCC) \cite{AC, Selinger} if it comes with a contravariant dagger functor
\[
(-)^\dagger:{\bf C}^{op}\to {\bf C}
\]
that coherently preserves the compact structure.  An ordinary  category that comes
with such a functor is called a \em dagger category \em (dC).

We call the composite of the transposed and the dagger the \em conjugate\em.  Explicitly, for a morphism $f:A\to B$ its conjugate is
\beq\label{def:conjugate}
\bar{f}:= (f^\dagger)^T=  (1_{B^*}\otimes \epsilon_A)\circ (1_{B^*}\otimes f^\dagger\otimes 1_{A^*})\circ(\eta_B\otimes 1_{A^*})
=\raisebox{-0.40cm}{\epsfig{figure=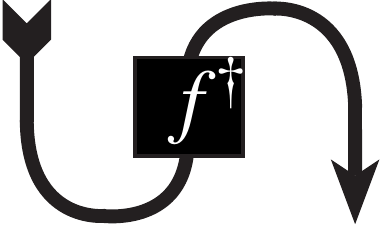,width=48pt}}
: A^*\to B^*\,.
\eeq
Coherence now requires that $\bar{\eta}_A=\eta_A$ and $\bar{\epsilon}_A=\epsilon_A$.  In the graphical calculus this condition can be derived  from the interpretation of the dagger as flipping pictures upside-down.

A \em dagger Frobenius structure \em \cite{CarboniWalters, CPV} on an object $A$ consists of an (internal)  multiplication
\beq
m =\raisebox{-0.20cm}{\epsfig{figure=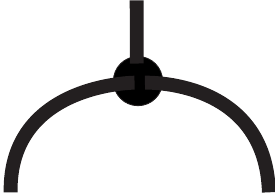,width=28pt}}\ :A\otimes A\to A
\eeq
which is associative, has a two-sided unit
\beq
u=\raisebox{-0.24cm}{\epsfig{figure=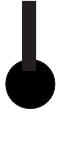,width=6pt}}\,: \II\to A\,,
\eeq
and satisfies the dagger Frobenius law. Diagrammatically these are, respectively,
\beq \label{eq:daggerFrobeniusstructure}
\raisebox{-0.60cm}{\epsfig{figure=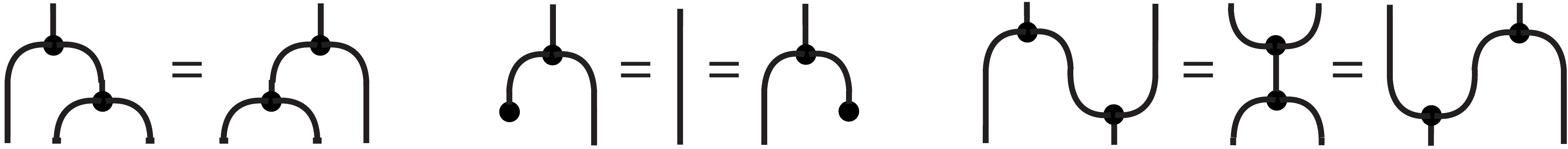,width=380pt}}
\eeq
The morphism $\delta:=m^\dagger= \raisebox{-0.20cm}{\epsfig{figure=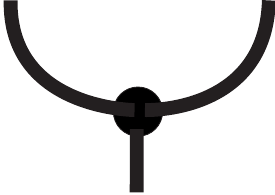,width=28pt}}\ :A\to A\otimes A$ is called a \em comultiplication \em and $\epsilon:=u^\dagger=\raisebox{-0.24cm}{\epsfig{figure=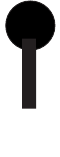,width=6pt}}: A\to\II$ its \em counit\em. A dagger Frobenius structure is \em commutative \em when we have
\beq\label{eq:commutativeFROB}
m=\raisebox{-0.36cm}{\epsfig{figure=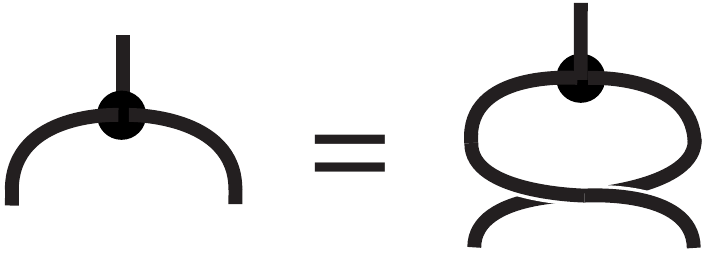,width=76pt}}=m\circ\sigma_{A,A}\,.
\eeq

A dagger Frobenius structure admits an elegant diagrammatic calculus in terms of `spiders' \cite{Kock,Lack,CPaq}.
More precisely, one can show that any morphism
\beq
f: \underbrace{A\otimes \ldots \otimes A}_n\to \underbrace{A\otimes \ldots \otimes A}_m
\eeq
obtained by composing and tensoring  $m, u, m^\dagger, u^\dagger, 1_A$ (and also $\sigma$ in the case that the multiplication is commutative), and of which  the diagrammatic representation is connected, only depends on $n$ and $m$.  We represent this unique morphism of that type as:
\beq
\raisebox{-0.25cm}{$\underbrace{\overbrace{\epsfig{figure=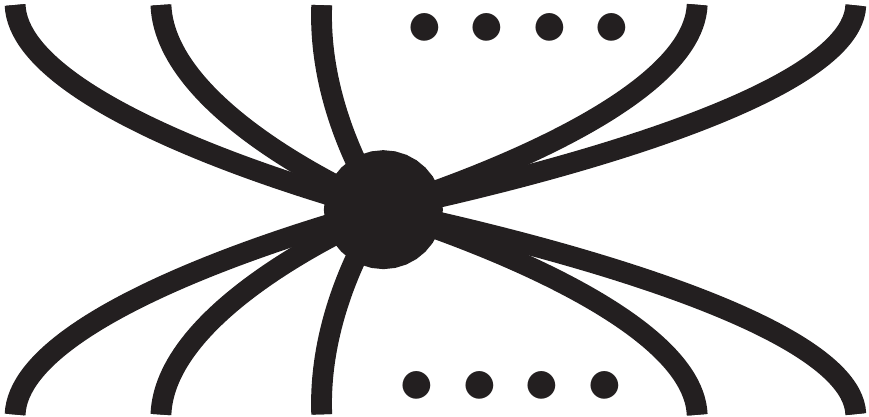,width=60pt}}^{\scriptstyle{m}}}_{\scriptstyle{n}}$}
\eeq
and it is then also immediately clear that these `spiders' compose as follows:
\beq
\raisebox{-1cm}{$\underbrace{\overbrace{\epsfig{figure=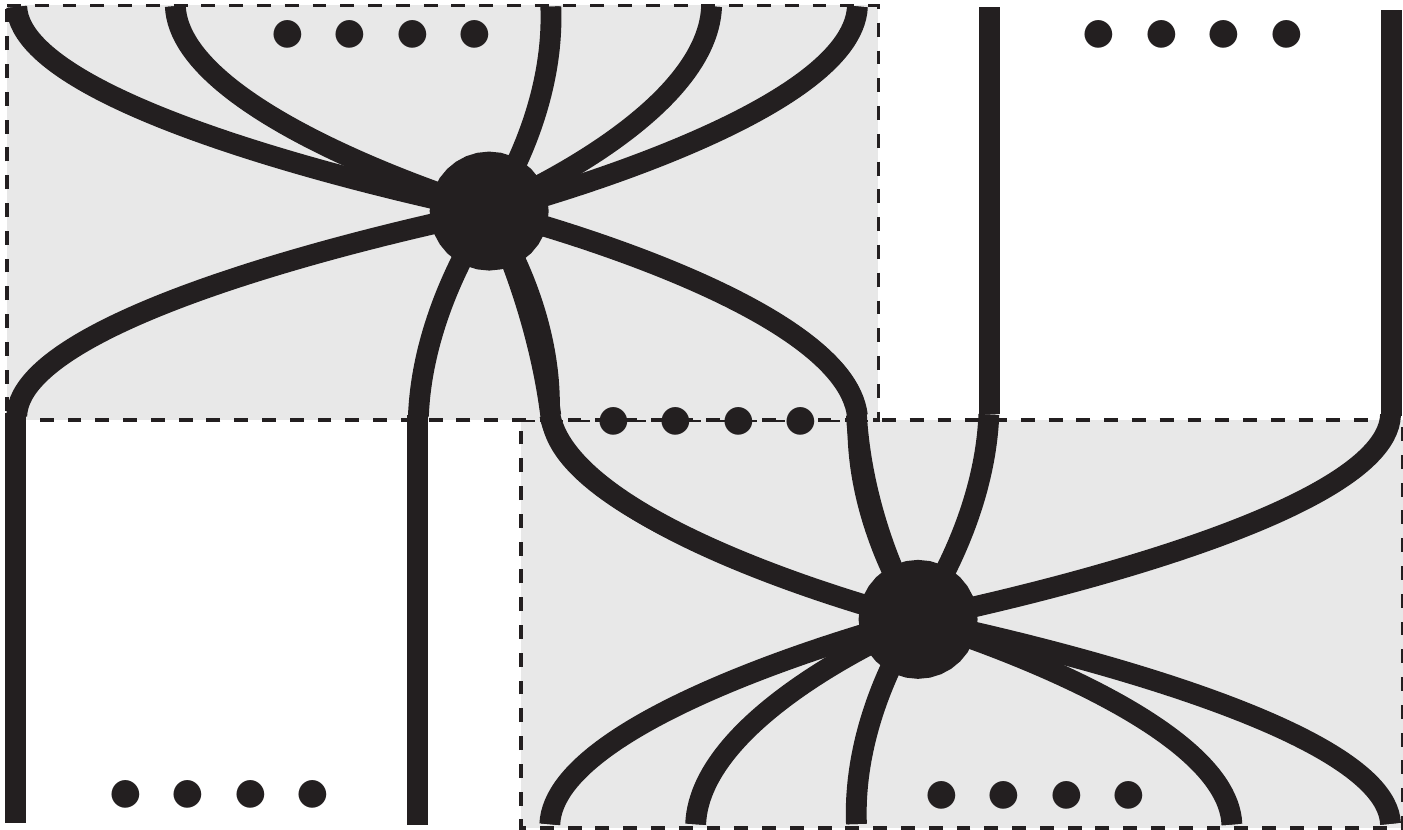,height=62pt}}^{\scriptstyle{m}}}_{\scriptstyle{n}}
   \ \  \epsfig{figure=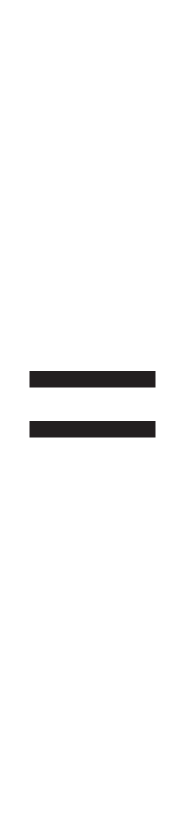,height=62pt}\ \
  \underbrace{\overbrace{\epsfig{figure=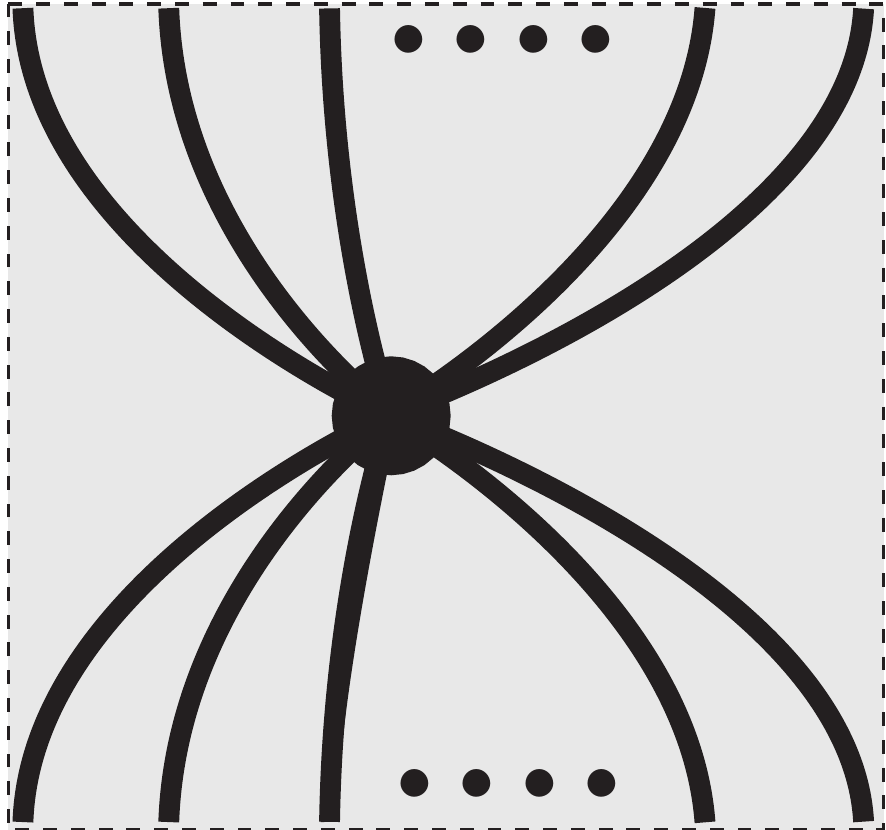,height=62pt}}^{\scriptstyle{m}}}_{\scriptstyle{n}}
 $  }
\eeq
This composition rule encapsulates all of the properties of a dagger Frobenius structure  in
Eq.~(\ref{eq:daggerFrobeniusstructure}), and is below referred to as the \em spider theorem\em.

Each Frobenius structure induces a \em self-dual \em (i.e.~$A=A^*$) compact structure
\beq
\eta^{Frob}=\raisebox{-0.24cm}{\epsfig{figure=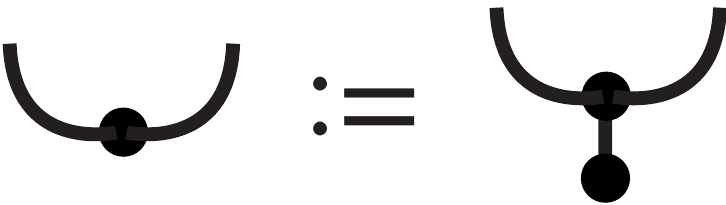,width=74pt}}=m^\dagger\circ u:I\to A\otimes A
\qquad
\epsilon^{Frob}=\raisebox{-0.24cm}{\epsfig{figure=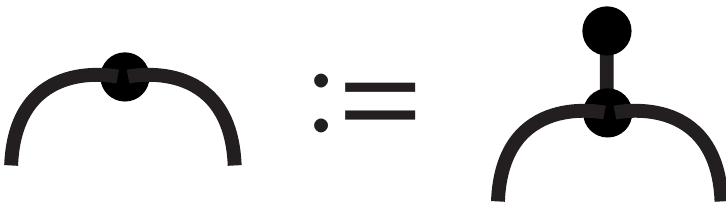,width=74pt}}=u^\dagger\circ m: A\otimes A\to I\,,
\eeq
for which we have:
\beq\label{eq:Frobcompactness}
\raisebox{-0.52cm}{\epsfig{figure=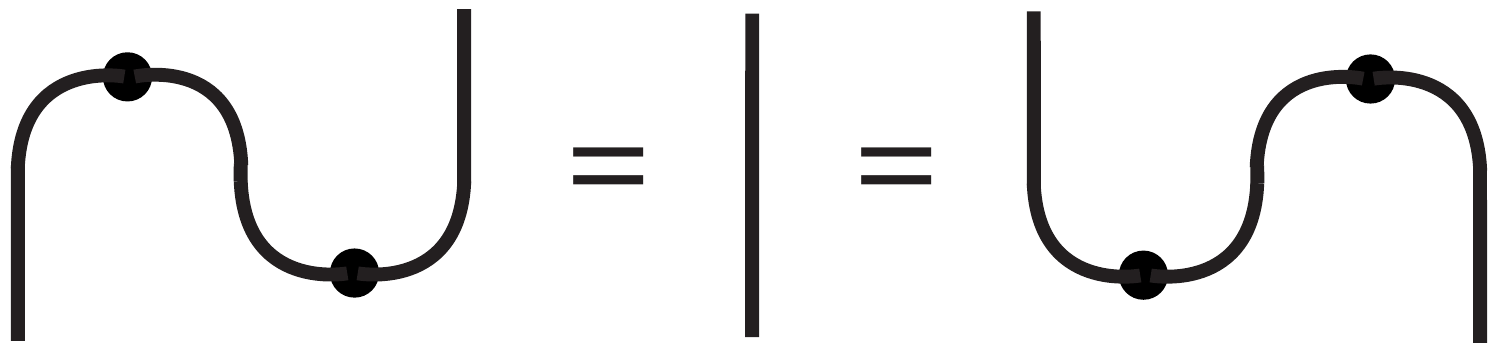,width=152pt}}\,.
\eeq
Because of the self-duality, we can omit the arrows in these diagrams, all of which would point upward. The dot in the cups and caps (where the arrows would change direction if we were to include them) denotes the fact that the compact structure is self-dual.
By self-duality we  also  have $A^{**}=A$, so coherence requires that the compact structure satisfies  (cf.~Eqs.~(\ref{eq:compactcoherence})):
\beq\label{eq:compactFROBcoherence}
\eta^{Frob}=\raisebox{-0.36cm}{\epsfig{figure=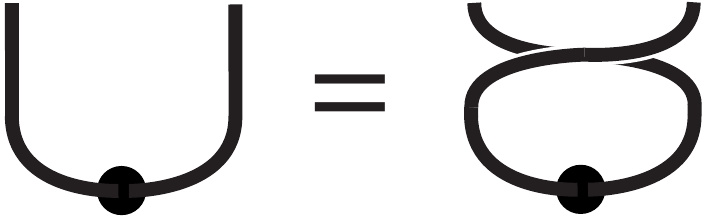,width=76pt}}=\sigma_{A,A}\circ \eta^{Frob}
\qquad
\epsilon^{Frob}=\raisebox{-0.36cm}{\epsfig{figure=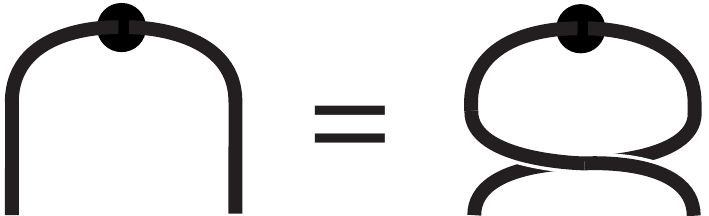,width=76pt}}=\epsilon^{Frob}\circ\sigma_{A,A}\,.
\eeq
We call such a compact structure \em commutative\em.  Obviously, in the case of a commutative Frobenius multiplication, the induced compact structure is always commutative.

For such a self-dual compact structure the convention of Eq.~(\ref{eq:contravarianttensor}) cannot be maintained since it leads to $A\otimes B=(A\otimes B)^*=B^*\otimes A^*=B\otimes A$, which is easily seen to cause a collapse of the structure.   Hence cups and caps
on compound objects now have to be denoted as:
\beq
\raisebox{-0.30cm}{\epsfig{figure=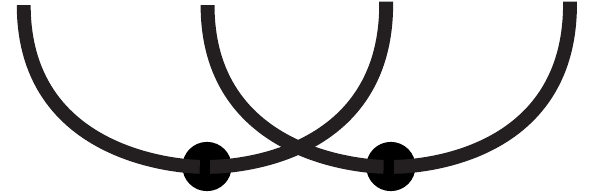,width=60pt}}\qquad\qquad\qquad\qquad
\raisebox{-0.30cm}{\epsfig{figure=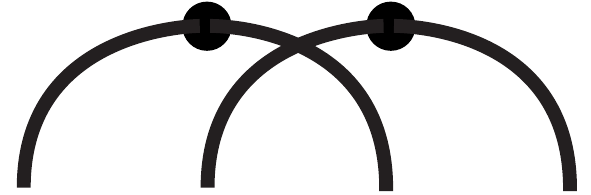,width=60pt}}\ .
\eeq
These arise from the \em canonical dagger Frobenius structure on $A\otimes B$ \em given one on both $A$ and $B$:
\beq
\raisebox{-0.40cm}{\epsfig{figure=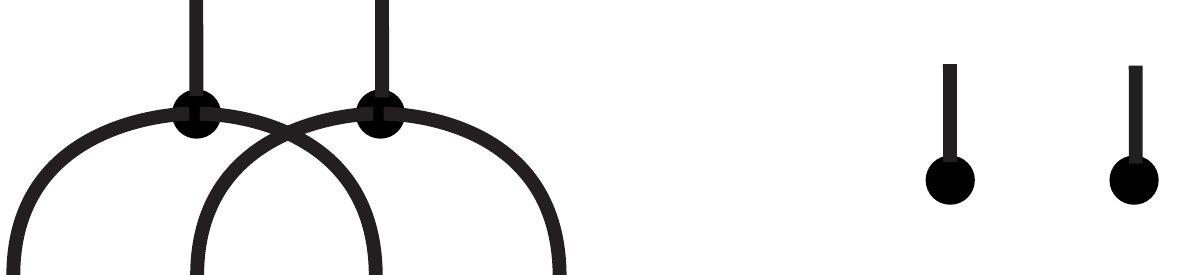,width=120pt}}\ .
\eeq


Frobenius structures have played a prominent role in characterizating classical structures within quantum theory.  It was shown that for the dCC of finite dimensional Hilbert spaces, ${\bf FdHilb}$, with linear maps as the morphisms, the tensor product as the monoidal tensor, and linear algebraic adjoint as the dagger, commutative dagger Frobenius structures are in bijective correspondence with orthogonal bases \cite{CPV}.  
When $m\circ\delta=1_A$, a condition referred to as \em specialness \em \cite{Kock}, we obtain a bijective correspondence with orthonormal bases.  Therefore, these Frobenius structures were called \em classical structures \em \cite{CPav, CPaqPav}. Explicitly,  given a basis $\{ |i\rangle \;|\; i\in\{1,\dots,n\} \}$ of a finite dimensional Hilbert space $\mathbb{C}^n$, the comultiplication is the linear map that `copies' these basis elements:
\beq
\delta: \mathbb{C}^n\to \mathbb{C}^n\otimes \mathbb{C}^n:: |i\rangle\mapsto |i\rangle\otimes |i\rangle\,.
\eeq
It has also been shown that in the dCC ${\bf FdHilb}$, noncommutative special dagger Frobenius structures are in bijective correspondence with noncommutative C*-algebras of linear operators acting on $\mathbb{C}^n$ \cite{Jamie}.   
From the algebraic point of view, classical structures in ${\bf FdHilb}$ are maximal commutative C*-algebras. 
Whether the Frobenius structure is commutative or not will also play an important role in distinguishing the theories of quantum and classical Bayesian inference.

\section{Bayesian graphical calculus}\label{sec:generalBayescal}


\subsection{Definition}

Consider a dSMC ${\bf C}$  in which each object comes with a dagger Frobenius structure.
\ben
\item[{\bf BC1}] For every object $A\in|{\bf C}|$, we assume the existence of a \em normalized state\em, that is, a point which when composed with the counit yields the morphism $1_\II:\II\to\II$ (the identity morphism on the trivial object), which we depict by the `empty picture':
\beq
\raisebox{-0.16cm}{\epsfig{figure=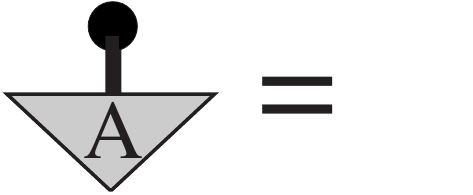,width=48pt}}:\II\to\II\,.
\eeq
A normalized state for a composite object $A\otimes B\in|{\bf C}|$,
\beq
\ \raisebox{-0.16cm}{\epsfig{figure=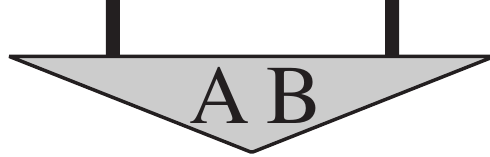,width=50pt}}\ :\II\to A\otimes B \quad\mbox{such that}\quad\raisebox{-0.16cm}{\epsfig{figure=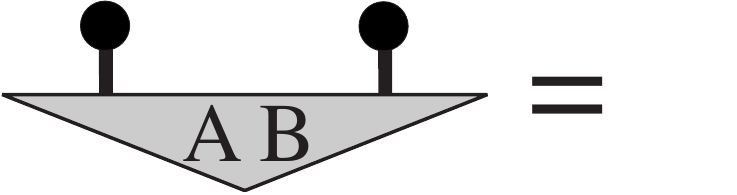,width=75pt}}:\II\to\II\,,
\eeq
 will be called a \em joint state\em.
Note that the composition of a joint state on $A\otimes B$ with the counit on $B$ is a state on $A$, which we call the \em marginal state \em on $A$,
\beq
\ \raisebox{-0.16cm}{\epsfig{figure=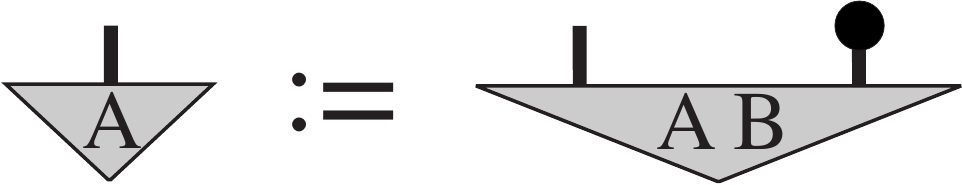,width=98pt}}\ : \II\to A\,.
\eeq
\een
For most of this article, we will be concerned with just a single joint state on a set of objects together with its marginals.  Consequently, it is adequate for our purposes to label states by the object on which they are defined.  On the few occasions in which it will be necessary to refer to two different states on a single object, we will distinguish these by a prime.
\ben
\item[{\bf BC2}] For every object $A\in|{\bf C}|$, we assume the existence of a \em modifier\em, that is, an endomorphism  
\beq
\raisebox{-0.28cm}{\epsfig{figure=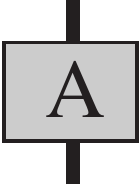,width=15.2pt}}\ :A\to A
\eeq
which is such that
\beq\label{eq:modifier}
\ \raisebox{-0.28cm}{\epsfig{figure=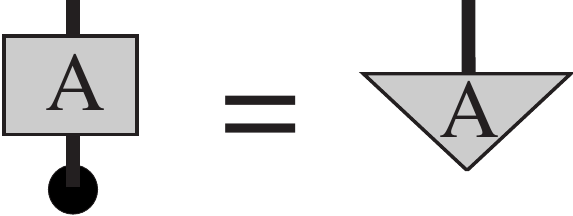,width=58.5pt}}\ ,
\eeq
and which is also \em self-transposed\em, that is,
\beq \label{eq:self-transposed}
\raisebox{-0.50cm}{\epsfig{figure=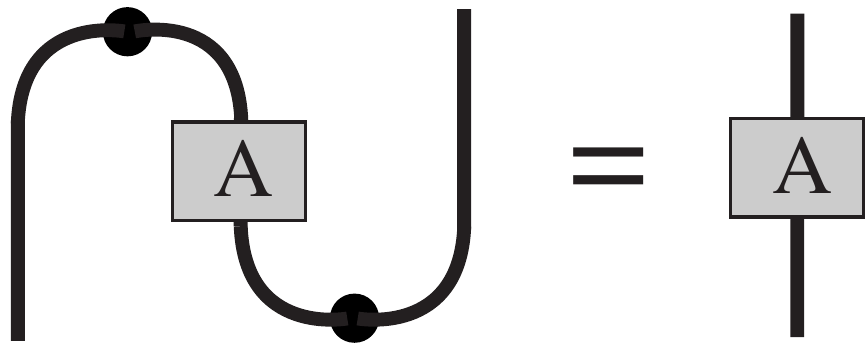,width=88pt}}\ .
\eeq
\een
These modifiers are calculus-specific. We give concrete examples in Sections \ref{sec:Qcalc} and \ref{sec:Ccalc} of how one can construct modifiers in terms of marginal states and the Frobenius multiplication.

\begin{proposition}\label{prop:modifiersmovecaps}
Since modifiers are self-transposed they can move along cups and caps:
\beq\label{eq:slidinggen}
\raisebox{-0.30cm}{\epsfig{figure=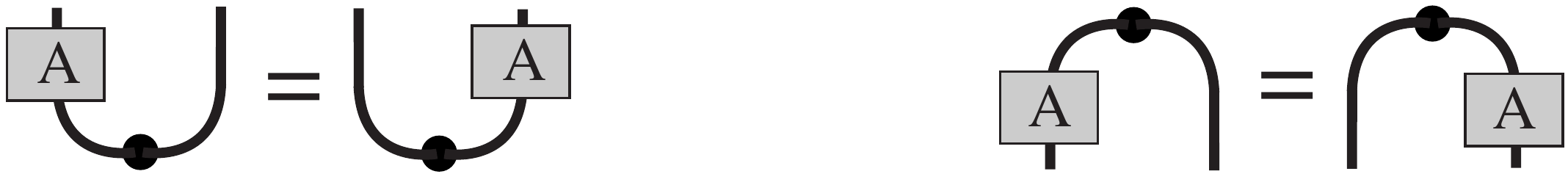,width=217pt}}\ .
\eeq
\end{proposition}
\bpf
By the definition of the transpose and Eq.~(\ref{eq:Frobcompactness}) we have:
\begin{center}
\raisebox{-0.40cm}{\epsfig{figure=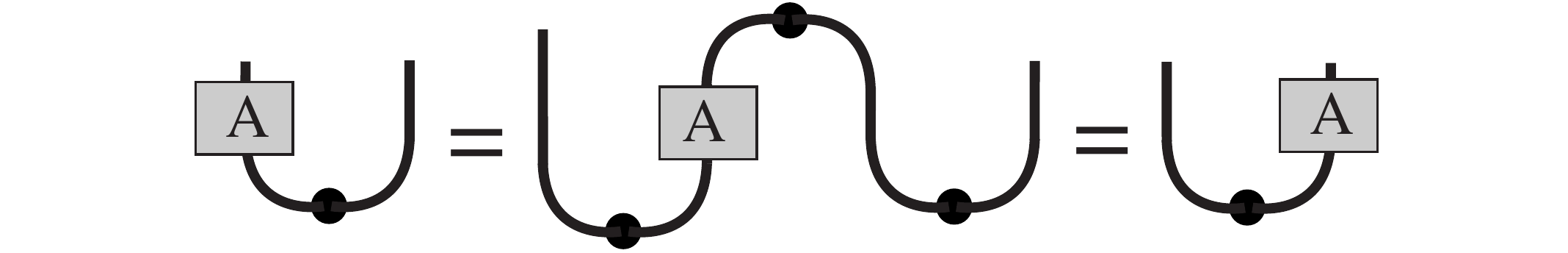,width=217pt}}\!\!\!\!\!\!\!\!\!\!\!\!.
\end{center}
\endproof

\begin{definition} The \em inverse of a modifier \em $\ \raisebox{-0.20cm}{\epsfig{figure=modifier1.pdf,width=14.2pt}}\ :A\to A$ is a process
\beq
\ \raisebox{-0.20cm}{\epsfig{figure=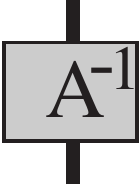,width=14.2pt}}\ :A\to A
\eeq
such that
\beq\label{def:catinverse}
\ \raisebox{-0.34cm}{\epsfig{figure=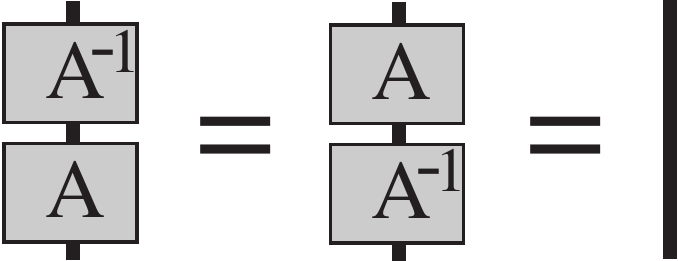,width=69pt}}\ .
\eeq
\end{definition}

Transpose invariance of modifiers also implies that  their inverses can move along cups and caps:
\begin{center}
\hspace{-3mm}\raisebox{-0.30cm}{\epsfig{figure=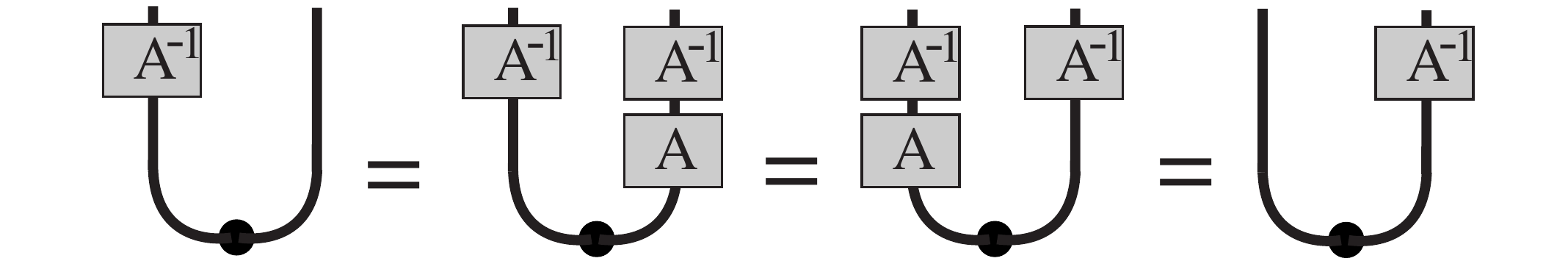,width=217pt}}\hspace{-3mm}\ .
\end{center}

\begin{definition}
The \em Frobenius inverse \em of $\raisebox{-0.16cm}{\epsfig{figure=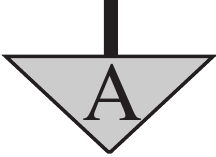,width=22pt}}\ $ relative to a Frobenius multiplication is  a point
\beq
\ \raisebox{-0.16cm}{\epsfig{figure=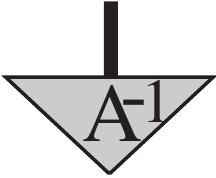,width=22pt}}:\II\to A
\eeq
such that
\beq\label{eq:sharedinverses}
\ \raisebox{-0.38cm}{\epsfig{figure=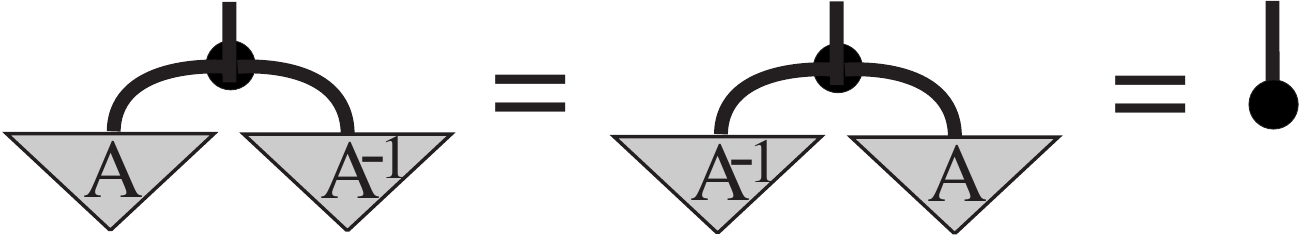,width=132pt}}\ \,.
\eeq
\end{definition}

As it is the case for inverses standardly, these Frobenius inverses are easily seen to be unique.

In our key examples, there will be marginal states and associated modifiers that do not have a Frobenius inverse or an inverse modifier respectively.  It turns out, however, that it suffices to have a more general notion of inverse, namely inverses relative to a support.

\begin{definition}
A \em support \em for $\raisebox{-0.16cm}{\epsfig{figure=marginalonly.pdf,width=22pt}}\ :\II\to A$ is a self-adjoint idempotent $\raisebox{-0.12cm}{\epsfig{figure=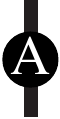,width=6pt}}:A\to A$ which is such that:
\ben
\item $\raisebox{-0.16cm}{\epsfig{figure=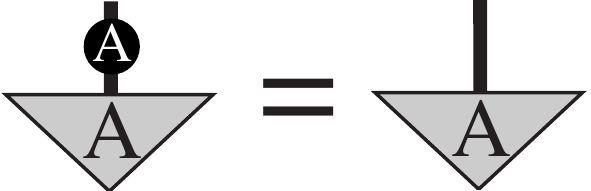,width=60pt}}\ $, and
 \item $\raisebox{-0.16cm}{\epsfig{figure=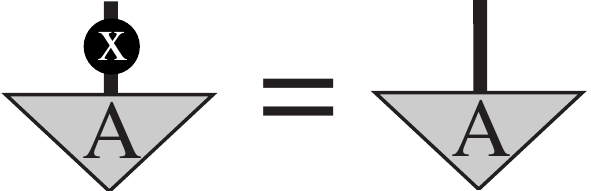,width=60pt}}$ for another self-adjoint idempotent  $\raisebox{-0.12cm}{\epsfig{figure=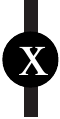,width=6pt}}\ :A\to A$ implies $\raisebox{-0.12cm}{\epsfig{figure=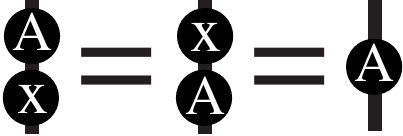,width=40pt}}$\,.
\een
We say that $\ \raisebox{-0.20cm}{\epsfig{figure=modifier1inv.pdf,width=14.2pt}}\ :A\to A$ is the \em inverse \em to
$\ \raisebox{-0.20cm}{\epsfig{figure=modifier1.pdf,width=14.2pt}}\ :A\to A$ \em relative to this support \em if we have that
\beq
\raisebox{-0.34cm}{\epsfig{figure=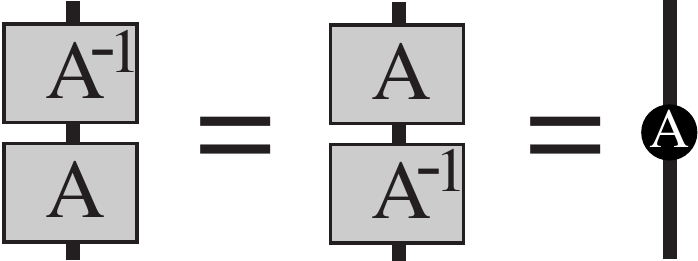,width=71pt}}\ ,
\eeq
and that $\ \raisebox{-0.16cm}{\epsfig{figure=Qinv.pdf,width=22pt}}\ :\II\to A$ is the \em Frobenius inverse \em to $\raisebox{-0.16cm}{\epsfig{figure=marginalonly.pdf,width=22pt}}\ :\II\to A$ \em relative to this support \em if we have that
\beq
\raisebox{-0.38cm}{\epsfig{figure=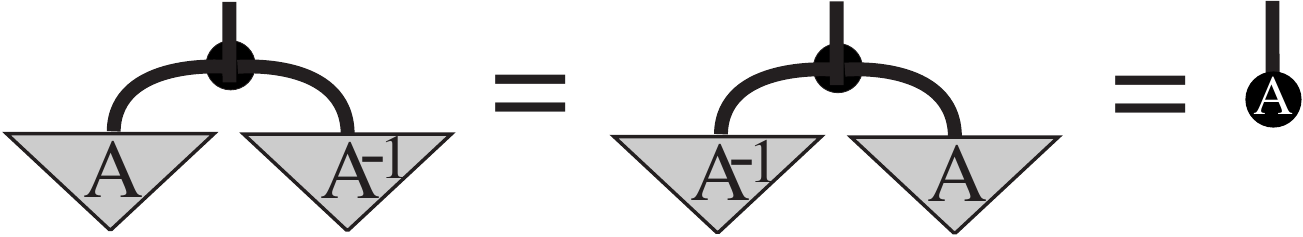,width=132.5pt}}\ .
\eeq
where $\raisebox{-0.16cm}{\epsfig{figure=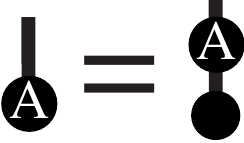,width=25pt}}\ $.
\end{definition}

We can always take the support of a joint state on a composite object to be the tensor product of the supports of its marginals.

Below all inverses are to be understood in this generalized sense i.e.~relative to a suitable support.  One can also incorporate the support within the Frobenius structure $\ \raisebox{-0.16cm}{\epsfig{figure=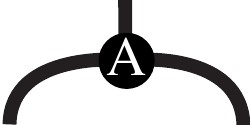,width=26pt}}\ $ by taking $\ \raisebox{-0.12cm}{\epsfig{figure=support.pdf,width=6pt}}\ $ as the identity.

The reason we don't need to indicate the support explicitly in the current work is that we will restrict our attention to a single joint state together with the marginals and conditional states  (cf.~below)  it defines and as such, we will never have need to consider states having different supports on the same object.

\ben
\item[{\bf BC3}]  We assume that each state admits of a Frobenius inverse relative to its support and each modifier admits an ordinary inverse relative to its support such that the latter is the modifier associated with the former:
\beq\label{eq:consistency}
\ \raisebox{-0.28cm}{\epsfig{figure=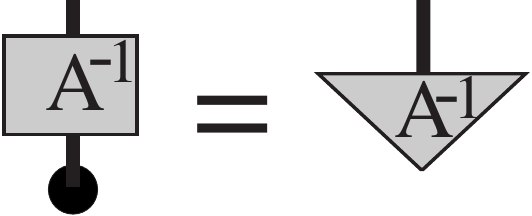,width=54pt}}\ \,.
\eeq
\een


\begin{definition}\label{def:ConditionalState}
For every joint state on a pair of objects, we can define a \em conditional state \em to be the point
\beq
\raisebox{-0.34cm}{\epsfig{figure=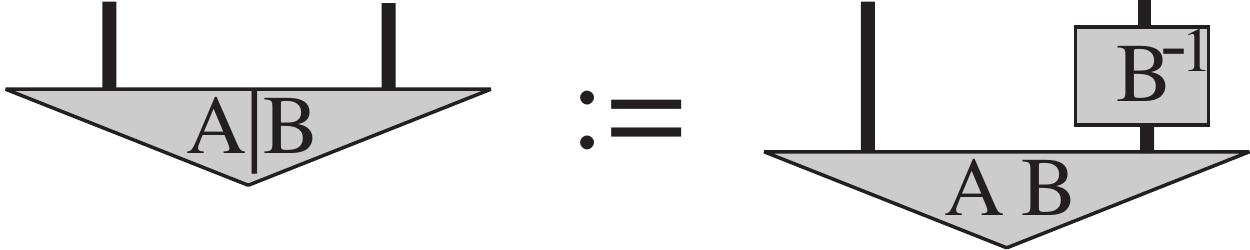,width=127pt}}\ : \II\to A\otimes B\,.
\eeq
\end{definition}

A conditional state is such that if we compose the conditioned object (the one on the left of the conditional bar) with the co-unit, we obtain the unit on the conditioning object (the one of the right of the conditional bar)
\beq\label{eq:condstatenorm}
\raisebox{-0.30cm}{\epsfig{figure=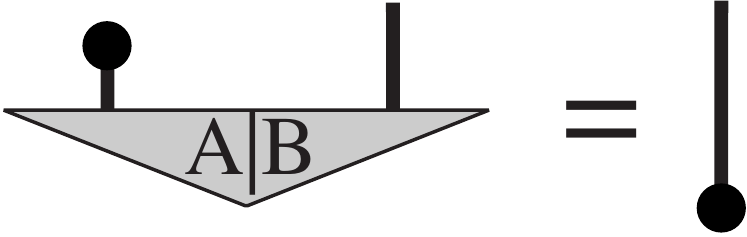,width=76pt}}\ .
\eeq

\begin{definition}\label{def:ByesCalc}
We call a graphical calculus with ingredients {\bf BC1}, {\bf BC2}, {\bf BC3} a \em Bayesian graphical calculus\em.
\end{definition}

This definition is motivated by the fact that with notions of joint states, marginal states, conditional states, modifiers and inverses, we have the minimal amount of structure required to describe basic concepts of Bayesian inference.
For example,  Bayes' rule depicts as:
\beq\label{eq:Bayes}
\raisebox{-0.44cm}{\epsfig{figure=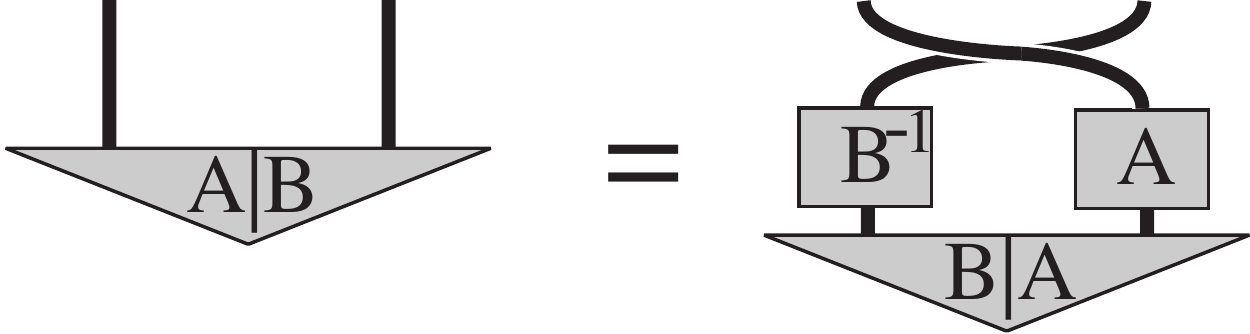,width=127pt}}\,.
\eeq
We can straightforwardly extend the above to multiple variables $A, B, C, \ldots$.   When setting:
\beq\label{eq:modswap}
\raisebox{-0.32cm}{\epsfig{figure=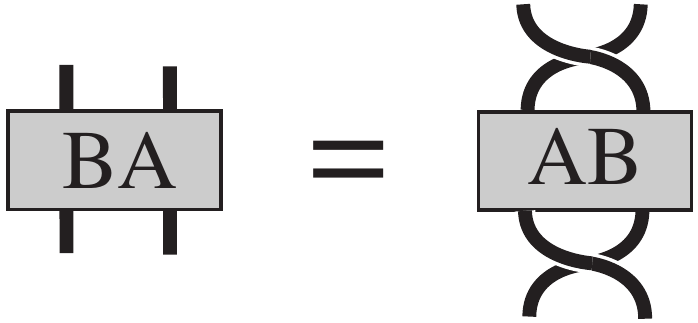,width=71pt}}
\eeq
it straightforwardly follows that:
\beq
\raisebox{-0.32cm}{\epsfig{figure=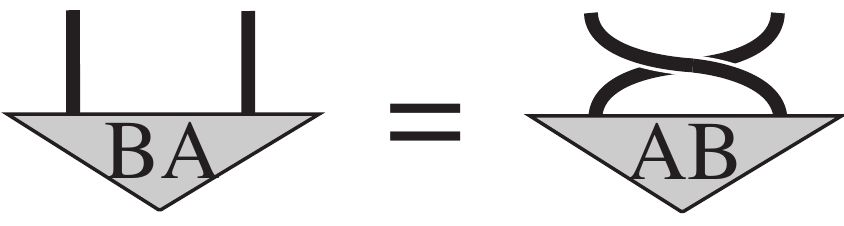,width=94pt}}
\eeq
and that
\beq\label{lm:givens commute}
\ \ \raisebox{-0.32cm}{\epsfig{figure=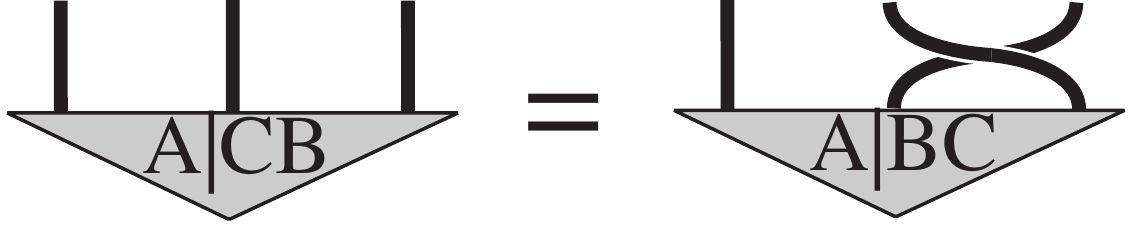,width=122pt}} \,.
\eeq

Many important concepts can now be defined at this high level of generality, most notably, conditional independence (cf.~Section \ref{ConditInd} below), and many results can be derived graphically, e.g.~pooling (cf.~Section \ref{sec:pooling} below).

\subsection{Classical Bayesian graphical calculus}\label{sec:Ccalc}

\begin{definition}\label{def:ClassicalByesCalc}
A Bayesian graphical calculus is called \em  classical \em  if it satisfies the following equivalent conditions:
\ben
\item[(a)]
modifiers can move through the Frobenius structure:
\beq\label{def:slidinginverseclass}
\raisebox{-0.30cm}{\epsfig{figure=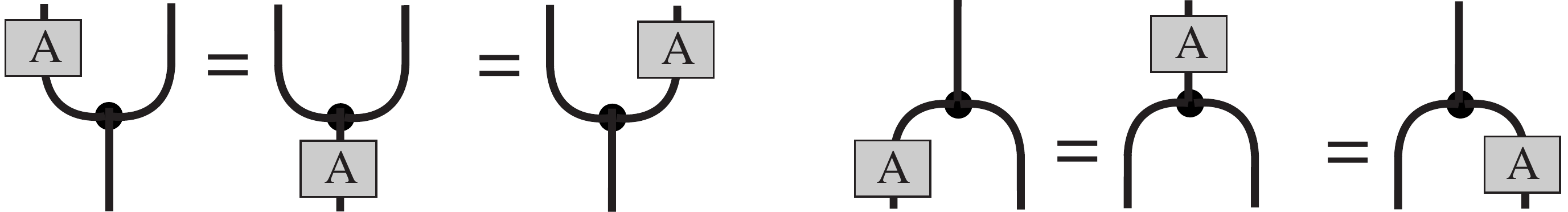,width=276pt}}\ .
\eeq
\item[(b)]
modifiers are of the form:
\beq\label{eq:Ccalcmofifier}
\raisebox{-0.20cm}{\epsfig{figure=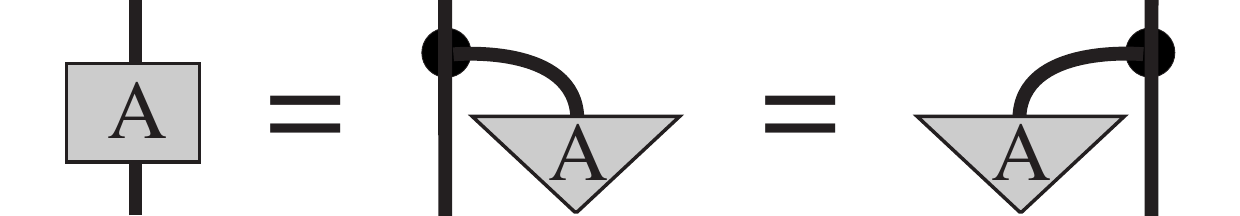,width=126pt}}\,,
\eeq
\een
\end{definition}

\bpf
We first show that condition (a) implies condition (b).  By the spider theorem (more specifically, the counit law in Eq.~(\ref{eq:daggerFrobeniusstructure})), Eq.~(\ref{def:slidinginverseclass}) and Eq.(\ref{eq:modifier}),
\begin{center}
\raisebox{-0.32cm}{\epsfig{figure=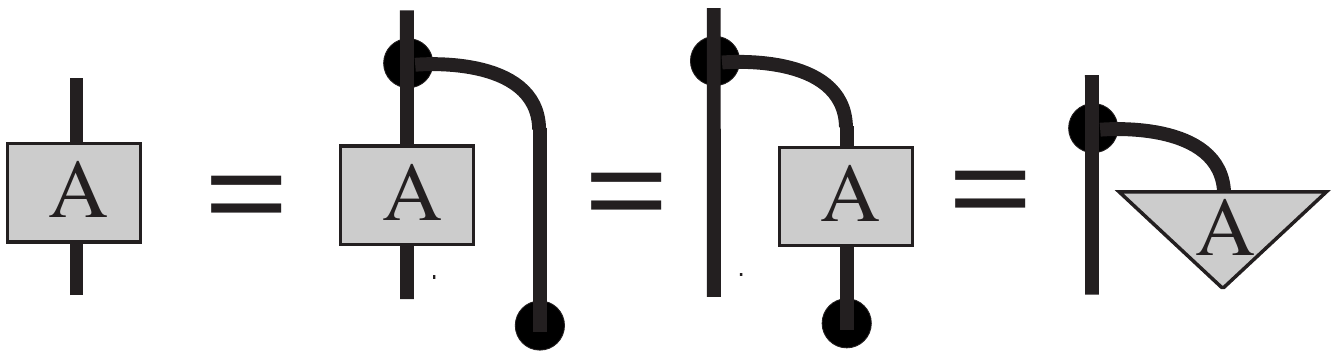,width=142pt}}\ .
\end{center}
The second equality in condition (b) is proven with the mirror image of this argument.
Condition (b) implies condition (a) since again by the spider theorem (more specifically, the Frobenius law in Eq.~(\ref{eq:daggerFrobeniusstructure})),
\begin{center}
\raisebox{-0.32cm}{\epsfig{figure=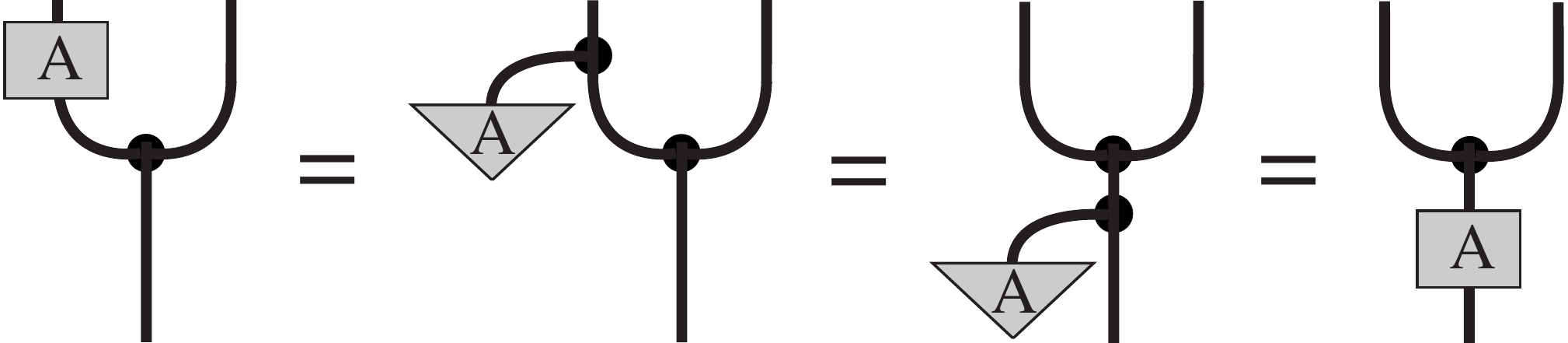,width=205pt}}\ .
\end{center}
The second equality in condition (a) is again proven with the mirror image of this argument.
Finally, note that the form of this modifier is consistent with the conditions BC2 and BC3, which are defining conditions for a Bayesian graphical calculus.  Clearly, the modifier acting on the unit gives the associated point, Eq.~(\ref{eq:modifier}), and it is self-transposed, Eq.~(\ref{eq:self-transposed}).  Also, the  consistency condition on inverses in Eq.~(\ref{eq:consistency}), that is, the equivalence of $\ \raisebox{-0.16cm}{\epsfig{figure=Qinv.pdf,width=22pt}}\ $ and
$\ \raisebox{-0.30cm}{\epsfig{figure=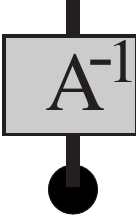,width=14pt}}\ $, is automatically satisfied because
\begin{center}
\raisebox{-0.30cm}{\epsfig{figure=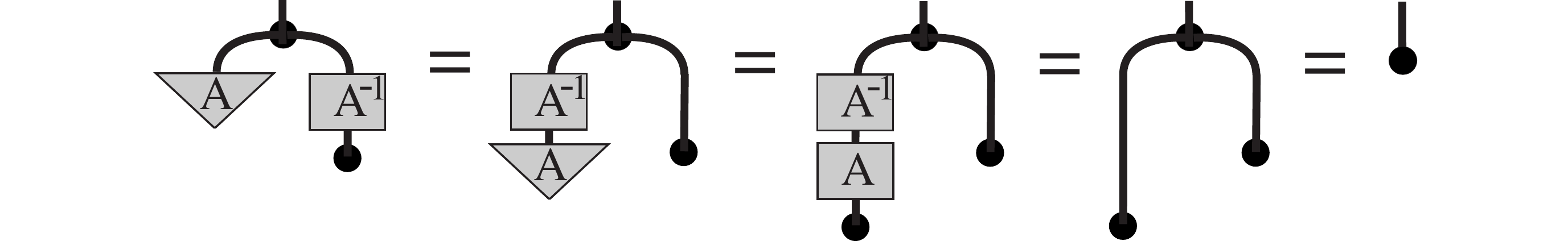,width=276pt}}\ .\vspace{-2mm}
\end{center}
\endproof\newline

So in classical Bayesian graphical calculi, in addition to moving along cups and caps (cf.~Proposition \ref{prop:modifiersmovecaps}), modifiers can move through the Frobenius structure, and hence, by the spider theorem,  in a classical Bayesian graphical calculus modifiers can move through arbitrary spiders.

Note that the conditions in Eq.~(\ref{def:slidinginverseclass}) and Eq.~(\ref{eq:Ccalcmofifier}) hold for states and modifiers of composite objects using the Frobenius structure for the latter.



It is useful to consider some of the features of such a calculus.
\begin{proposition}
In a classical Bayesian graphical calculus, modifiers on composite objects move through the Frobenius structure of one of the objects:
\beq\label{lem:slidinginverseclass}
\raisebox{-0.30cm}{\epsfig{figure=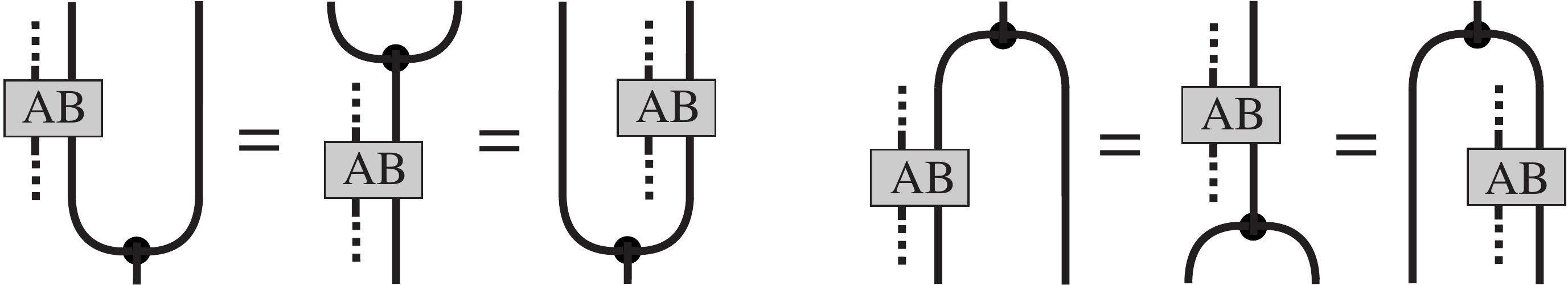,width=276pt}}\ .
\eeq
\end{proposition}
\bpf
We have:
\[
\raisebox{-0.30cm}{\epsfig{figure=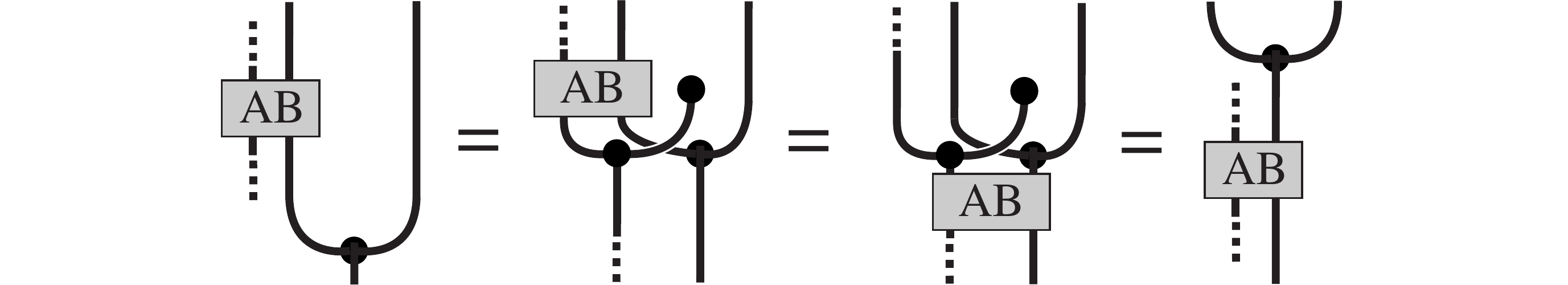,width=276pt}}
\]
by Eq.~(\ref{eq:daggerFrobeniusstructure}) and Eq.~(\ref{def:slidinginverseclass}) applied to $A\otimes B$.
The other equalities are proven similarly.
\endproof

\begin{proposition}
In a classical Bayesian graphical calculus, the Frobenius multiplication always acts commutatively on states, that is:
\beq\label{eq:Bayes2}
\raisebox{-0.44cm}{\epsfig{figure=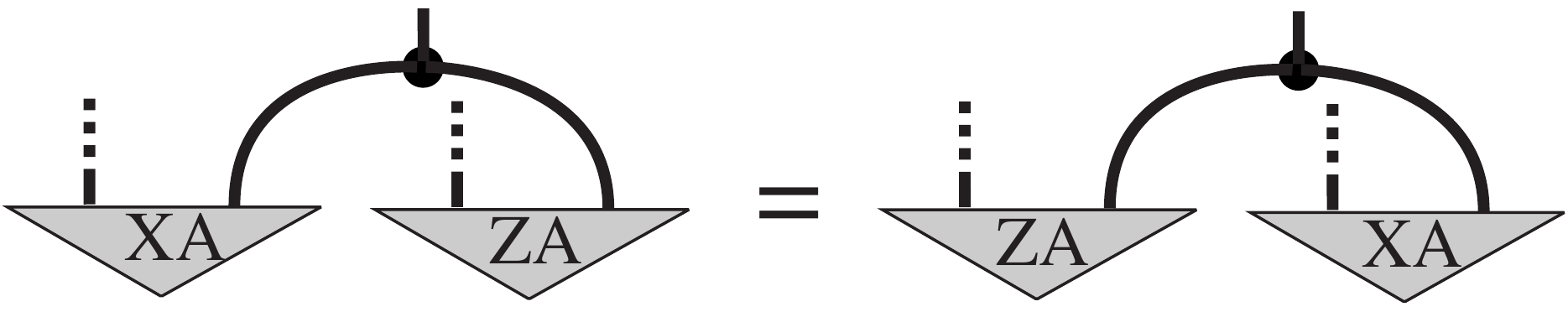,width=190pt}}\,.
\eeq
Multiplication is also commutative if one or both of the states are replaced by conditional states.
\end{proposition}
\bpf
By the spider theorem and Eq.~(\ref{eq:Ccalcmofifier}) we have
\[
\raisebox{-0.44cm}{\epsfig{figure=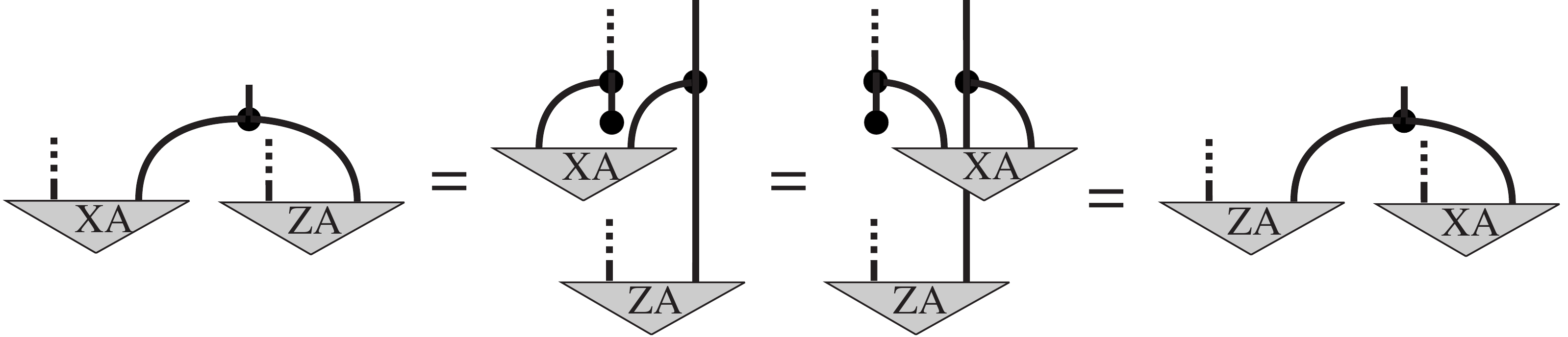,width=325pt}}\,.
\]
\endproof\newline

 Note that it could however still be possible that the Frobenius structure itself is not commutative, but just acts commutative on the joint and marginal probabilities under consideration.  E.g.~the Q${}_{1/2}$-calculus of  Section \ref{Sec:PresQonehalf} below  when all relevant density operators commute.

  \begin{proposition}\label{Prop:ModCom}
In a classical Bayesian graphical calculus, composition of modifiers on an object is commutative:
\beq
\raisebox{-0.34cm}{\epsfig{figure=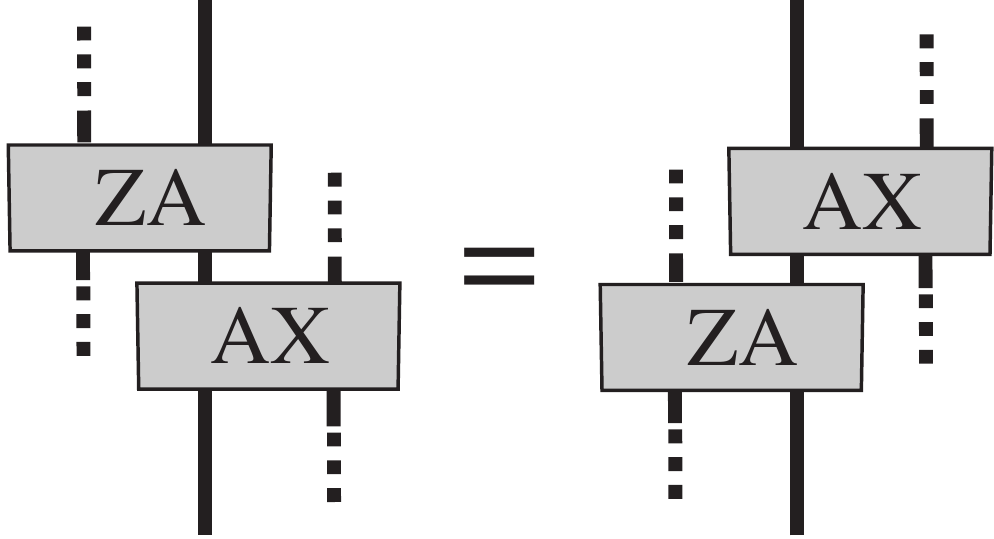,width=101pt}}\!\!\!\! .
\eeq
\end{proposition}
\bpf
By Eqs.~(\ref{eq:daggerFrobeniusstructure}), (\ref{eq:slidinggen}), (\ref{def:slidinginverseclass}) we have:
\begin{center}\raisebox{-0.30cm}{\epsfig{figure=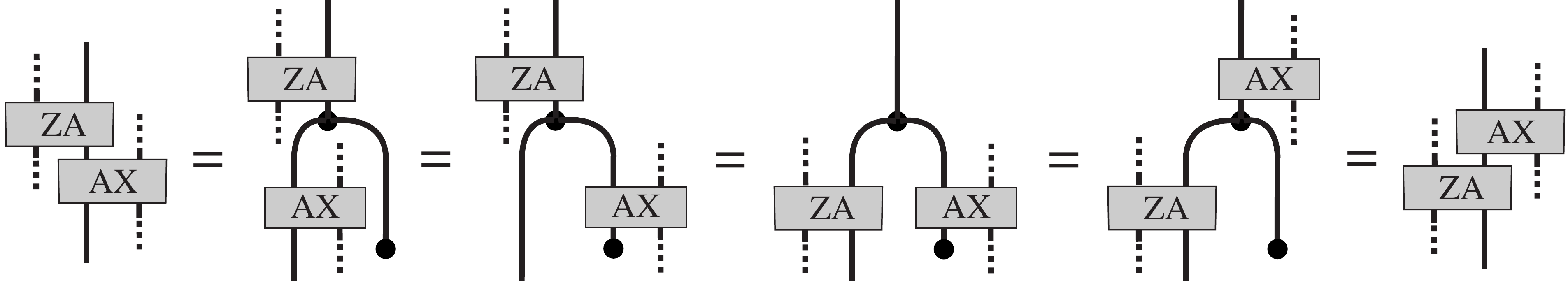,width=387pt}}\ .\end{center}
\endproof\newline

For a classical Bayesian calculus,  conditional states  have the form:
\beq\label{eq:C-Bayespre}
\raisebox{-0.40cm}{\epsfig{figure=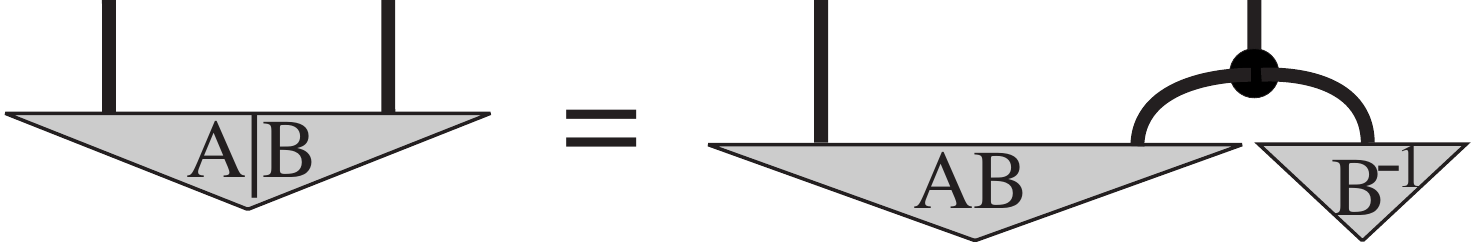,width=150pt}}\ ,
\eeq
and   Bayes' theorem, Eq.~(\ref{eq:Bayes}), has the form
\beq\label{eq:C-Bayes}
\raisebox{-0.40cm}{\epsfig{figure=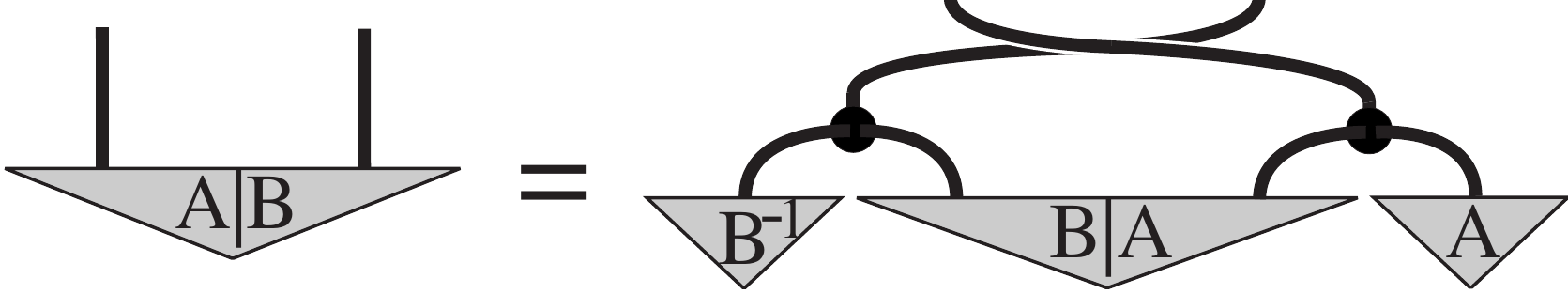,width=170pt}}\ .
\eeq
By virtue of the multiplicative commutativity, the order in which the states are  `Frobenius-multiplied'  doesn't matter (unlike the quantum generalization, as we will see).

This is an abstract characterization of classical Bayesian inference.  We now present a couple of concrete realizations of this calculus.  We shall thereby see how the abstract characterization avoids the conventional elements of the concrete realizations.

\subsection{Representations of the classical Bayesian graphical calculus}
\begin{example}\textbf{Standard probability theory.} \label{ex:clascalc}
Standard probability theory constitutes a special case of a classical Bayesian calculus.
The objects are natural numbers and the morphisms from $n$ to $m$ are the $m\times n$ positive-valued matrices (consequently the points are column vectors and their daggers are row vectors).  Composition is matrix product, and the tensor product is the matrix tensor product.
It follows that we have
\beq
\raisebox{-0.16cm}{\epsfig{figure=marginalonly.pdf,width=22pt}}\ :\II \to A=\textbf{p}:= (p_1,p_2,\dots,p_n)\, ,
\eeq
where $(p_1,p_2,\dots,p_n)$ denotes a column vector. The unit is
\beq
\raisebox{-0.24cm}{\epsfig{figure=unit.pdf,width=6pt}}\,: \II\to A =\textbf{u}:=(1,1,\dots,1)\, ,
\eeq
which implies that the co-unit must be
\beq
\raisebox{-0.24cm}{\epsfig{figure=unitdag.pdf,width=6pt}}:A\to\II =\textbf{u}^T = \textrm{row}(1,1,\dots,1),
\eeq
where $T$ denotes matrix transposition.
The counit acting on a point gives the sum of the coefficients of the associated vector
\beq
\raisebox{-0.16cm}{\epsfig{figure=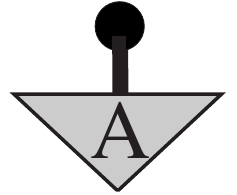,width=24pt}}\ = \textbf{u}^T \textbf{p}
=\sum_{j=1}^n p_j.
\eeq
It follows that a normalized state in the Bayesian graphical calculus (cf.~condition BC1) here corresponds to a positive vector with coefficients that sum to 1:
\beq
\raisebox{-0.16cm}{\epsfig{figure=marginalonly.pdf,width=22pt}}\ :\II \to A=(p_1,p_2,\dots,p_n) \textrm{  such that  } \sum_{j=1}^n p_j=1.
\eeq
In other words, normalized states for an object are probability distributions over the set $\{1,\dots,n\}$.  Normalized states on a composite object (nm) are simply probability distributions over the set $\{1,\dots,nm\}$,
\beq
\raisebox{-0.16cm}{\epsfig{figure=joint.pdf,width=50pt}}:\II \to A\otimes B=\textbf{p}:=
(p_{i,j} |i\in\{1,\dots,n\},j\in\{1,\dots,m\})
\textrm{  such that  } \sum_{i=1}^n \sum_{j=1}^m p_{i,j}=1.
\eeq
when the joint state is a tensor product of a state $(p_1,p_2,\dots,p_n)$ on $A$ and a state $(q_1,q_2,\dots,q_m)$ on $B$,
\beq
\raisebox{-0.16cm}{\epsfig{figure=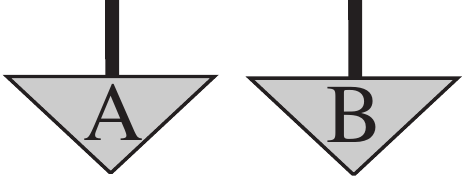,width=50pt}}\ :\II \to A\otimes B = (p_i q_j |i\in\{1,\dots,n\},j\in\{1,\dots,m\})\,,
\eeq
we say that it is \em uncorrelated \em.

The Frobenius multiplication is the $n\times n^2$ matrix $M:=\textrm{row}(M^{(1)},M^{(2)},\dots,M^{(n)})$ where $M^{(k)}$ is the $n\times n$ matrix which is zero everywhere except at the $k$th diagonal element, where it is one, 
\beq
\raisebox{-0.20cm}{\epsfig{figure=multiplication.pdf,width=28pt}}\ :A\otimes A\to A =M:=\left(
  \begin{array}{cccccccccccccc}
1 & 0 & \dots & 0 && 0 & 0 & \dots & 0 &  & 0 &  0 & \dots & 0 \\
  0 & 0 & \dots & 0 && 0 & 1 & \dots & 0 &  & 0 &  0 & \dots & 0 \\
  \vdots &  &  &  && \vdots &  &  &  & \dots  & \vdots &  &  &  \\
  0 & 0 & \dots & 0 && 0 & \dots & 0 & 0 &  & 0 &  0 & \dots & 1
   \end{array}
\right).
\eeq
Therefore, composing an arbitrary point on $A\otimes A$ with the Frobenius multiplication yields
\beq
\raisebox{-0.42cm}{\epsfig{figure=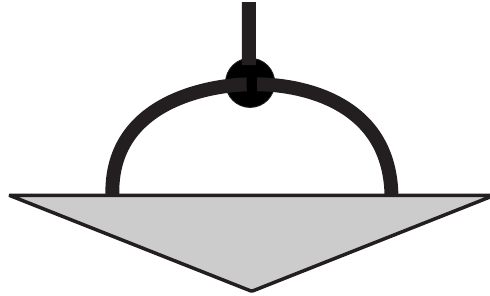,width=50pt}}\ =M(p_{i,i'}|i,i'\in\{1,\dots,n\})=(p_{i,i}|i\in\{1,\dots,n\}).
\eeq
If the point on $A\otimes A$ is a product of a state $(p_1,p_2,\dots,p_n)$ on $A$ and a different state $(p'_1,p'_2,\dots,p'_n)$ also on $A$, then composing with the Frobenius multiplication yields
\beq
\raisebox{-0.16cm}{\epsfig{figure=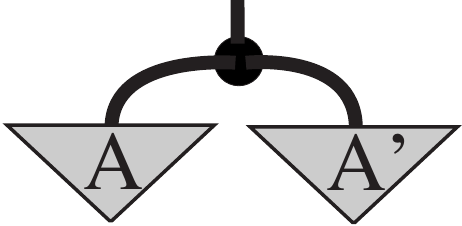,width=50pt}}\ =M(p_{i}p'_{i'}|i,i'\in\{1,\dots,n\})=(p_{i}p'_{i}|i\in\{1,\dots,n\}).
\eeq
This is simply the component-wise product of the input vectors.

It follows from the above that the Frobenius co-multiplication is the $n^2\times n$ matrix which is the matrix transpose of $M$.
\beq
 \raisebox{-0.20cm}{\epsfig{figure=comultiplication.pdf,width=28pt}}\ :A\to A\otimes A =M^T.
\eeq
Composing an arbitrary state on $A$ with the comultiplication yields
\beq
\raisebox{-0.30cm}{\epsfig{figure=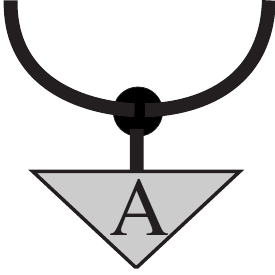,width=28pt}}\ =M(p_{i}|i\in\{1,\dots,n\})=(p_{i}\delta_{i,i'}|i,i'\in\{1,\dots,n\}).
\eeq
where
\beq
\delta_{i,i'}:=\left\{\begin{array}{cll}
                1 & \textrm{  if } & i=i'\\
                0 & \textrm{  if } & i\ne i'
              \end{array}\right.
\eeq
is the Kronecker delta. The comultiplication can therefore be understood as a classical \em broadcasting map \em \cite{Broadcast}  (see Section \ref{Sec:PresQonehalf} below).
It is tedious but straightforward to verify that these definitions of unit, co-unit, multiplication and co-multiplication yield a Frobenius structure.

The cups and caps of the compact structure induced by this Frobenius structure are as follows:
\beq
\raisebox{-0.18cm}{\epsfig{figure=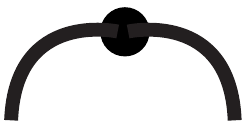,width=28pt}}  :A\otimes A\to I = \textbf{e} := \left(
  \begin{array}{cccccccccccccc}
1 & 0 & \dots & 0 && 0 & 1 & \dots & 0 &  & 0 &  0 & \dots & 1 \\
   \end{array}
\right),
\eeq
and
\beq
\raisebox{-0.18cm}{\epsfig{figure=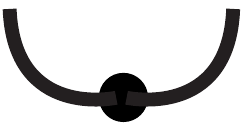,width=28pt}} :I \to A\otimes A = \textbf{e}^T = (\delta_{i,i'} | i,i'\in\{1,\dots,n\}).
\eeq
The latter can be interpreted, modulo normalization, as the probability distribution expressing perfect correlation.

We now demonstrate that standard probability theory is a representation of the classical Bayesian graphical calculus.

The marginal state on $A$ of a joint state on $A \otimes B$ associated with probability distribution
\[
(p_{i,j}|i\in\{1,\dots,n\},j\in\{1,\dots,m\})
\]
 is simply the marginal distribution on $A$, that is,
\beq
\raisebox{-0.16cm}{\epsfig{figure=marginal.pdf,width=98pt}}:\II \to A=(p_i=\sum_j p_{i,j}|i\in\{1,\dots,n\}).
\eeq

If one defines the modifier associated with a state $(p_1,\dots,p_n)$ on $A$ through Eq.~(\ref{eq:Ccalcmofifier}), then it is represented by the $n\times n$ matrix
\beq \label{eq:probmodifier}
\raisebox{-0.20cm}{\epsfig{figure=Ccalc.pdf,width=126pt}}:A \to A = M[(p_1,p_2,\dots,p_n)\otimes I]=\left(
                                                         \begin{array}{cccc}
                                                           p_1 & 0 & \dots & 0 \\
                                                           0 & p_2 & \dots & 0 \\
                                                           \vdots &  &  &  \\
                                                           0 & 0 & \dots & p_n \\
                                                         \end{array}
                                                       \right)
\eeq

The Frobenius inverse of a marginal state $(p_1,\dots,p_n)$ on $A$ is the vector
\beq
\raisebox{-0.16cm}{\epsfig{figure=Qinv.pdf,width=22pt}}:\II \to A = (r_1,\dots,r_n)
\eeq
where
\beq
r_i:=\left\{\begin{array}{lcc}
      p_i^{-1} &\textrm{  if  } & p_i\ne 0 \\
      0 & \textrm{  if  } & p_i=0
    \end{array}\right.\, .
\eeq
Furthermore, one easily verifies that the inverse of the modifier in Eq.~(\ref{eq:probmodifier}) is simply the matrix $diag(r_1,r_2,\dots,r_n)$, so that property BC3 is indeed satisfied.

It follows that the conditional state on $A\otimes B$ that arises from the joint state on $A\otimes B$ is simply the ordered set of conditional probability distributions that arise from the joint distribution $p_{i,j}$,
that is,
\beq
\raisebox{-0.34cm}{\epsfig{figure=conditional.pdf,width=127pt}}:\II \to A\otimes B= (p_{i|j}|i\in\{1,\dots,n\},j\in\{1,\dots,m\})
\eeq
where
\beq
p_{i|j}:=  \left\{\begin{array}{cll}
            p_{i,j}p_j^{-1} &\textrm{  if  } & p_j\ne 0 \\
            0 &\textrm{  if  } & p_j=0
          \end{array}\right.\, ,
\eeq
is a probability distribution over $i\in \{1,\dots,n\}$, labeled by $j\in\{1,\dots,m\}$. Note that  because $p_j:=\sum_i p_{i,j}$ in this expression, we can infer the normalization of $p_{i|j}$,
\beq
\sum_{i=1}^n p_{i|j}=1 \textrm{  for all  } j\in\{1,\dots,m\}.
\eeq
Consequently the composition of the conditional state with the co-unit on $A$ is indeed the unit on $B$,
\beqn
\raisebox{-0.30cm}{\epsfig{figure=condstatenorm.pdf,width=76pt}}
&\!\!=\!\!&[\textrm{row}(1,1,\dots,1)\otimes I] (p_{i|j}|i\in\{1,\dots,n\},j\in\{1,\dots,m\}) \\
&\!\!=\!\!&(\sum_i p_{i|j}|j\in\{1,\dots,m\})=(1,1,\dots,1)\, .
\eeqn

In a slight abuse of notation, we can use $A$, $B$ and $C$ not only to denote the objects in our category but also to denote random variables associated with these.  For instance, we take $A$ to denote the random variable taking values from the set $\{1,\dots,n\}$ where $n$ is the natural number associated with the categorical object $A$.  We can also follow a standard notation and write
\beqn
p(A)&\!\!:=\!\!& (p_a|a\in\{1,\dots,n\})\, ,\\
p(A,B)&\!\!:=\!\! &(p_{a,b}|a\in\{1,\dots,n\},b\in\{1,\dots,m\})\, ,\\
p(A|B)&\!\!:=\!\! &(p_{a|b}|a\in\{1,\dots,n\},b\in\{1,\dots,m\})\, ,
\eeqn
 etcetera.  We can then write many equations in a simple form.  For instance,  the Bayes' rule for classical Bayesian graphical calculi as in Eq.~(\ref{eq:C-Bayes}),  takes the form
\beq
p(A|B) =  \frac{p(B|A) p(A)}{p(B)}\,,
\eeq
where this is understood to be an equality that holds component by component.

%
%
\end{example}





\newcommand{\boxcirc}{{\,\square\hspace{-2.45mm}\mbox{\raisebox{0.35mm}{$\circ$}}\hspace{1.2mm}}}

\begin{example}\textbf{Alternative representations.}
Here everything is defined as it was before -- objects are natural numbers, morphisms are positive-valued matrices, composition is the matrix product and tensor product is the matrix tensor product -- except that the underlying notions of scalar addition and multiplication are modified.  The new operations, denoted by $\boxplus$ and $\boxdot$ respectively, can be defined for an arbitrary pair $s,t$ of scalars as follows.  For any function $f$ that is bijective and hence invertible on the positive reals, they are
\beq
s \boxplus t = f(f^{-1}(s)+f^{-1}(t)), \quad \quad
s \boxdot t = f(f^{-1}(s)f^{-1}(t)).
\eeq
One easily verifies that these two operations are commutative and associative and obey the distributive law:
\[
s\boxdot(t_1\boxplus t_2)=(s\boxdot t_1)\boxplus(s\boxdot t_2)\,.
\]
The unit for the new notion of addition, denoted $0_{\boxplus}$ and satisfying $s\boxplus 0_{\boxplus} = s$ for all $s$, is
\beq
0_{\boxplus} = f(0),
\eeq
while the unit for the new notion of multiplication, denoted $1_{\boxdot}$ and satisfying $s\boxdot 1_{\boxdot} = s$ for all $s$, is
\beq
1_{\boxdot} = f(1).
\eeq
The new product of two matrices $M$ and $N$, denoted $M \boxcirc N$, is defined accordingly:
\begin{eqnarray}
[M \boxcirc N]_{ij} = \boxplus_k ([M]_{ik} \boxdot [N]_{kj}),
\end{eqnarray}
as is the new tensor product of two matrices, $M$ and $N$, denoted $M \boxtimes N$,
\begin{eqnarray}
[M \boxtimes N]_{ik,jl} = [M]_{ij} \boxdot [N]_{kl}.
\end{eqnarray}
The Frobenius multiplication, co-multiplication, unit and co-unit are defined as before, but with the scalars 0 and 1 replaced by $0_{\boxplus}$ and $1_{\boxdot}$.  By construction, for every monotonic function $f$, we obtain a representation of a Bayesian graphical calculus.
\end{example}

It is useful to consider an example of this sort of alternative to the standard probability representation.

\begin{example}\textbf{The negative logarithm of probability representation.}\label{ex:clascalc}
Consider the case where the monotonic function $f$ is the negative natural logarithm (the generalization to an arbitrary base is straightforward),
\beq
f(s)=-\ln{s}, \quad \quad f^{-1}(s)=e^{-s},
\eeq
so that
\beq
s \boxplus t = -\ln (e^{-s}+e^{-t}), \quad \quad
s \boxdot t = s+t.
\eeq
We then have
\footnote{As an aside, there is often a subtelty concerning inverse.
The new multiplicative inverse of a scalar $s$, denoted $s^{\boxminus 1}$, must satisfy $s \boxdot s^{\boxminus 1}= 1_{\boxdot}$.  It follows that
\beq
s^{\boxminus 1}= - s.
\eeq
However, the new additive inverse of a scalar $s$, denoted $\boxminus s$ must satisfy $s \boxplus \boxminus s= 0_{\boxplus}$, which implies that $\boxminus s = s - \ln (-1)$, which is undefined.  Consequently, there are no additive inverses in this new calculus.}
\beqn
{[M \boxcirc N]_{ij}} \!\!&=&\!\! -\ln [\sum_k e^{-([M]_{ik}+[N]_{kj})}], \\
{[M \boxtimes N]_{ik,jl}} \!\!&=&\!\! [M]_{ij} + [N]_{kl}, \\
0_{\boxplus} \!\!&=&\!\! \infty, \\
1_{\boxdot} \!\!&=&\!\! 0\,.
\eeqn

Now consider a state $(s_1,s_2,\dots,s_n)$.  For it to be normalized, it must satisfy the condition
\beq
(1_{\boxdot},1_{\boxdot},\dots,1_{\boxdot})^T \boxcirc (s_1,s_2,\dots,s_n)= 1_{\boxdot},
\eeq
which implies that
\begin{eqnarray}
-\ln [\sum_k e^{-s_k}]=0\\
\sum_k e^{-s_k}=1.
\end{eqnarray}
It follows that the components of the vector $(s_1,s_2,\dots,s_n)$ are the negative logarithms of the components of a probability distribution $(p_1,p_2,\dots,p_n)$,
\beq
\forall k: s_k = -\ln p_k.
\eeq
In this new calculus, an impossible value of $k$ (one for which $p_k=0)$ is represented by $s_k=\infty$, while a certain value (one for which $p_k=1)$ is represented by $s_k=0$.

We can represent these vectors as $s(A), s(A,B), s(A|B)$ and so forth.  We find that we have
\beq
s(A|B)=s(A,B)-s(B),
\eeq
which is understood component-wise, that is,
\beq
s(A|B):=(s_{a|b}|a\in\{1,\dots,n\},b\in\{1,\dots,m\})
\eeq
where
\beq \label{condneglogprob}
s_{a|b}:= - \ln p_{a|b}.
\eeq
The Bayes' rule takes the form
\beq \label{neglogprobBayes}
s(A|B)=s(B|A)+s(A)-s(B).
\eeq
\end{example}

One has a choice in representing degrees of belief.  It can be done with probabilities, but it can also be accomplished with negative logarithms of probabilities, or indeed any monotonic function of probabilities.  It is a matter of convention only which is chosen.  An argument to this effect was made by R.~T.~Cox in the context of an axiomatization of Bayesian inference  \cite{Cox}.  We have supported Cox's conclusion by demonstrating that an abstract graphical characterization of Bayesian inference shows certain aspects of the standard probability calculus to be merely conventional.

Finally, note that by taking the usual inner product of the vector $s(A):=(s_1,s_2,\dots,s_n)$ of negative logarithms of probabilities with the vector $p(A):=(p_1,p_2,\dots,p_n)$ of probabilities, one obtains the \em Shannon entropy \em of the probability distribution $p(A)$, denoted $S(A)$,
\beq
S(A):= \sum_k p_k s_k = -\sum_k p_k \ln p_k.
\eeq
One can similarly obtain the joint entropy as
\beq
S(A,B):=\sum_{i,j} p_{i,j} s_{i,j} = -\sum_{i,j} p_{i,j} \ln p_{i,j},
\eeq
and the conditional entropy as
\beq
S(A|B):=\sum_{i,j} p_{i,j} s_{i|j} = -\sum_{i,j} p_{i,j} \ln p_{i|j}.
\eeq
Noting the marginal entropy can also be obtained by averaging over the joint distribution,
\beq
\sum_{i,j} p_{i,j} s_{i} =\sum_{i} p_{i} s_{i}=S(A),
\eeq
it follows that any expression that holds among joints, marginals and conditionals for negative logarithms of distributions (i.e. among $s_{i,j}$, $s_{i}$, $s_{i|j}$ etcetera) also holds among the joint, marginal and conditional \em entropies. \em  For instance, Bayes' rule in terms of negative logarithms of probabilites, Eq.~(\ref{neglogprobBayes}), implies the analogous relation among entropies
\beq \label{neglogprobBayes}
S(A|B)=S(B|A)+S(A)-S(B).
\eeq
Thus the classical Bayesian graphical calculus has the power to represent relations among classical entropies.

 In more abstract terms one realizes this by considering the $p$- and the $s$-calculi as two distinct  composition and tensor structures on morphisms, above denoted by $(\circ, \otimes)$ and $(\!\boxcirc, \boxtimes)$,  were $\otimes$ and $\boxtimes$  do coincide on objects.  One then post-composes both sides in Eq.~(\ref{eq:C-Bayes}), realized in the  $s$-calculus,  with the normalized joint state of the $p$-calculus by means of the $\circ$-composition. That is,
\beq\label{eq:entropy}
\raisebox{-0.860cm}{\epsfig{figure=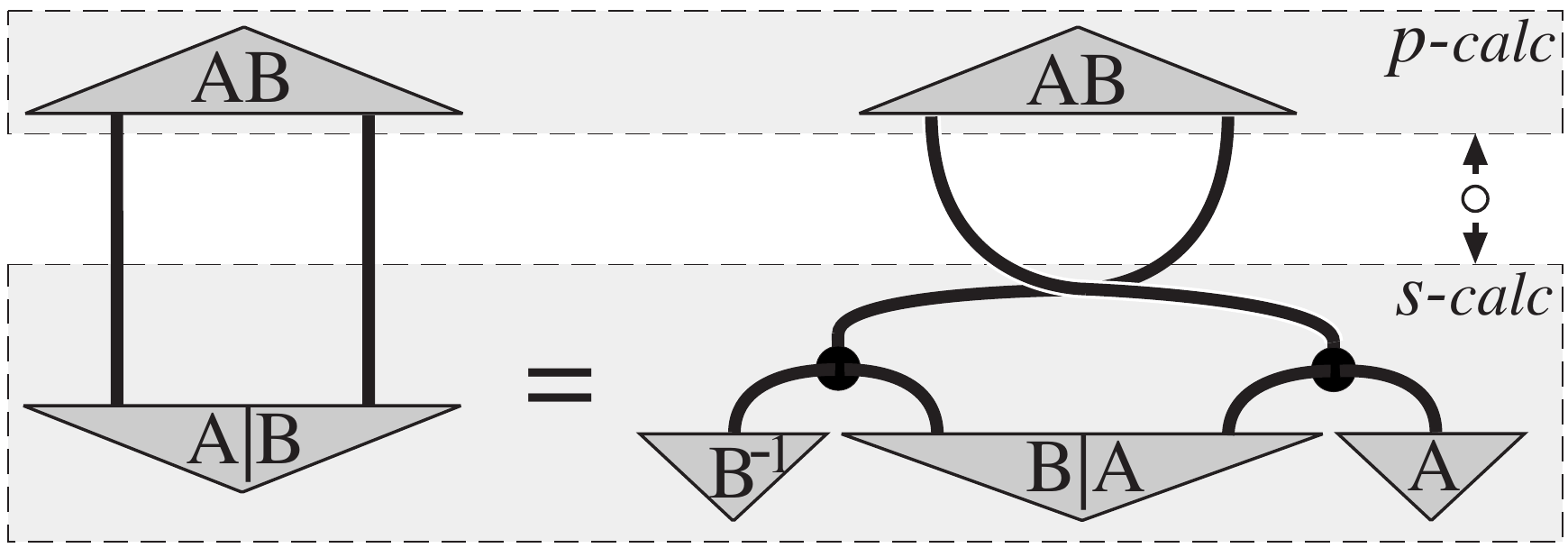,width=170pt}}\ .
\eeq
In other words, a `$p$-operation'
\[
p(A, B)^T\circ-: {\bf C}(A\otimes B, \II)\to {\bf C}(\II, \II)
\]
is applied to both sides of an equation between $s$-terms in ${\bf C}(A\otimes B, \II)$. Since such a  p-operation can be applied to both sides of any equation between $s$-terms in classical Bayesian calculus, such an equation always results in a corresponding statement about classical entropies.

\subsection{Q${}_{1/2}$-calculus}\label{sec:Qcalc}

Particular cases of Bayesian graphical calculi arise by choosing a specific construction of the modifiers.

\begin{definition}\label{def:Qcal}
A Bayesian graphical calculus is a \em  Q${}_{1/2}$-calculus \em when modifiers are of the form:
\beq\label{eq:formQmodifiers}
\epsfig{figure=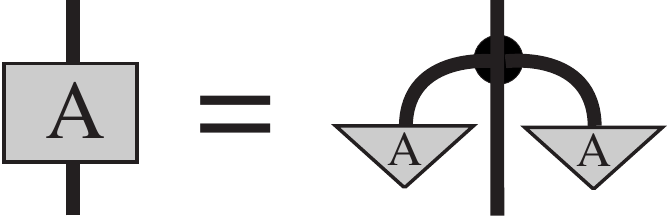,width=68pt}\,.
\eeq
\end{definition}

In this definition we introduced points $\ \raisebox{-0.16cm}{\epsfig{figure=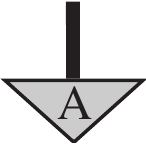,width=15pt}}\ :\II\to A $ that are distinguished from the marginal states by being denoted by smaller triangles.
By Eq.~(\ref{eq:modifier}) and the spider theorem, these must obey:
\[
\ \raisebox{-0.16cm}{\epsfig{figure=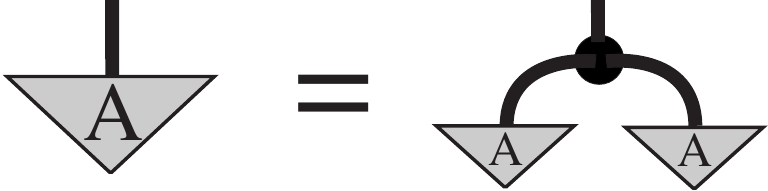,width=78pt}}\ ,
\]
that is, they are the \em square roots of marginal states \em relative to the multiplication operation of the Frobenius structure.  Again by the spider theorem
we also have the following lemma:

\begin{lemma}\label{lm:consistencysqrt}
If Eqs.~(\ref{eq:formQmodifiers}) and (\ref{eq:modifier}) hold, and if $\ \raisebox{-0.16cm}{\epsfig{figure=Qinvsmall1.pdf,width=15pt}}\ $ has an inverse $\ \raisebox{-0.16cm}{\epsfig{figure=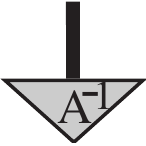,width=15pt}}\ $,  then
\beq
\ \raisebox{-0.26cm}{\epsfig{figure=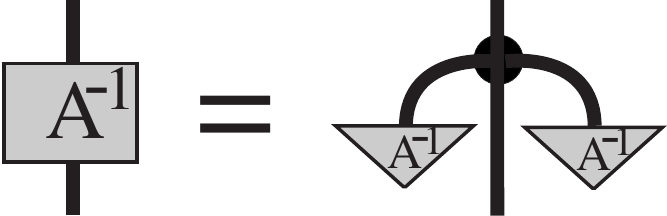,width=68pt}}\ \qquad\qquad
\mbox{and}\qquad\qquad\ \raisebox{-0.16cm}{\epsfig{figure=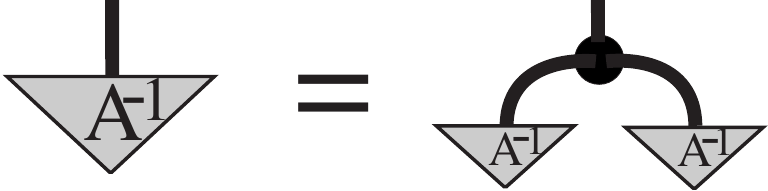,width=78pt}}\ ,
\eeq
and hence, the consistency condition Eq.~(\ref{eq:consistency}) is also automatically satisfied.
\end{lemma}

Also that modifiers are self-transposed now comes for free:
\begin{lemma}\label{prop:Qtranspinv}
Modifiers of the form Eq.~(\ref{eq:formQmodifiers}) are automatically self-transposed.
\end{lemma}
\bpf
We have:
\begin{center}
\epsfig{figure=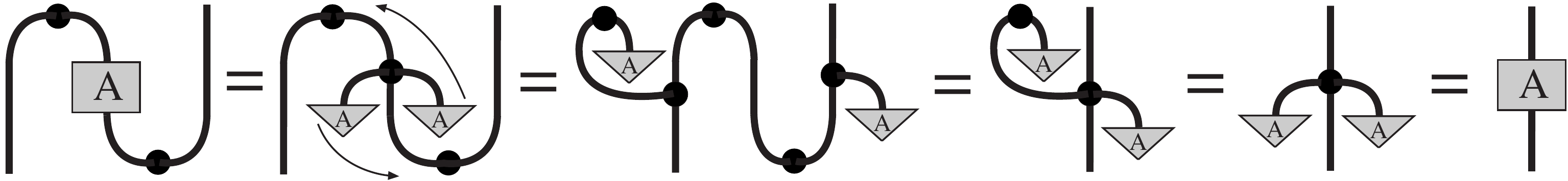,width=300pt}
\end{center}
where the 2nd and 3rd step use the spider theorem, and the 4th one uses commutativity of the caps.
\endproof\newline

In terms of the canonical dagger Frobenius structure on $A\otimes B$ we have:
\beq
\raisebox{-0.24cm}{\epsfig{figure=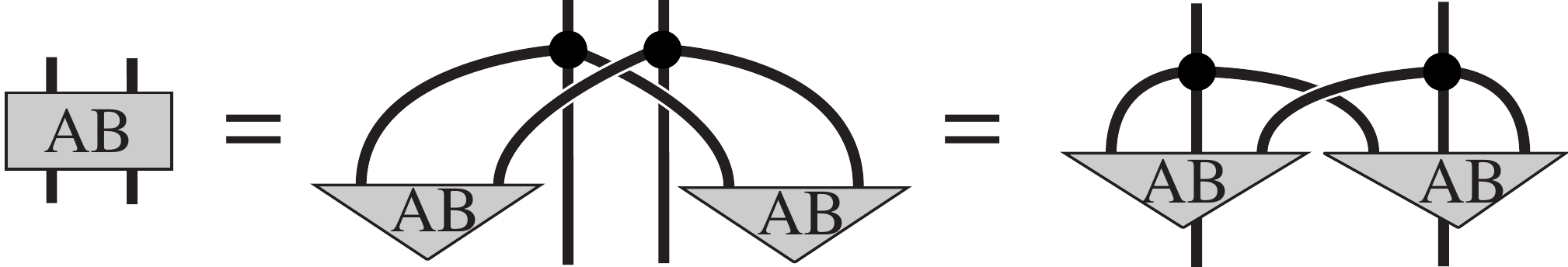,width=200pt}}
\eeq
where:
\beq
\raisebox{-0.24cm}{\epsfig{figure=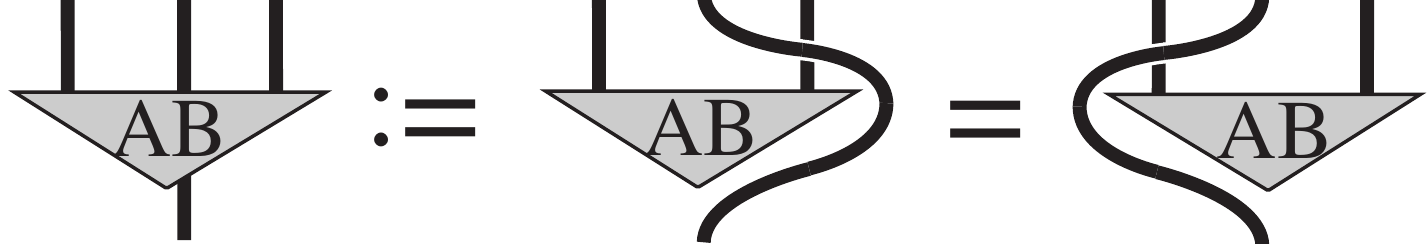,width=150pt}}\ ,
\eeq
which follows by naturality of symmetry.

We will assume the existence of inverses of the square roots of marginals for Q${}_{1/2}$-calculi, and consequently, by Lemma \ref{lm:consistencysqrt}, inverses of the marginals themselves will also exist in Q${}_{1/2}$-calculi.

For Q${}_{1/2}$-calculi, the Bayesian update law Eq.~(\ref{eq:Bayes}) becomes:
\beq\label{eq:Q-Bayes}
\raisebox{-0.64cm}{\epsfig{figure=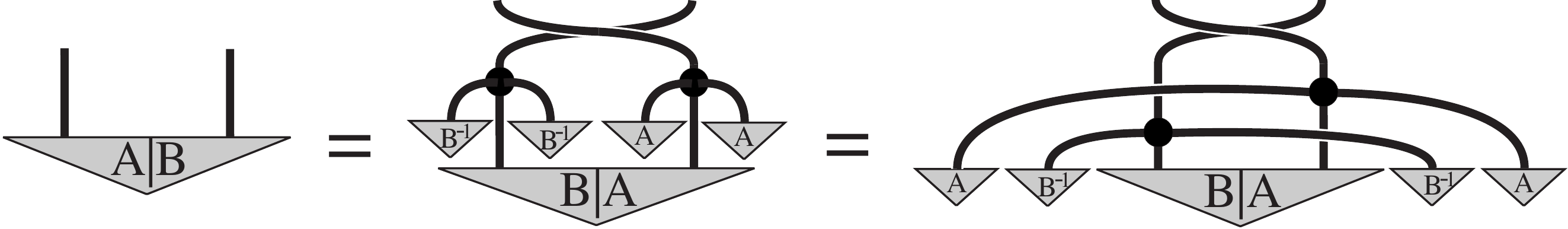,width=269pt}}\,.
\eeq
In the final expression of Eq.~(\ref{eq:Q-Bayes}), the order of the two small triangles on the left could be reversed because they are not connected to each other by a spider.  The same is true of the two small triangles on the right.

\subsection{A representation of the Q${}_{1/2}$-calculus}\label{Sec:PresQonehalf}


Leifer's conditional density operator calculus \cite{Leifer1,Leifer2,LeiferSpekkens} provides an example of a  Q${}_{1/2}$-calculus.   The objects are natural numbers and the morphisms from $n$ to $m$ are the linear maps from $\mathcal{L}(\mathbb{C}^m)$ to $\mathcal{L}(\mathbb{C}^n)$ where $\mathcal{L}(\mathbb{C}^m)$ is the space of linear operators on an $m$-dimensional complex Hilbert space.  Composition and tensor product of these maps are just the normal such notions. (Note that each such linear map can be represented as an $n^2\times m^2$ matrix, in which case composition is matrix product, and the tensor product is the matrix tensor product.)

We take  the point $\raisebox{-0.16cm}{\epsfig{figure=marginalonly.pdf,width=22pt}}$ to be a density operator $\rho_A \in \mathcal{L}(\mathbb{C}^n)$ and the point $\raisebox{-0.16cm}{\epsfig{figure=joint.pdf,width=50pt}}$ to be the joint density operator $\rho_{AB} \in \mathcal{L}(\mathbb{C}^n)\otimes \mathcal{L}(\mathbb{C}^m)$.   We take the Frobenius multiplication $\ \raisebox{-0.20cm}{\epsfig{figure=multiplication.pdf,width=28pt}}\ $ to be the (non-commutative) operator product of density operators, and hence the
identity operator $1_A$ is its unit $\ \raisebox{-0.24cm}{\epsfig{figure=unit.pdf,width=6pt}}\ $.   The co-unit $\ \raisebox{-0.24cm}{\epsfig{figure=unitdag.pdf,width=6pt}}\ $ is the trace operation (which is indeed the adjoint  to the identity operator when taken in a suitable manner \cite{Selinger}).  It follows that a normalized state in this graphical calculus is simply a density operator with trace one.

Given that the co-unit is the trace operation and recalling that the Frobenius co-multiplication must satisfy co-unitality,
\beq
\raisebox{-0.60cm}{\epsfig{figure=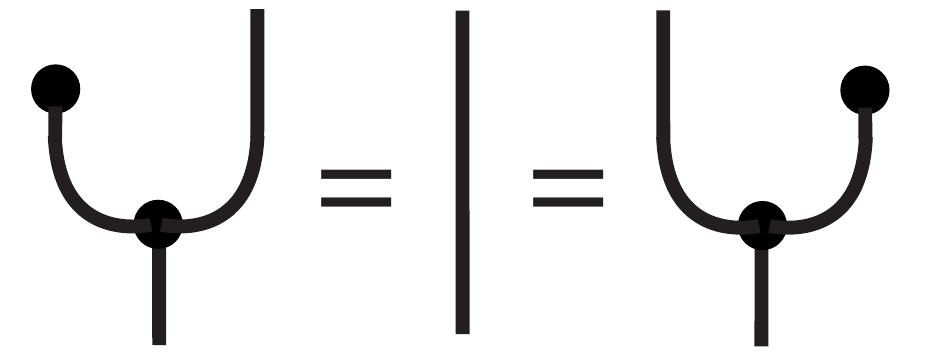,width=95pt}},
\eeq
we conclude that the Frobenius co-multiplication is a broadcasting operation, that is, a linear map $\mathcal{B}: \mathcal{L}(\mathbb{C}^n) \to \mathcal{L}(\mathbb{C}^n) \otimes \mathcal{L}(\mathbb{C}^n)$ satisfying
\beq\label{def:broadcasting}
(\mathrm{tr}_A\otimes \textrm{id}_A)\circ \mathcal{B} = \textrm{id}_A =(\textrm{id}_A \otimes \mathrm{tr}_A)\circ \mathcal{B}\,,
\eeq
where $\textrm{id}_A$ is the identity map on $\mathcal{L}(\mathbb{C}^n)$.
Given the quantum no-broadcasting theorem \cite{Broadcast}, this mathematical operation is necessarily non-physical, i.e.~it is not a completely positive map.

 If $\{ |i\rangle \;|\; i\in\{1,\dots,n\} \}$ is a basis of $\mathbb{C}^n$ so that $\{ |i\rangle \langle j| \mid i,j\in\{1,\dots,n\} \}$ is a basis for the operator space $\mathcal{L}(\mathbb{C}^n)$, then $\mathcal{B}$ can be defined by
\beq
\mathcal{B}(|i\rangle \langle j|)= \sum_k |i\rangle \langle k| \otimes |k\rangle \langle j|,
\eeq
which, despite appearances, is basis-independent (In Section \ref{sec:Qcalc_in_CPM}, we present a diagrammatic representation of this operation in the dCC \textbf{FdHilb}.).  It is straightforward to verify that this broadcasting operation is associative and that the interplay between it and the operator product satisfies the Frobenius law.  Consequently, we have a representation of a non-commutative dagger Frobenius structure.

The `cup' of the compact structure induced by this Frobenius structure is the partial transpose of the maximally entangled state, that is, the positive operator $\frac{1}{n}\sum_{i,j} |i\rangle \langle j| \otimes |j\rangle \langle i|$.

The Frobenius inverse of a state on $A$, $\ \raisebox{-0.16cm}{\epsfig{figure=Qinv.pdf,width=22pt}}\ $,  is the inverse of the associated density operator (more precisely, the inverse over the support), namely $\rho_A^{-1}$.  The square root of a state on $A$,
    $\ \raisebox{-0.12cm}{\epsfig{figure=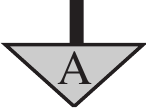,width=15pt}}\ $, is the square-root of the associated density operator, namely $\sqrt{\rho_A}$.  The modifier $\ \raisebox{-0.25cm}{\epsfig{figure=Qcalc.pdf,width=68pt}}\ $ is the completely positive map $\sqrt{\rho_A}( - )\sqrt{\rho_A}$.   The consistency condition Eq.~(\ref{eq:consistency}) is now also clearly satisfied.

 Given that the co-unit is the trace operation, marginal states arise by tracing out a system on a joint density operator and therefore correspond to reduced density operators.  The conditional state $\raisebox{-0.16cm}{\epsfig{figure=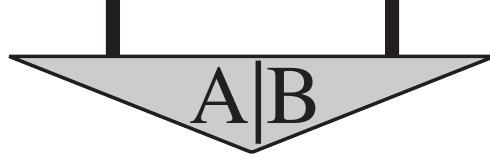,width=50pt}}$ is Leifer's conditional density operator \cite{Leifer1,Leifer2}, that is, a positive operator $\rho_{A|B} \in \mathcal{L}(\mathbb{C}^n)\otimes \mathcal{L}(\mathbb{C}^m)$ such that $ \mathrm{Tr}_A[\rho_{A|B}]=1_B. $
   Note that the commutation of the compact structure, Eq.~(\ref{eq:compactFROBcoherence}), corresponds to the cyclic property of the trace, i.e.~$\mathrm{tr}(\rho_A \rho'_A)=\mathrm{tr}(\rho'_A \rho_A)$.

Applying this translation to the diagrammatic Eq.~(\ref{eq:Q-Bayes}), we obtain the Bayesian update rule as an identity between operators,
\beq\label{eq:quantumBayes}
\rho_{A|B}=(1_B\otimes\sqrt{\rho_A})(\sqrt{\rho_B}^{-1}\!\!\otimes 1_A)
\rho_{B|A}
(\sqrt{\rho_B}^{-1}\!\!\otimes 1_A)(1_B\otimes\sqrt{\rho_A})
\eeq
where $\rho_{B|A}$ is the conditional density operator associated with $\raisebox{-0.16cm}{\epsfig{figure=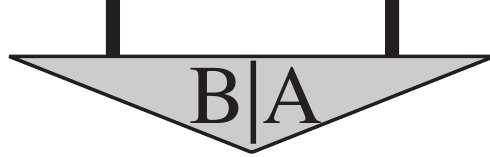,width=50pt}}$.  To see that this is indeed the translation, it is useful to consider the diagram that incorporates each element of Eq~(\ref{eq:quantumBayes}) explicitly, namely,
\beq\label{eq:DetailedquantumBayes}
\raisebox{-1.2cm}{\epsfig{figure=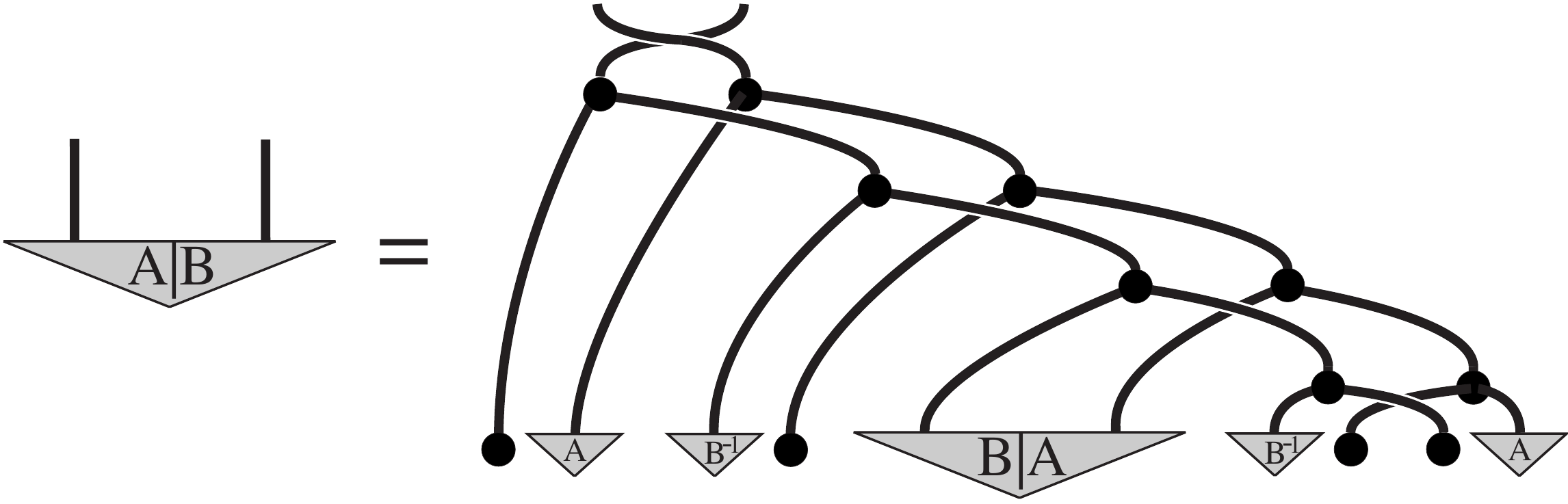,width=240pt}}\ ,
\eeq
then note that the latter can be reduced to Eq.~(\ref{eq:Q-Bayes}) by application of the spider theorem.
The fact that we require a swap map is due to our diagrammatic convention to interpret in $\raisebox{-0.16cm}{\epsfig{figure=jointAB.pdf,width=50pt}}$ the left wire as $A$ and the right wire as $B$ while in
$\raisebox{-0.16cm}{\epsfig{figure=jointBA.pdf,width=50pt}}$ it is the other way around.
By leaving implicit the identity operators and the product symbols in Eq.~(\ref{eq:quantumBayes}), it can be expressed as
\beq
\rho_{A|B}=\sqrt{\rho_A}\sqrt{\rho_B}^{-1}\!\!
\rho_{B|A}
\sqrt{\rho_B}^{-1}\!\!\sqrt{\rho_A},
\eeq
which makes the equivalence with the diagrammatic expression Eq.~(\ref{eq:Q-Bayes}) more evident.
This \em quantum Bayes rule \em was introduced in this form in \cite{LeiferSpekkens}.

In Section~\ref{sec:Qcalc_in_CPM}, we show that for every dCC one can construct a model with a non-commutative Frobenius structure and from this a graphical Bayesian calculus.  By doing so for the dCC \textbf{FdHilb}, one can recover the conditional density operator calculus described here.



\section{Inferential presentation of Bayesian graphical calculus}\label{sec:causal}

Above, we represented both joint and conditional states by the same triangles, only distinguishing them in terms of their labeling.  We will now rely on the compact structure induced by the Frobenius structure to clearly distinguish between givens (objects on the right of the conditional bar ``$\mid$'' in our notation) and conclusions (objects on the left of the conditional) by representing the first as inputs (appearing at the bottom of the diagram) and the latter as outputs (appearing at the top).  We do so by defining the following process, which we call a \em conditional process\em:
\beq \label{eq:diagCJ1}
\raisebox{-0.20cm}{\epsfig{figure=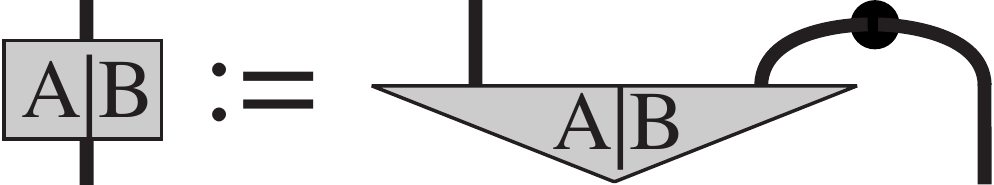,width=101pt}}\ .
\eeq
We can recover the conditional state from the conditional process by acting it upon the cup of the compact structure:
\beq \label{eq:diagCJ2}
\raisebox{-0.30cm}{\epsfig{figure=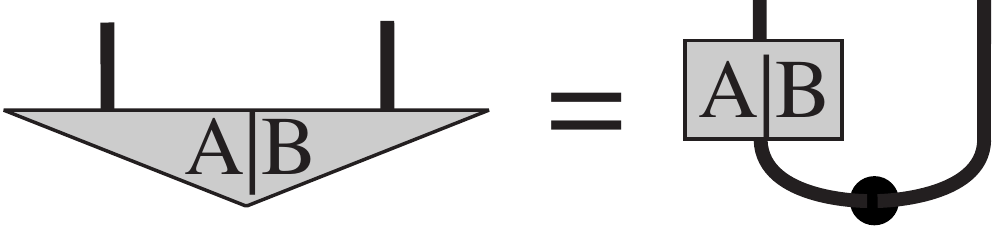,width=101pt}}\ .
\eeq
(In the context of the conditional density operator calculus, this isomorphism between conditional processes and conditional states corresponds to the version of the Choi-Jamiolkowski isomorphism described in \cite{Leifer1}.)  For multiple givens  we set:
\beq
\raisebox{-0.40cm}{\epsfig{figure=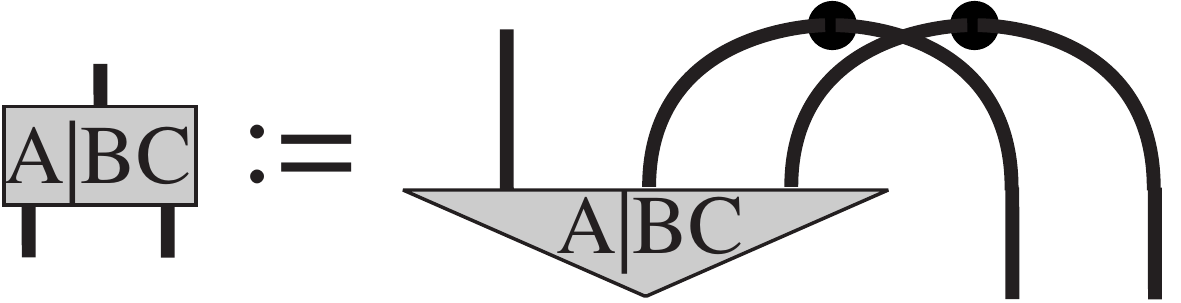,width=120pt}\qquad\qquad
\epsfig{figure=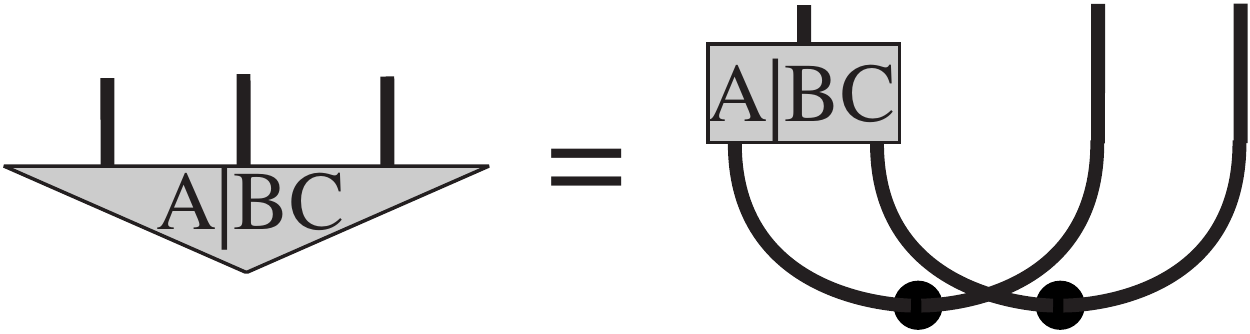,width=127pt}}\ .
\eeq

The normalization condition for conditional states, Eq.~(\ref{eq:condstatenorm}), is expressed in terms of conditional processes as
\beq
\raisebox{-0.30cm}{\epsfig{figure=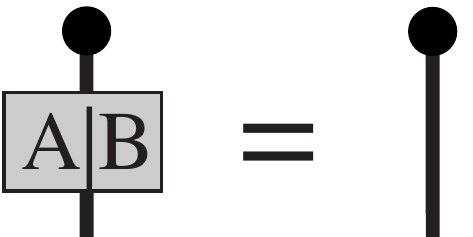,width=48pt}}\ .
\eeq

Using this dictionary, results that were previously expressed in terms of states may be expressed in terms of conditional processes.  For instance, the commutativity of multiplication of conditional states in the classical Bayesian graphical calculus, described in Prop.~(\ref{eq:Bayes2}) is equivalent to the commutativity of comultiplication of conditional processes.

\begin{proposition}\label{prop:Condindswap}
In a classical Bayesian graphical calculus:
\beq
\raisebox{-0.44cm}{\epsfig{figure=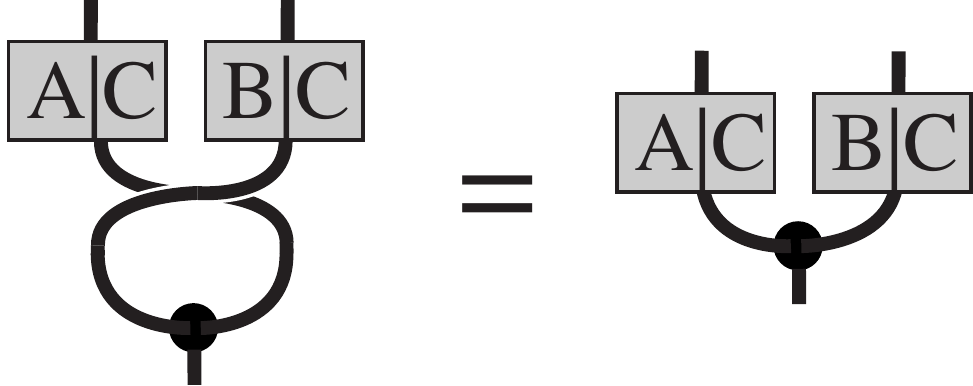,width=99pt}}
\eeq
\end{proposition}
\bpf
Follows from the fact that in a classical Bayesian graphical calculus, the Frobenius structure on states is commutative, Eq.~(\ref{eq:Bayes2}), and from the definition of conditional processes in terms of conditional states, Eq.~(\ref{eq:diagCJ1}).
\endproof\newline

We shall refer to the diagrammatic representation of an expression wherein every conditional state is replaced by its isomorphic process as the \em inferential presentation \em because by reading the diagram from bottom to top one follows a chain of inferences.

Note that one should not interpret the morphisms in a Bayesian graphical calculus as transformations of a physical system, but as the steps of a computation that a theorist might make in reasoning about the physical system.  It is useful to emphasize this point.  The classical Bayesian graphical calculus \em does not \em model the evolution of random variables undergoing stochastic maps but rather the mathematical operations (i.e. the belief propagation algorithm) that a statistician applies in drawing conclusions about one random variable from information about another.  Similarly, a \em quantum \em Bayesian graphical calculus does not model the evolution of density operators under completely positive maps (in contrast to the graphical calculi that have been introduced in other works  e.g.~\cite{ContPhys}), but rather the mathematical operations that a quantum theorist applies in a quantum analogue of a belief propagation algorithm.


Bayes' rule for a general Bayesian calculus, described in Eq.~(\ref{eq:Bayes}), has a particularly nice form in the inferential presentation.  We simply replace the conditional states in Eq.~(\ref{eq:Bayes}) with their associated modifiers using Eqs.~(\ref{eq:diagCJ1}) and (\ref{eq:diagCJ2}) to obtain:
\beq\label{BayesRuleWithModifiers}
\raisebox{-0.56cm}{\epsfig{figure=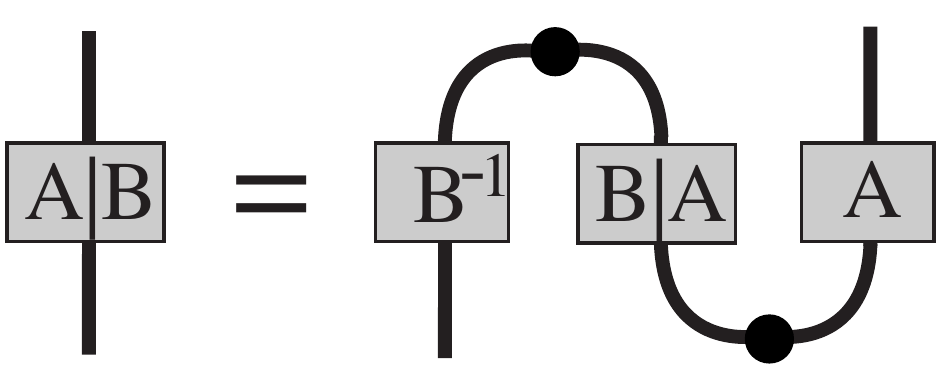,width=96pt}}\ .
\eeq

This form can be simplified further.  One easily verifies that the morphisms
\beq\label{eq:modifiedcompactstructure}
\raisebox{-0.3cm}{\epsfig{figure=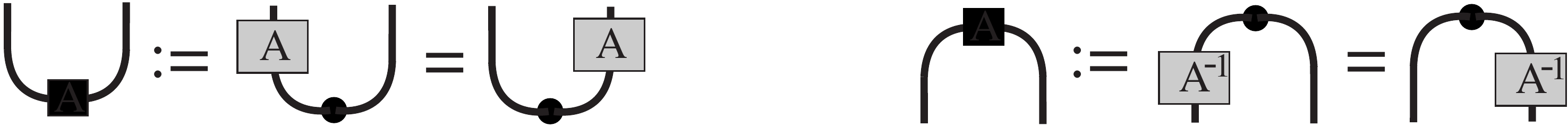,width=300pt}}
\eeq
define another compact structure on $A$, which we will refer to as the \em modified compact structure\em.  Note that, like the original compact structure, it is commutative and self-dual.

This modified compact structure simplifies diagrams considerably.  For instance, the isomorphism between conditional processes and conditional states, Eqs.~(\ref{eq:diagCJ1}) and (\ref{eq:diagCJ2}), can be expressed elegantly in terms of conditional processes and \em joint \em states using the new compact structure, as follows:
\beq \label{eq:diagCJ12bis}
\raisebox{-0.34cm}{\ \epsfig{figure=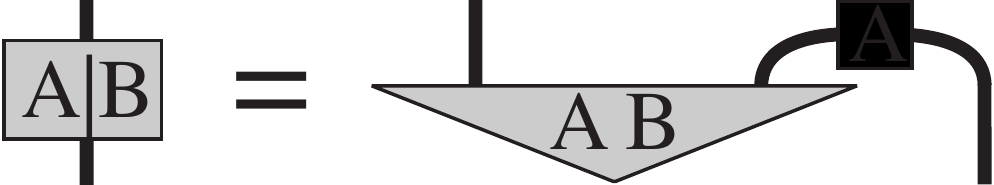,width=101pt}\qquad\qquad
\ \epsfig{figure=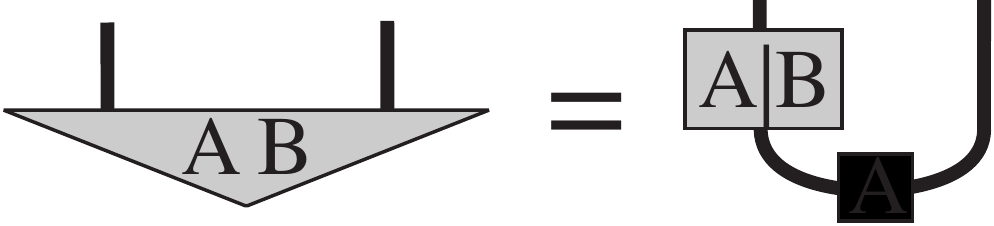,width=101pt}}\ .
\eeq
We do not decorate the black box of the modified compact structure with the label of the modifier, since this label can be inferred from the object to which the black  box is connected within an inferential scenario.

The modified compact structure also provides a very simple formulation of Bayes' rule for general Bayesian calculi.  It is simply the statement that  $\ \raisebox{-0.18cm}{\epsfig{figure=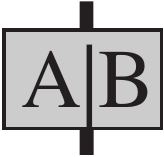,width=16.7pt}}\ $  is the \em modified transpose \em of  $\ \raisebox{-0.18
cm}{\epsfig{figure=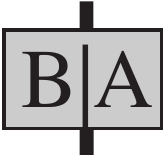,width=16.7pt}}\ $:
\beq\label{BayesTransposition}
\raisebox{-0.60cm}{\epsfig{figure=Bayestranspose.pdf,width=88pt}}\ .
\eeq

It is straightforward to generalize these results to an arbitrary number of objects.  For simplicity, we consider pairs of objects; the general case is analogous. A modifier containing a pair of object labels is simply the modifier defined by the joint state for those labels.  One then introduces a modified compact structure for a pair of objects in a manner analogous to Eq.~(\ref{eq:modifiedcompactstructure}), namely,
\beq
\raisebox{-0.60cm}{\epsfig{figure=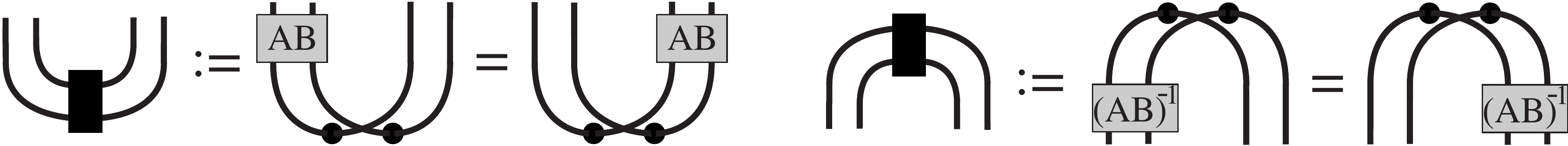,width=364pt}}\ .
\eeq
Our diagrammatic convention is that the objects on the left of the modifier are in the same order as the objects on the right.  In other words, we `hide' the crossing of wires within the black box.  This convention maintains the diagrams as planar as possible and in the cases where non-commutativity plays a role, it minimizes the number of swap operations one must display simultaneously.

It follows, for instance, that a conditional process of the form $\ \raisebox{-0.18
cm}{\epsfig{figure=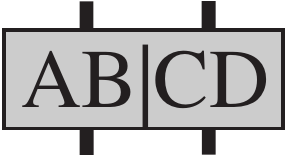,width=29pt}}\ $ can be expressed in terms of the joint state $ABCD$ using the modified compact structure on $CD$:
\beq
\raisebox{-0.40cm}{\epsfig{figure=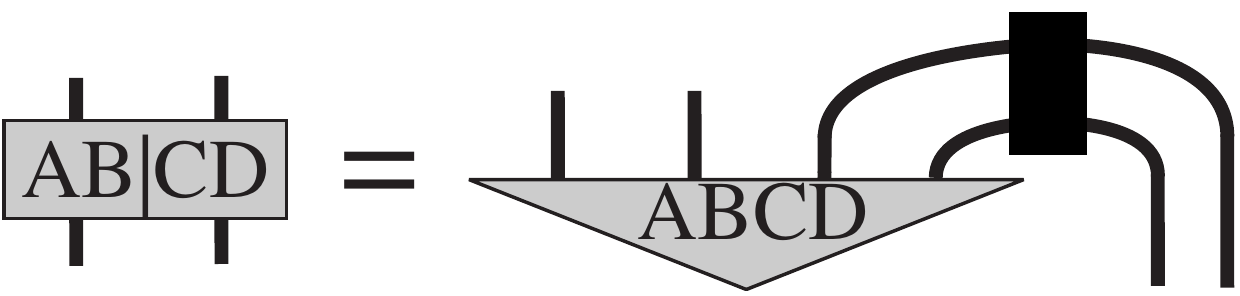,width=126pt}}\ .
\eeq

\begin{remark}
The canonical natural isomorphism $u_{A,B}$ --cf.~\cite{KellyLaplaza}\S6-- in the diagram
\begin{diagram}
\II &\rTo^{\eta_B} & B^{(*)}\otimes B\\
\dTo^{\eta_{A\otimes B}} & & \dTo_{1_{B^{(*)}}\otimes\eta_A\otimes 1_B}\\
A^{(*)}\otimes B^{(*)}\otimes A\otimes B & \rTo_{u_{A,B}\otimes 1_{A\otimes B}} & B^{(*)}\otimes A^{(*)}\otimes A\otimes B
\end{diagram}
is crucially non-trivial --i.e.~not just $\sigma_{A, B}$-- for the modified compact structure.  It is
\beqn
u_{A,B}
&=& (1_{B\otimes A}\otimes \epsilon_{A\otimes B})\circ
(((1_B\otimes \eta_A\otimes 1_B)\circ\eta_B)\otimes 1_{A\otimes B})\vspace{4mm}\\
&=& \raisebox{-0.68cm}{\epsfig{figure=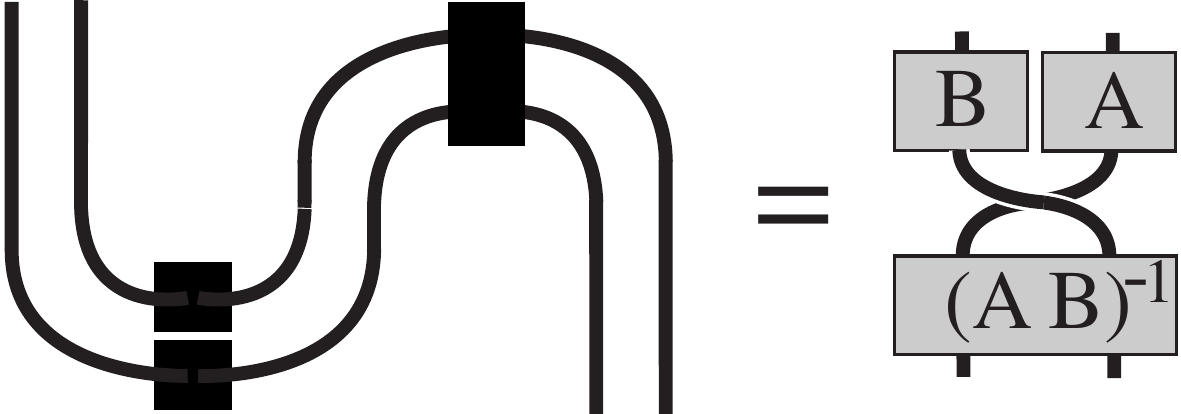,width=120pt}}: A\otimes B\to B\otimes A\,.
\eeqn
\end{remark}

\begin{remark}[generalized transposition]
The \em transposition rule \em in Eq.~(\ref{BayesTransposition}) can be generalized to arbitrary numbers of objects, but this requires some caution.  For instance, suppose one wants to express the conditional process $ACD|BE$, which we call the \em target conditional,\em in terms of the conditional process $AB|CDE$, which we call the \em source conditional\em.  It is done as follows:
\beq\label{eq:jj0}
\raisebox{-0.86cm}{\epsfig{figure=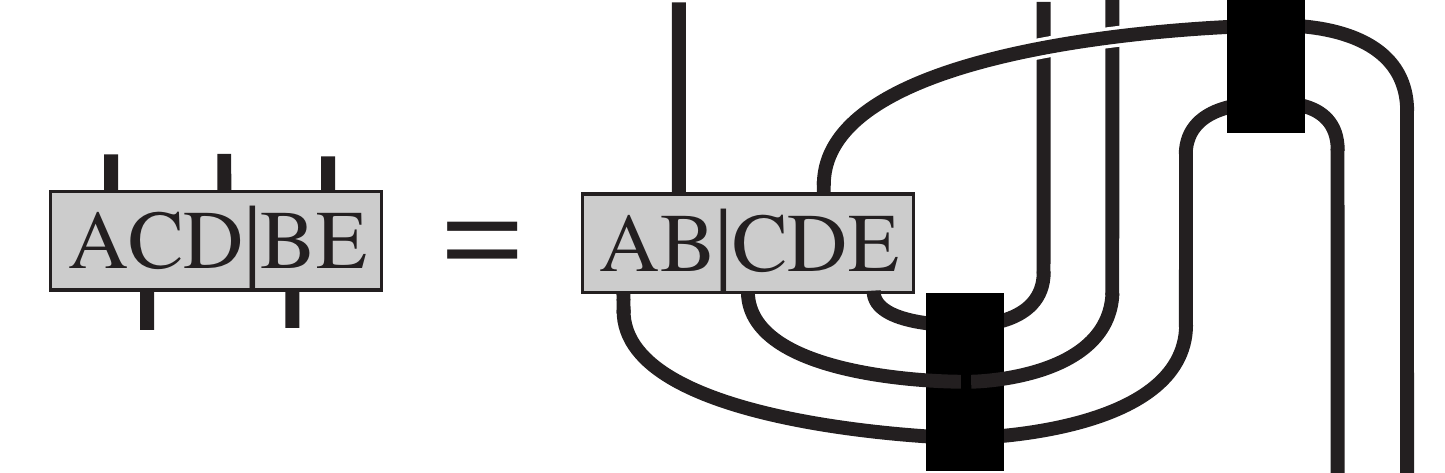,width=147pt}}
\eeq
The general prescription for how to act upon the source conditional with the modified compact structure to obtain the target conditional is as follows (we illustrate with our example):
\ben
\item[(1)] one transposes all inputs into outputs:
\beq
\raisebox{-0.60cm}{\epsfig{figure=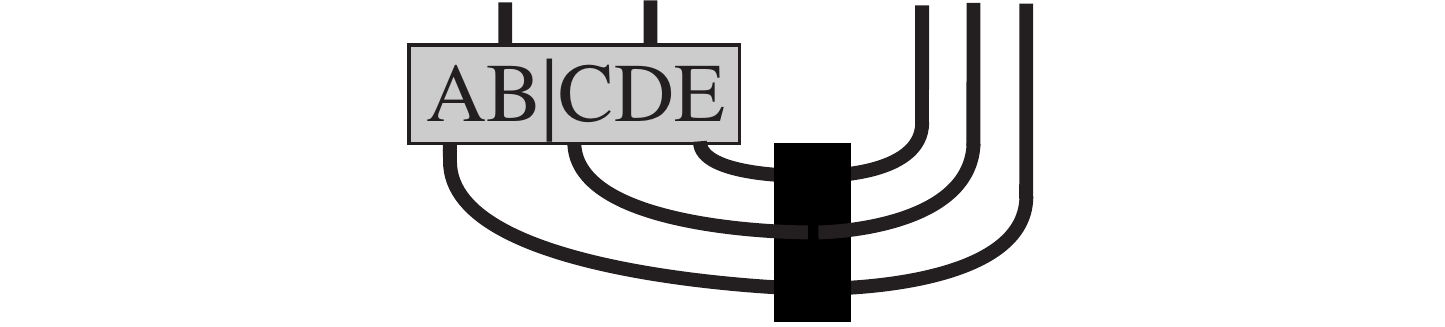,width=147pt}}
\eeq
\item[(2)] one transposes those of the outputs which initially were inputs and that one wants to retain as inputs back to inputs, together with those of the initial outputs one wants to transpose into inputs:
\beq
\raisebox{-0.86cm}{\epsfig{figure=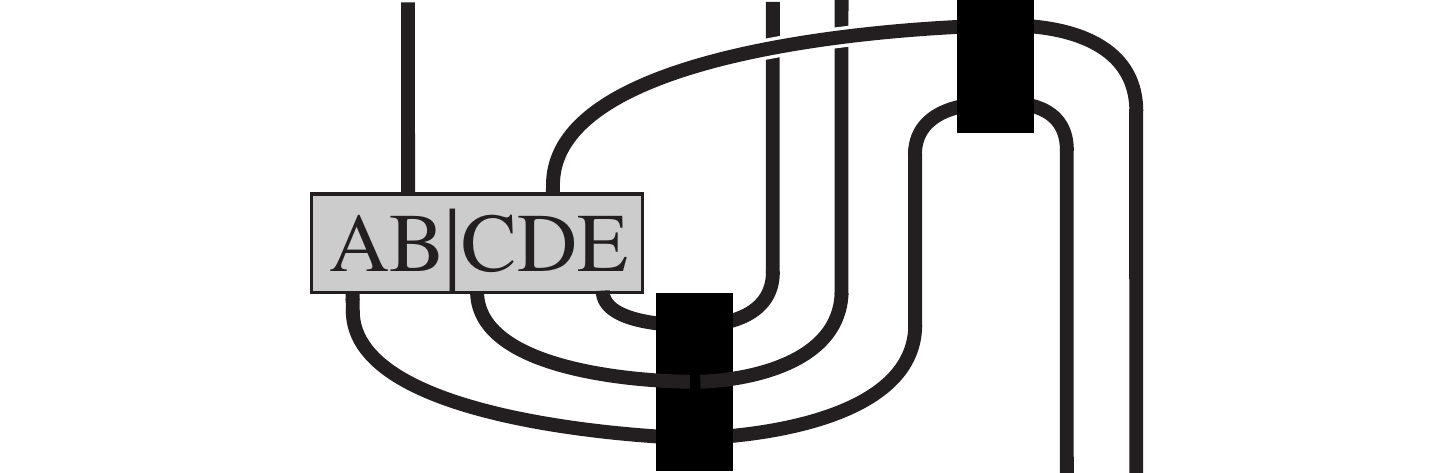,width=147pt}}
\eeq
\een
To see that this is the correct prescription, note simply that if one expresses the joint state in terms of the source conditional using the modified compact structure\,
\beq\label{eq:jj1}
\raisebox{-0.60cm}{\epsfig{figure=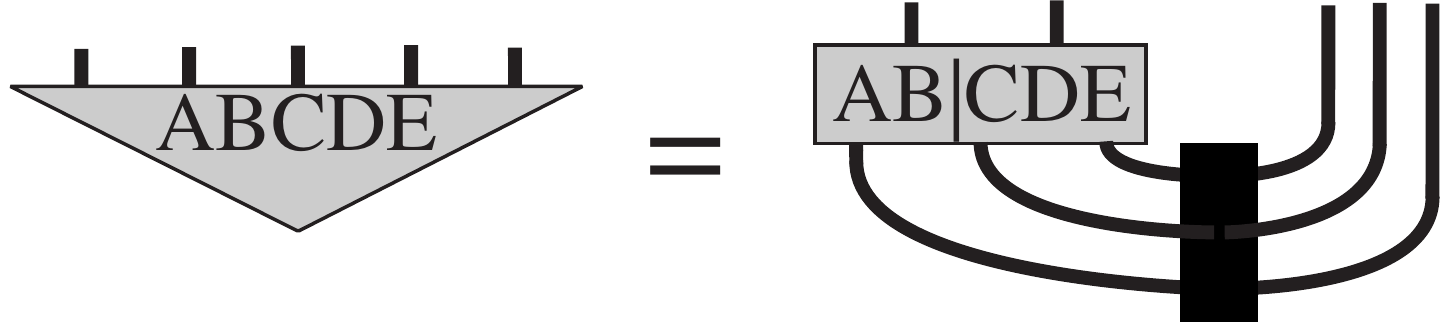,width=147pt}}
\eeq
and one expresses the target conditional in terms of the joint state using the modified compact structure\,
\beq\label{eq:jj2}
\raisebox{-0.54cm}{\epsfig{figure=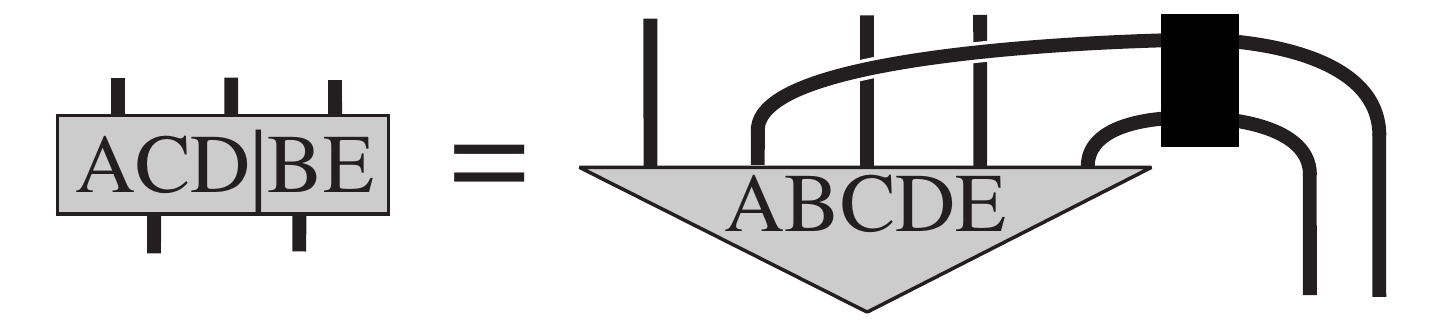,width=147pt}}
\eeq
then by substituting Eq.~(\ref{eq:jj1}) into Eq.~(\ref{eq:jj2}), one obtains Eq.~(\ref{eq:jj0}).
\end{remark}

\section{Conditional independence}\label{ConditInd}

\subsection{Definition}
One of the most important notions in the theory of Bayesian inference is that of conditional independence.  In classical probability theory, a set of random variables $X$ and another set $Y$ are said to be conditionally independent given a third set $Z$ if the following equivalent conditions hold:
\begin{enumerate}\label{eq:classicalCI}
  \item[(a)] $p(X|Y,Z)=p(X|Z)$ \label{eq:cCI1L}
  \item[(b)] $p(Y|X,Z)=p(Y|Z)$ \label{eq:cCI1R}
  \item[(c)] $p(X,Y|Z)=p(X|Z)p(Y|Z)$ \label{eq:cCI2}
\end{enumerate}

In the general Bayesian graphical calculus, there are analogues of each of these conditions, but they are no longer equivalent.  We therefore distinguish two pairs of notions of conditional independence.  The first pair are the analogues of Eqs.~(a) and (b)
respectively, while the second pair constitute analogues of Eq.~(c)
where one differs from the other by an interchange of the roles of $A$ and $B$ (which yields a different condition due to the non-commutativity of the Frobenius structure):
\ben
\item[{\bf CI1$_L$}]\quad\raisebox{-0.44cm}{\epsfig{figure=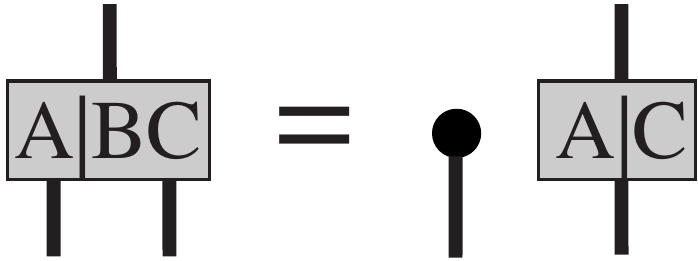,width=71pt}}\label{eq:CI1L} \vspace{2mm}
\item[{\bf CI1$_R$}]\quad\raisebox{-0.44cm}{\epsfig{figure=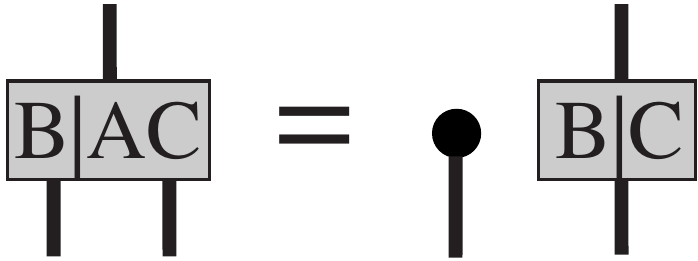,width=71pt}}\label{eq:CI1R}\vspace{2mm}
\item[{\bf CI2$_L$}]\quad\raisebox{-0.38cm}{\epsfig{figure=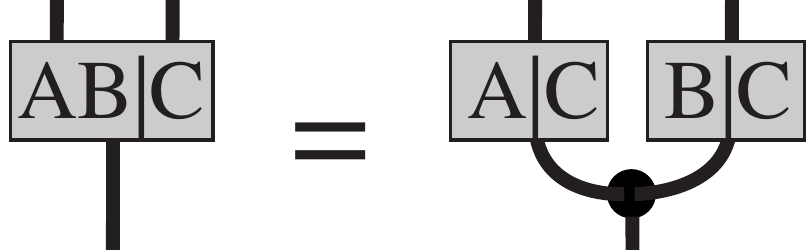,width=82pt}}
\label{eq:CI2L}
\vspace{2mm}
\item[{\bf CI2$_R$}]\quad\raisebox{-0.38cm}{\epsfig{figure=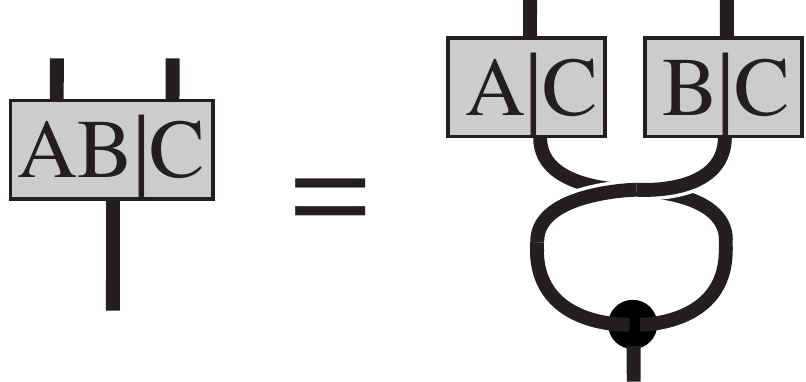,width=82pt}}    \label{eq:CI2R} \vspace{2mm}
\een

\subsection{Results}
\begin{proposition}\label{prop:condind}
In a Bayesian graphical calculus, if any two of the following three equalities hold then the third one also holds:
\ben
\item[{\bf CI1$_L$}]\quad\raisebox{-0.44cm}{\epsfig{figure=CIalternative.pdf,width=71pt}}\vspace{2mm}
\item[{\bf CI2$_L$}]\quad\raisebox{-0.38cm}{\epsfig{figure=conditiontranspose.pdf,width=82pt}}\vspace{2mm}
\item[{\bf F$_L$}]\quad\raisebox{-0.84cm}{\epsfig{figure=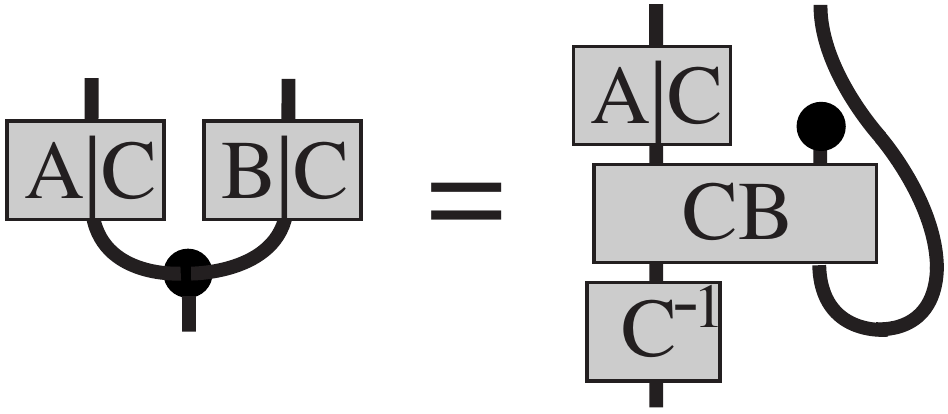,width=96.5pt}}\vspace{2mm}\,.
\een
And similarly for the case where one interchanges $A$ and $B$ (where a condition F$_R$ is defined in the obvious way).
\end{proposition}

To see this, we make use of the following lemma:
\begin{lemma}
The condition {\bf CI1$_L$} is equivalent to
\ben
\item[{\bf CI1$_L'$}]\quad\raisebox{-0.84cm}{\epsfig{figure=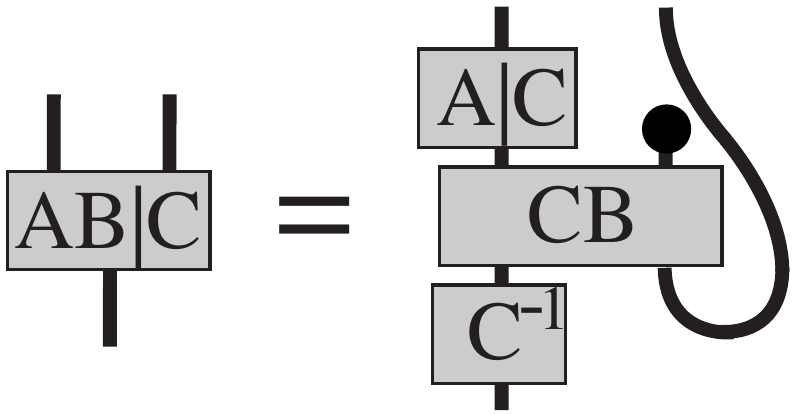,width=82pt}} \vspace{2mm}
\een
\end{lemma}
\bpf
Using Bayesian inversion together with Eq.~(\ref{eq:modswap}) and CI1$_L$, we have:
\beq
\raisebox{-0.44cm}{\epsfig{figure=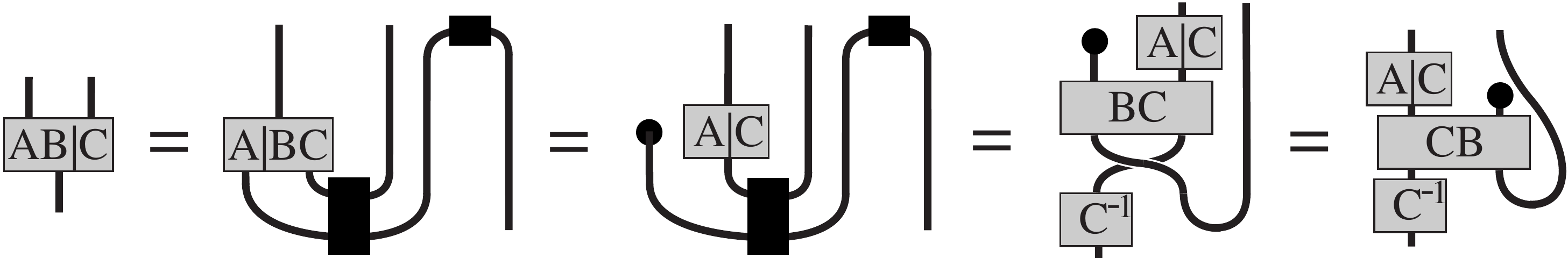,width=296pt}}
\eeq
\endproof
\newline

\bpf [proposition \ref{prop:condind}]
Since:
\begin{center}
\raisebox{-0.44cm}{\epsfig{figure=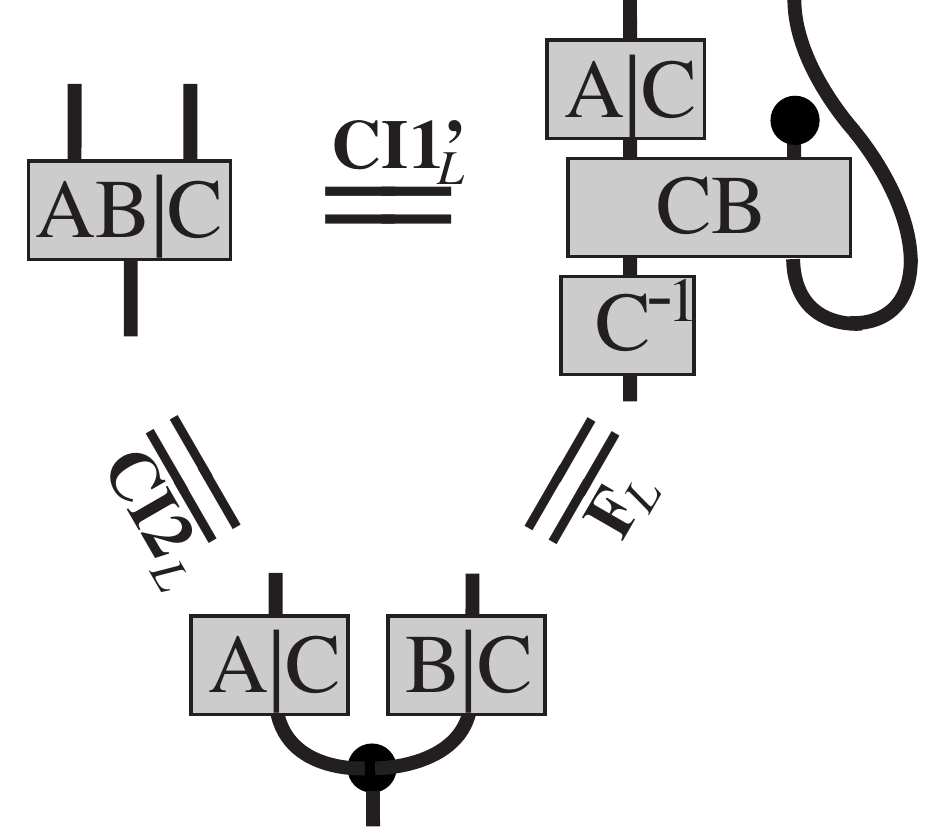,width=95pt}}
\end{center}
validity of any two of these equalities implies that the third also holds.   The analogous equalities hold if one interchanges $A$ and $B$
\endproof\newline

It is straightforward to recover the classical notion of conditional independence, as follows.
\begin{proposition}
In a classical  Bayesian graphical calculus, the four notions of conditional independence, CI1$_L$, CI1$_R$, CI2$_L$, CI2$_R$, are all equivalent.
\end{proposition}

\bpf  First, note that the equality F$_L$ always holds in a classical Bayesian graphical calculus.  This is proven using Eqs.~(\ref{def:slidinginverseclass}) and (\ref{eq:Ccalcmofifier}) and the spider theorem:
\begin{center}
\raisebox{-0.44cm}{\epsfig{figure=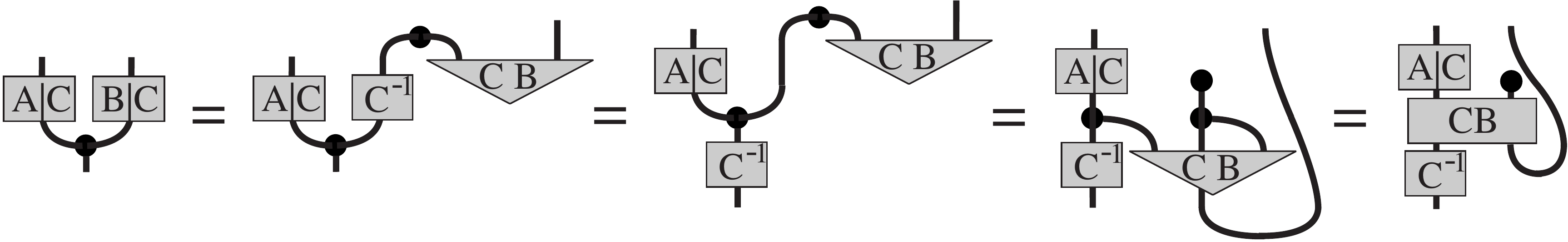,width=351pt}}.
\end{center}
Similarly, one can prove the equality F$_R$, wherein $A$ and $B$ are interchanged relative to F$_L$.
Given these equalities, Prop.~(\ref{prop:condind}) implies that CI1$_L$ is equivalent to CI2$_L$ and that  CI1$_R$ is equivalent to CI2$_R$.  Finally, the commutativity of the comultiplication of conditional processes, Prop.~(\ref{prop:Condindswap}), implies that CI2$_L$ and CI2$_R$ are equivalent.  Consequently, all four conditions are equivalent.
\endproof\newline

What is more difficult is to recover a quantum notion of conditional independence.   An open question is whether specifying that the form of the modifiers is as given in Eq.~(\ref{eq:formQmodifiers}) is sufficient to prove everything that can be proven within the conditional density operator calculus.  In particular, it is not clear how to derive that CI2$_L$ and CI2$_R$ are equivalent.

\begin{example}
In \cite{Leifer2}, by relying on results in \cite{HJPW}, which in turns rely on a Theorem by Uhlmann  \cite{Uhlmann}, it was established that in  the case the Q${}_{1/2}$-calculus of  Section \ref{Sec:PresQonehalf}  {\bf CI1$_L$} implies {\bf F$_L$}, and hence, by Prop.~(\ref{prop:condind}]), {\bf CI1$_L$} is equivalent to the pair {\bf CI2$_L$} and {\bf F$_L$}.  The analogue holds if we interchange $A$ and $B$. It would be interesting to establish whether there is a weakening to our definition of classical  Bayesian graphical calculus which also establishes this.  Note here also that the assumption made in \cite[Thm 3.8]{Leifer2} to derive {\bf CI2$_L$} from  {\bf CI1$_L$} translates in graphical language to the condition that
\beq
\raisebox{-1.6mm}{\epsfig{figure=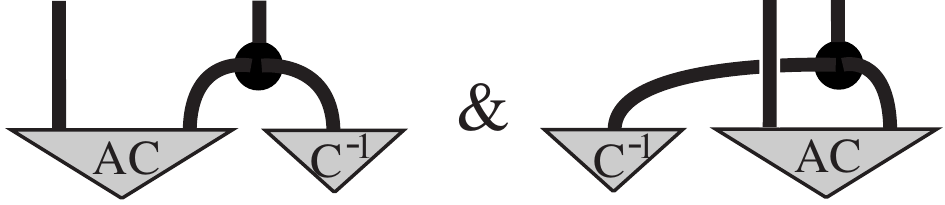,width=105pt}}
\qquad\mbox{commute with}\qquad \raisebox{-1.6mm}{\epsfig{figure=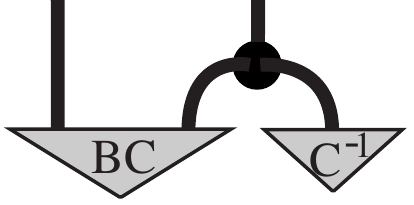,width=42pt}} \\
\eeq
relative to  the Frobenius multiplication, that is, for example:
\beq
\raisebox{-1.6mm}{\epsfig{figure=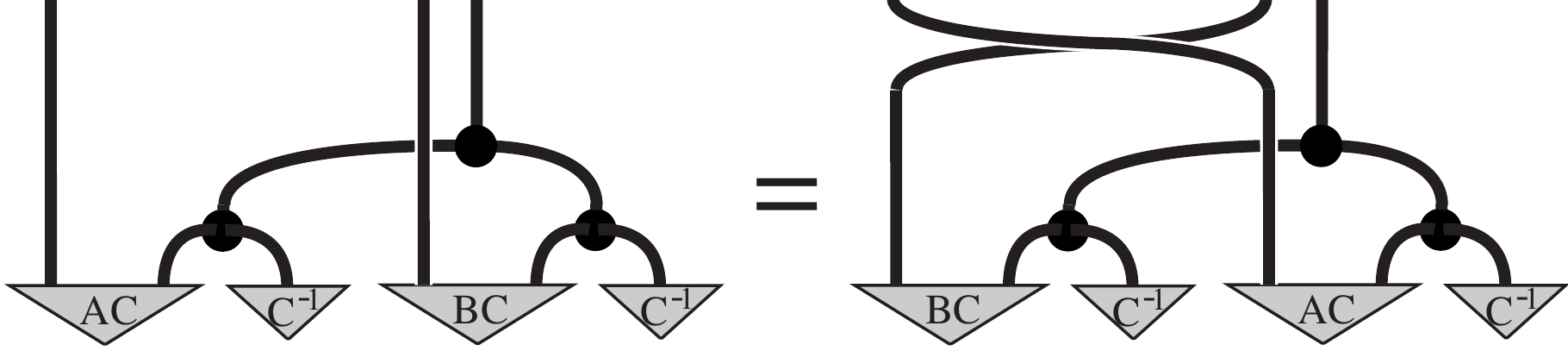,width=215pt}}  \ .
\eeq
By a tedious calculation, it can be shown that these conditions imply the weaker condition {\bf F$_L$}, which suffices for this purpose.
\end{example}

\subsection{Example application: generalized pooling}\label{sec:exampapl}
\label{sec:pooling}

A simple example of what one can derive from the notion of conditional independence, we consider the problem of pooling.  Here, one seeks to assign a conditional state to $C$ given $A,B$ and the question is whether this state can be expressed in terms of a conditional state for $C$ given $A$ and a conditional state for $C$ given $B$.  In the classical case (which we shall describe below), a sufficient condition for this to be possible is that $A$ and $B$ are conditionally independent given $C$.  We here consider an analogue for general Bayesian graphical calculi.

\begin{proposition}
If $A$ and $B$ are conditionally independent relative to $C$,  in the sense of {\bf CI2$_L$},  then we have
\beq\label{eq:pooling}
\raisebox{-1.2cm}{\epsfig{figure=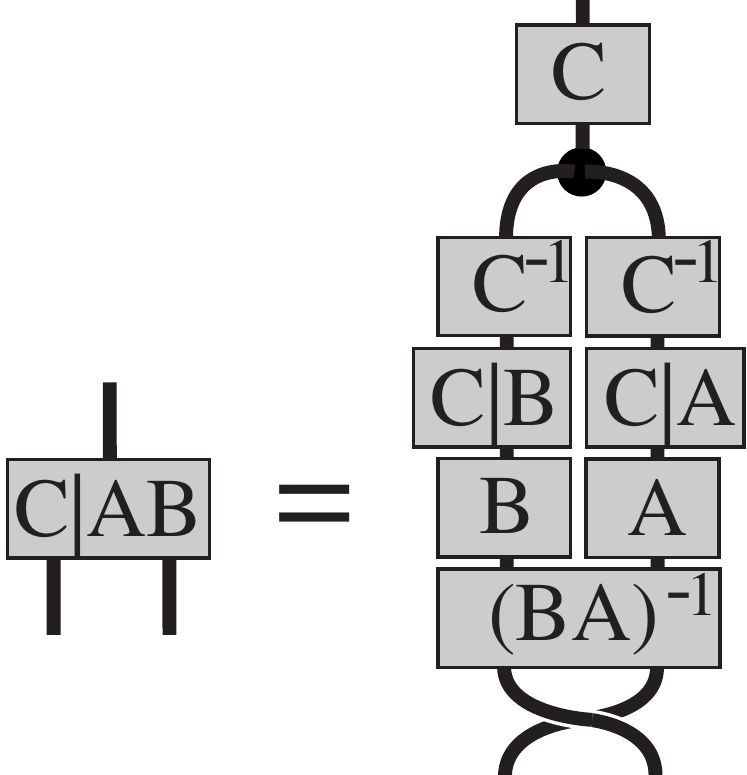,width=76pt}}\ .
\eeq
\end{proposition}
\bpf
\begin{center}
\epsfig{figure=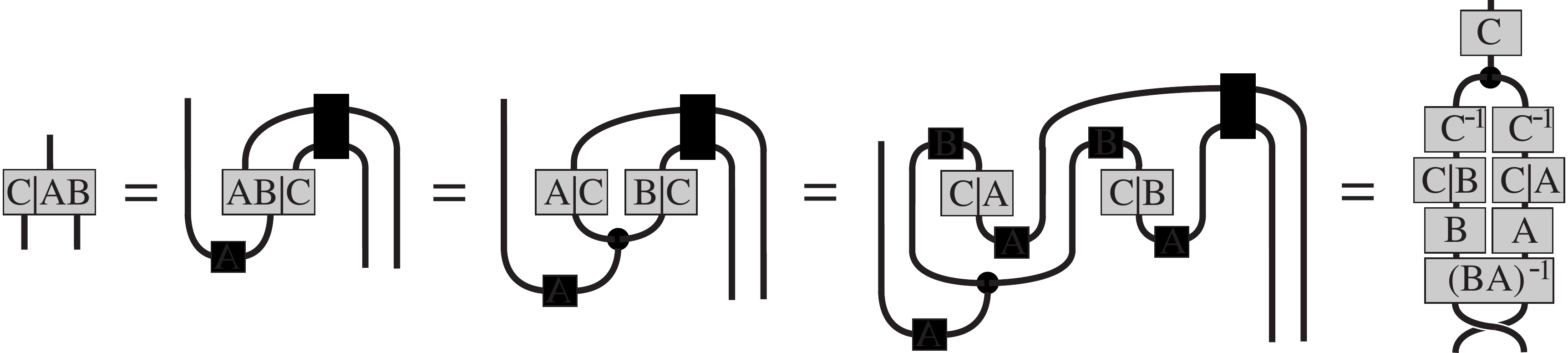,width=350pt}
\end{center}
\endproof\newline

The case where $A$ and $B$ are conditionally independent relative to $C$ in the sense of {\bf CI2$_R$} differs by a swap:
\beq\label{eq:poolingbis}
\raisebox{-0.84cm}{\epsfig{figure=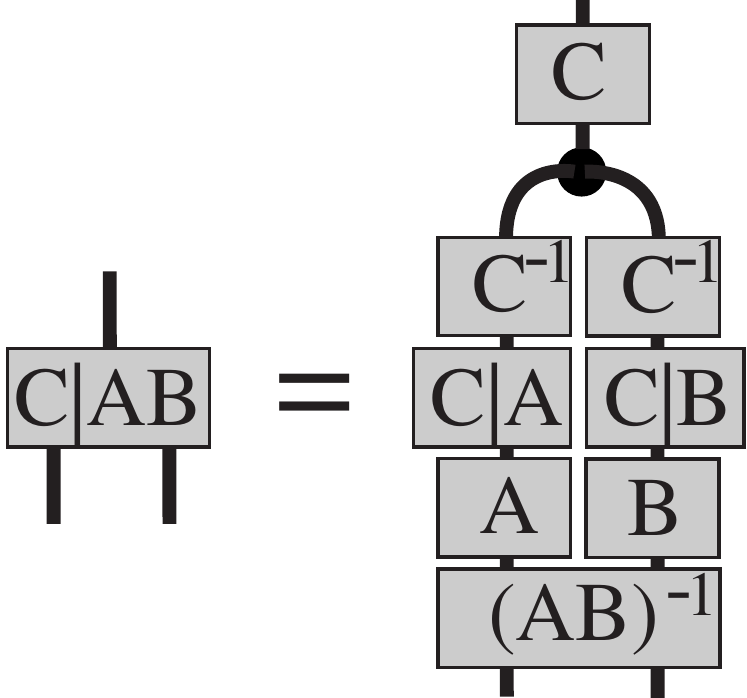,width=76pt}}\ .
\eeq

\begin{example}
For Q${}_{1/2}$-calculi, when expressing Eq.~(\ref{eq:pooling}) in terms of conditional states rather than in the inferential form we obtain:
\beq \label{eq:Q-pooling}
  \raisebox{-1.16cm}{\epsfig{figure=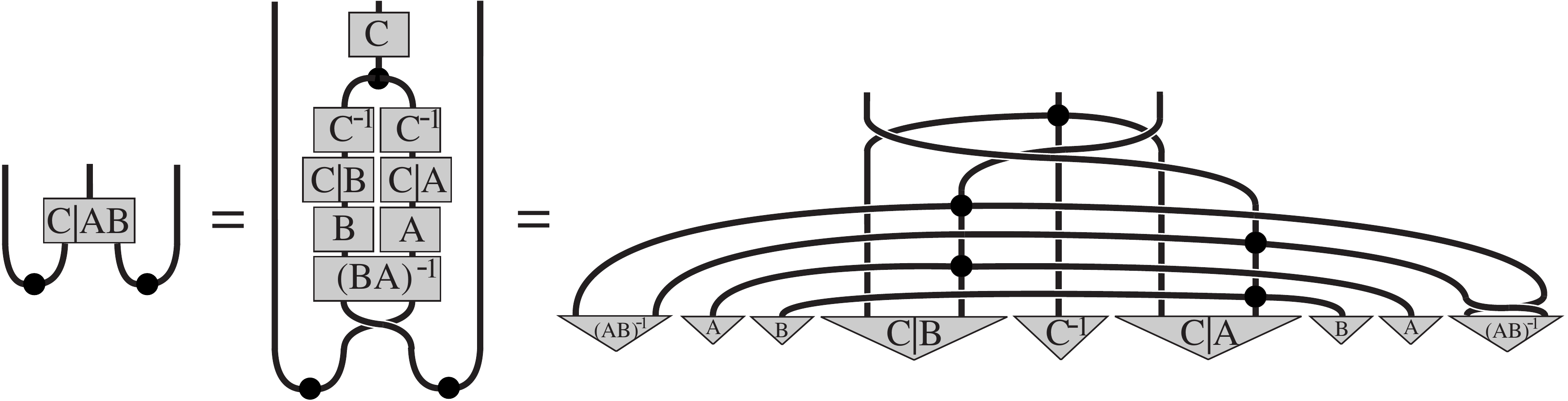,width=355pt}} \ .
\eeq
For density operators, Eq.~(\ref{eq:Q-pooling}) is equivalent to
\beq \label{eq:quantumpooling}
\rho(C|AB)=\sqrt{\rho(A, B)}^{-1}\!\!\sqrt{\rho(A)} \sqrt{\rho(B)}\rho(C|B)\, \rho(C)^{-1}\rho(C|A)\sqrt{\rho(B)}\sqrt{\rho(A)}\sqrt{\rho(A, B)}^{-1}\!.
\eeq
For classical probability distributions, we obtain
\beq\label{eq:classicalpooling}
P(C|AB)=\frac{P(A)\,P(B)}{P(A,B)} \cdot \frac{P(C|A)\,P(C|B)}{P(C)} \, .
\eeq
This result is known as the \em pooling formula \em because if $A$ and $B$ are conditionally independent given $C$, the posterior $P(C|AB)$ can be reconstructed from the posteriors $P(C|A)$ and $P(C|B)$ and the prior $P(C)$ (the dependence on $A$ and $B$ is inferred from normalization).  As such, it is sufficient to ``pool'' the information contained in the two posteriors. Eq.~(\ref{eq:quantumpooling}) generalizes this to a \em quantum pooling formula\em,  and Eq.~(\ref{eq:pooling}) generalizes this further, to arbitrary Bayesian calculi.
\end{example}

\subsection{The semi-graphoid axioms}
One of the reasons for identifying relationships of conditional independence among objects is to have the ability to describe their mutual dependencies without providing a full specification of their joint state. Thus, it is useful to consider what implications hold among statements of conditional independencies.  These conditions are well known in the classical case as the semi-graphoid axioms \cite{Pearl88}.  Let $U, W, X$ and $Y$ denote sets of random variables and let $X\cup Y$ denote the set-theoretic union of $X$ and $Y$.  In a standard notation, $I(U, W\mid X)$
 is taken to express the statement that the variables in $U$ and the variables in $W$ are conditionally independent given $X$.  The semi-graphoid axioms, which are easily derived from the definition (cf. \ref{eq:classicalCI}) of conditional independence, are:
\ben
\item Symmetry:		$I(U,W|X) \Leftrightarrow I(W,U|X)$	
\item Decomposition:	$I(U,W\cup Y|X) \Rightarrow I(U,W|X)$	
\item Weak Union: 	$I(U,W\cup Y|X) \Rightarrow I(U,W|X\cup Y)$	
\item Contraction: 		$I(U,W|X) \textrm{ and } I(U,Y|X\cup W)\Rightarrow I(U,W\cup Y|X)$
\een
The semi-graphoid axioms are important because their satisfaction implies the possibility of a representation of (certain facts about) the mutual dependencies of sets of random variables in terms of a directed acyclic graph known as a Bayesian network.

It is interesting to explore the extent to which these axioms hold true for a general Bayesian graphical calculus when objects play the role of sets of random variables, tensor product plays the role of set-theoretic union, and $I(A,B\mid C)$ expresses the statement that the ordered pair of objects $A,B$ are conditionally independent given $C$.  Because we have four distinct notions of conditional independence in a general Bayesian graphical calculus, one can ask about the satisfaction of the axioms for any of these.  As it turns out, few of the axioms hold for any of the notions of conditional independence in a general Bayesian graphical calculus.  We leave for future work the question of what additional ingredients are required of a Bayesian graphical for the axioms to be satisfied.  We note, however, that they are all satisfied by the classical Bayesian graphical calculus.  In this sense, our formalism for classical Bayesian inference is at least as powerful as the graphoid axiomatization.

Significantly, Leifer and Poulin have shown in Ref.~\cite{Leifer2} that the conditional density operator calculus satisfies the semi-graphoid axioms, so that one may apply the tools of Bayesian networks to quantum belief propagation.  Consequently, finding axiomatic graphical conditions implying the semi-graphoid axioms will presumably go hand-in-hand with finding an axiomatic graphical characterization of quantum Bayesian inference.

If the semi-graphoid axioms are satisfied within a Bayesian graphical calculus, the topology of our graphical representation of a set of correlations will reproduce the topology of the Bayesian network (with objects being mapped to nodes, and morphisms being mapped to sets of directed edges).  It is our hope that by understanding how Bayesian networks can be embedded within the diagrammatic calculus of dCCs, a bridge might be built between these two fields such that insights from one might be adapted to the other.

\section{Bayesian graphical calculi for arbitrary dagger compact categories}\label{sec:Qcalc_in_CPM}

\subsection{A graphical concretely non-commutative dagger Frobenius structure}

We now provide a class of models, one for every dCC, each coming with a canonical non-commutative Frobenius structure that can be used to construct graphical Bayesian calculi, for example Q${}_{1/2}$-calculi.  These include the conditional  density operator calculus of Section \ref{Sec:PresQonehalf}  as a special case, namely the one  that arises for the dCC ${\bf FdHilb}$. The diagrammatic presentation of mixed quantum states and completely positive maps in terms of dCCs is due to Selinger \cite{Selinger}. But here we cannot restrict ourselves to completely positive maps, since,  as shown above in Section \ref{Sec:PresQonehalf}, the Frobenius comultiplication cannot be a completely positive map.
In this context, the concrete graphical form of this non-completely positive map which we provide in this section  will be insightful.

\begin{definition}\label{defn:operatorcategory}
Given a dCC ${\bf C}$ we define another dagger category ${\rm D}({\bf C})$ as follows:
\bit
\item $|{\rm D}({\bf C})|:=|{\bf C}|$\,  i.e. the set of objects is the same for the two dCCs;
\item ${\rm D}({\bf C})(A,B):={\bf C}(A\otimes A^*, B\otimes B^*)$\,  i.e.~every morphism from $A\otimes A^*$ to $B\otimes B^*$ in ${\bf C}$, is a morphism from $A$ to $B$ in ${\rm D}({\bf C})$;
\item composition and  dagger are inherited from ${\bf C}$ via the embedding
\beq
E: {\rm D}({\bf C})\hookrightarrow {\bf C}
::\left\{\begin{array}{l}
A\mapsto A\otimes A^*\vspace{2mm}\\
f\mapsto f
\end{array}\right.\,.
\eeq
\eit
\end{definition}

Since ${\rm D}({\bf C})$ is a dCC in its own right it comes with its own graphical language.  It is useful to see how various elements of ${\rm D}({\bf C})$ are represented both in the graphical language of ${\rm D}({\bf C})$ and in the graphical language of ${\bf C}$.  Some examples are provided in the table below.  The first three columns depict morphisms on a single object: a general morphism, identity, and composition of two morphisms.  Note that in the graphical language of ${\bf C}$ we adopt the convention that the dual objects will be represented by wires \em to the right \em of the primal objects.

\[
\begin{tabular}{c|c|}
\hline ${\rm D}({\bf C})$ & \raisebox{-0.48cm}{\epsfig{figure=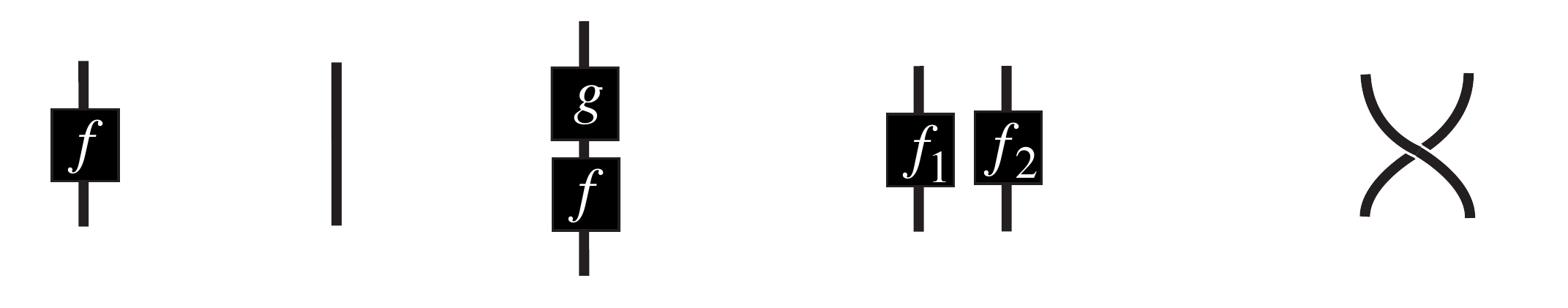,width=210pt}}\\
\hline
${\bf C}$ & \raisebox{-0.82cm}{\epsfig{figure=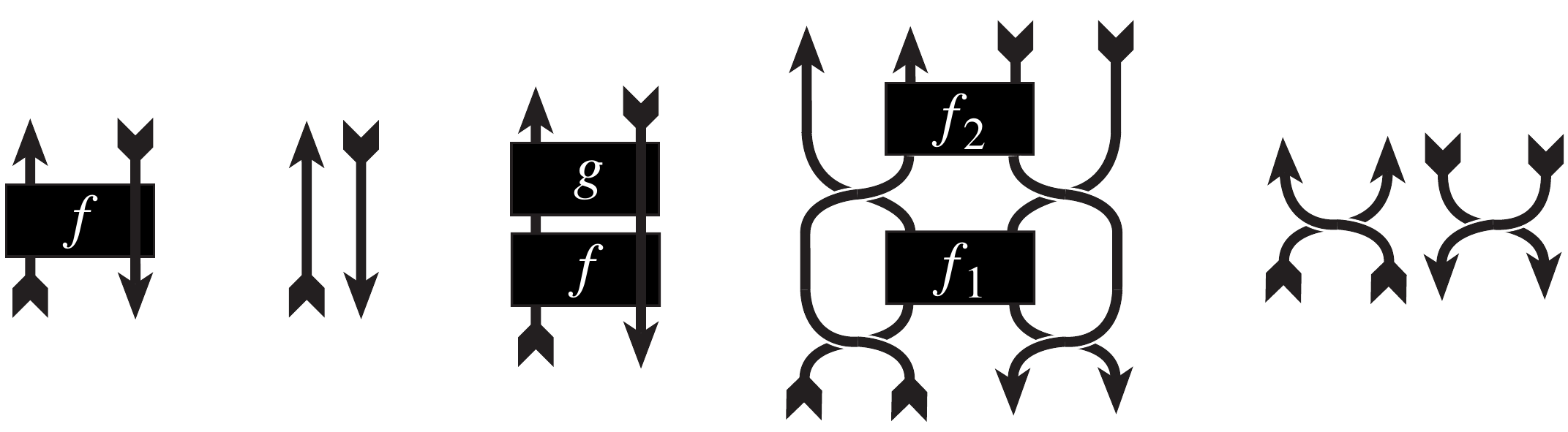,width=210pt}}\\
\hline \end{tabular}
\]

We now consider tensor products.
\begin{definition}
For
\beq
f_i\in {\rm D}({\bf C})(A_i,B_i):={\bf C}(A_i\otimes A^*_i, B_i\otimes B^*_i)\,,
\eeq
we define a tensor $\otimes_{\rm D}$ on ${\rm D}({\bf C})$ as
\beq
f_1\otimes_{\rm D}\! f_2:=(1_{B_1}\otimes \sigma_{B_1^*, B_2\otimes B_2^*})
\circ(f_1\otimes f_2)\circ
(1_{A_1}\otimes \sigma_{A_2\otimes A_2^*, A_1^*})
=\raisebox{-0.66cm}{\epsfig{figure=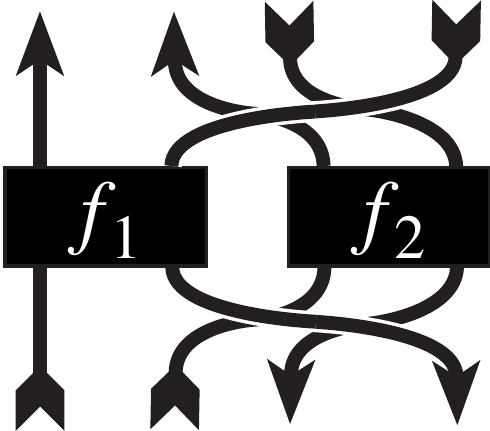,width=50pt}}
=\raisebox{-0.86cm}{\epsfig{figure=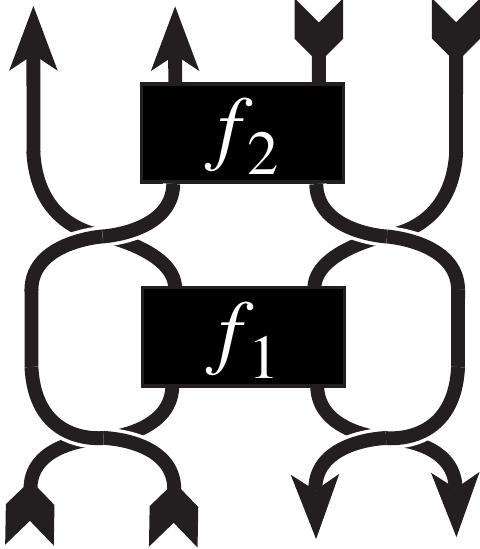,width=49pt}}.
\eeq
\end{definition}

\begin{proposition}
Recalling that $\bar{f}$ is the \em conjugate \em to $f$ i.e.~the transpose of $f^\dagger$,
an SMC-structure and compactness arises on ${\rm D}({\bf C})$ from the
SMC-structure and compactness of ${\bf C}$ via the functor
\beq\label{Functor:Dinclusion}
F:{\bf C}\to{\rm D}({\bf C})::
\left\{\begin{array}{l}
A\mapsto A\\
f\mapsto f\otimes \bar{f}
\end{array}\right.
\eeq
which maps the tensor $\otimes$ of ${\bf C}$ on the tensor $\otimes_{\rm D}$ of ${\rm D}({\bf C})$.
\end{proposition}
\bpf
This is a trivial  generalization of Theorem 4.20 in \cite{Selinger}.
\endproof\newline

We recall that  (cf.~Eq.~(\ref{eq:contravarianttensor})) in the graphical language of ${\bf C}$ we adopt another useful convention: for the case where there is more than one object, the wires for the dual objects (in addition to appearing on the right) will appear \em in the opposite order \em to those of the primal objects.

The table above presents some additional examples of elements of ${\rm D}({\bf C})$ represented both in the graphical language of ${\rm D}({\bf C})$ and in the graphical language of ${\bf C}$, in particular, the last four columns depict a tensor product of morphisms, the swap (symmetry), and the cups and caps of the compact structure.


\begin{notation}
To avoid confusion, below all $1$'s, $\otimes$'s, $\sigma$'s, $\epsilon$'s and $\eta$'s refer to the dCC ${\bf C}$, except for when explicitly stated otherwise.  We write
$f:A\to_{{\rm D}({\bf C})}B$ for a morphism made up of these components to stipulate  its type in the dCC  ${\rm D}({\bf C})$.
\end{notation}

\begin{proposition}\label{prop:noncomFrob}
For every object  $A\in |{\rm D}({\bf C})|$ the morphism
\[
{\cal F}= \raisebox{-0.30cm}{\epsfig{figure=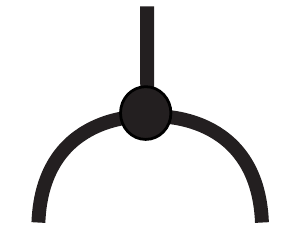,width=30pt}} :A\otimes_{\rm D} A\to_{{\rm D}({\bf C})} A
\]
 defined by
\beq
{\cal F}:= (1_A\otimes \epsilon_A\otimes 1_{A^*})\circ (1_{A\otimes A}\otimes \sigma_{A^*\!,A^*})=
\raisebox{-0.66cm}{\epsfig{figure=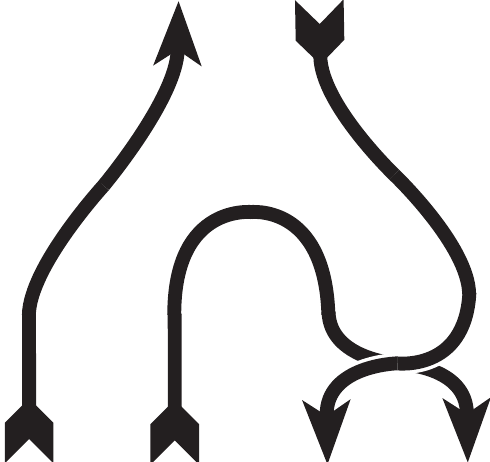,width=50pt}}: A \otimes A \otimes A^* \otimes A^* \to_{{\bf C}} A \otimes A^*
\eeq
is the multiplication of a dagger Frobenius structure
with unit $\eta_{A^*}=\raisebox{-0.28cm}{\epsfig{figure=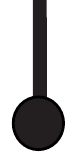,width=8pt}}:\II \to_{{\rm D}({\bf C})} A$ defined by
\beq
\eta_{A^*}=\ \raisebox{-0.20cm}{\epsfig{figure=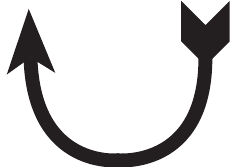,width=24pt}} :\II \to_{\bf C} A \otimes A^*\, .
\eeq
\end{proposition}

The following table depicts the multiplication, its unit, the comultiplication and its counit of the dagger Frobenius structure in the respective graphical languages of ${\rm D}({\bf C})$ and ${\bf C}$.
\[
\begin{tabular}{c|c|}
\hline ${\rm D}({\bf C})$ & \raisebox{-0.48cm}{\epsfig{figure=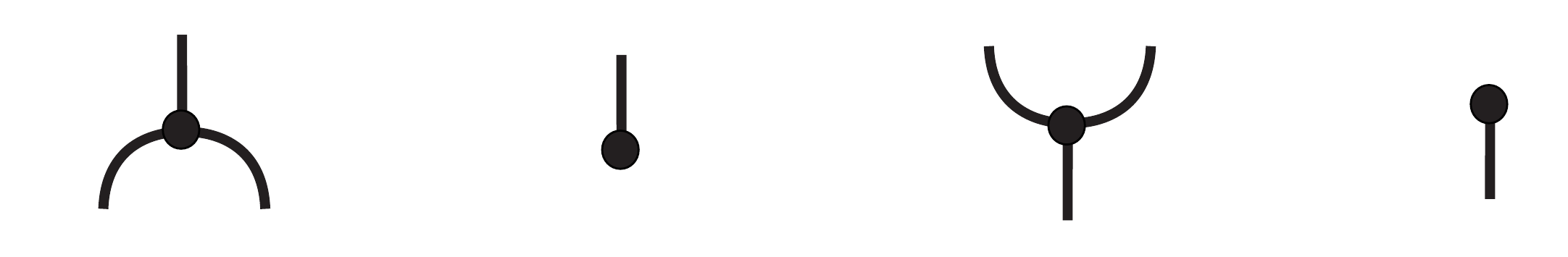,width=220pt}}\\
\hline
${\bf C}$ & \raisebox{-0.82cm}{\epsfig{figure=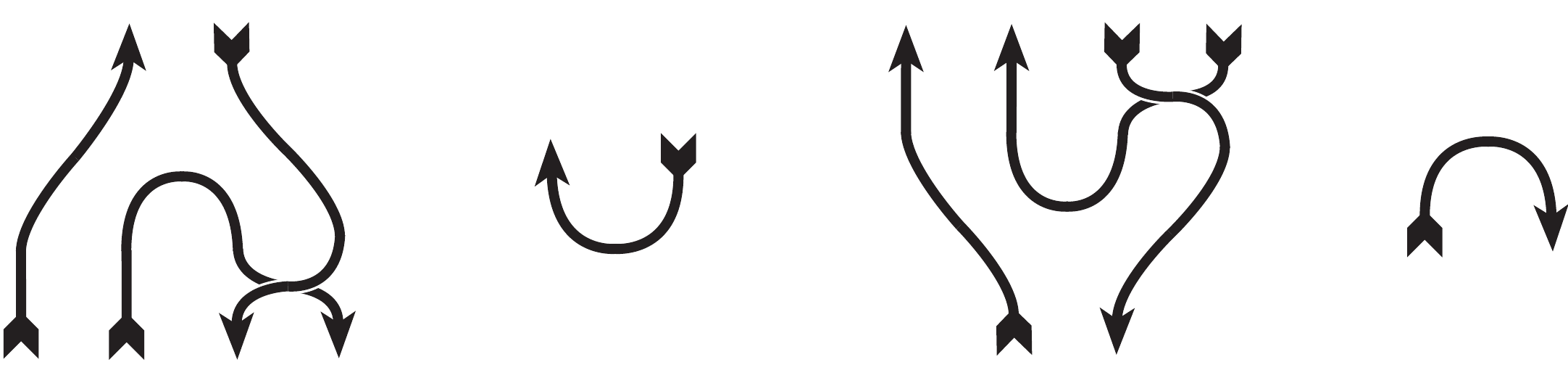,width=220pt}}\\
\hline \end{tabular}
\]
Because the two graphical representations of the multiplication have a similar shape, it is easy to misinterpret the mapping between these.  By our convention, the left leg of $\raisebox{-0.30cm}{\epsfig{figure=FrobeniusAllDx.pdf,width=30pt}}$ is \em not \em associated with the left pair of legs of $\raisebox{-0.66cm}{\epsfig{figure=Frobenius.pdf,width=50pt}}$, but rather with the outermost pair of legs, while the right leg of the former is associated with the innermost pair of the latter.  It is useful to imagine a central left-right partition for the diagrams in ${\bf C}$ which divides the primal objects on the left from the oppositely-ordered dual objects on the right.  The shape of the diagram in ${\rm D}({\bf C})$ should be compared with the left-hand side of the diagram in ${\bf C}$.

Note also that in the graphical language of ${\rm D}({\bf C})$ we use a dot decorated by an `F' to denote the Frobenius structure just defined.  We do so to distinguish it from a Frobenius structure native to ${\bf C}$ (although we will not need to make use of such a structure in this article).

\bpf
We must verify that ${\cal F}$ is associative and satisfies the dagger Frobenius law, and that $\eta_{A^*}$ is indeed a two-sided unit.  Representing ${\cal F}$ and $\eta_{A^*}$ in the graphical language of ${\rm D}({\bf C})$, these properties are given diagrammatically as Eq.~(\ref{eq:daggerFrobeniusstructure}).  The  tedious but straightforward proof  proceeds by recasting each identity within the graphical language of ${\bf C}$ and verifying graph isomorphism for each. For example, associativity of the multiplication is verified as follows:
\[
\begin{tabular}{c|c|}
\hline ${\rm D}({\bf C})$ & \raisebox{-0.64cm}{\epsfig{figure=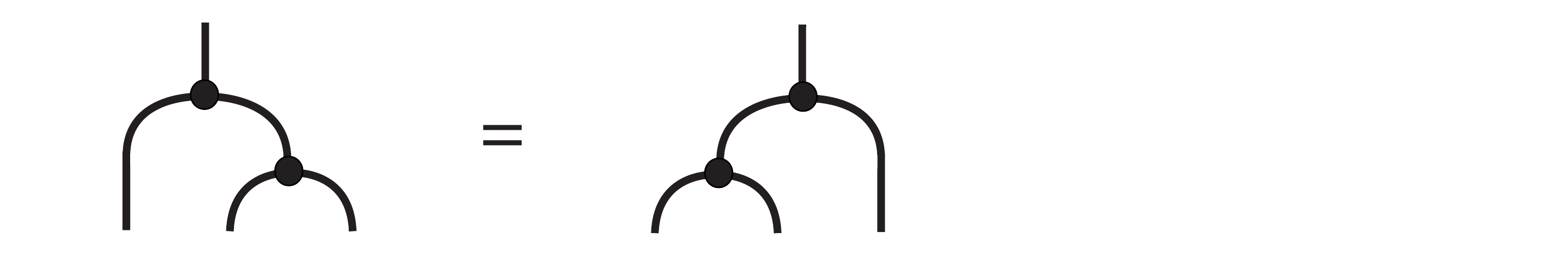,width=290pt}}\\
\hline
${\bf C}$ & \raisebox{-1.32cm}{\epsfig{figure=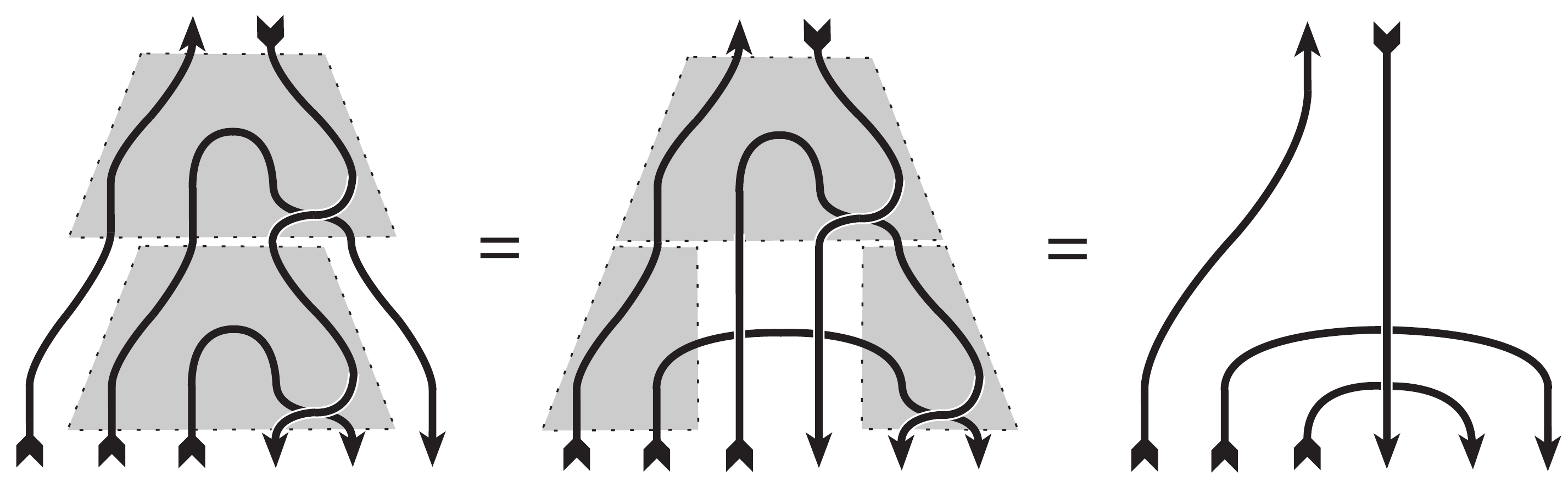,width=290pt}}\\
\hline \end{tabular}
\]
The other properties are verified similarly.
\endproof\newline

The above also illustrates how a non-commutative Frobenius multiplication can be constructed from commutative compact structures.
The dagger Frobenius structure ${\cal F}$ induces a self-dual compact structure, which depicts as follows:
\[
\begin{tabular}{c|c|}
\hline ${\rm D}({\bf C})$ & \raisebox{-0.34cm}{\epsfig{figure=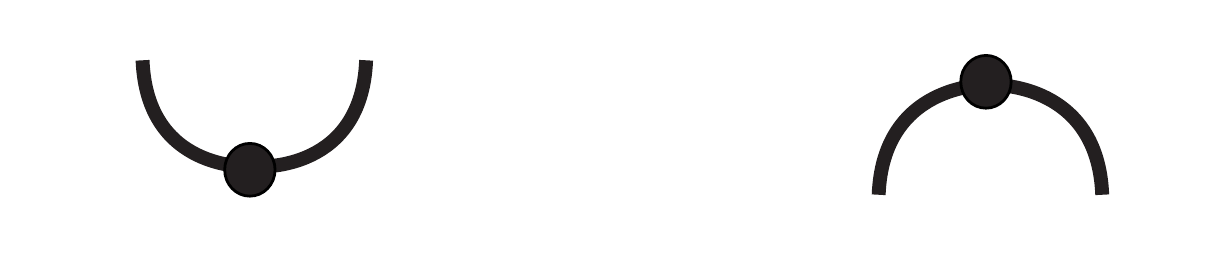,width=125pt}}\\
\hline
${\bf C}$ & \raisebox{-0.62cm}{\epsfig{figure=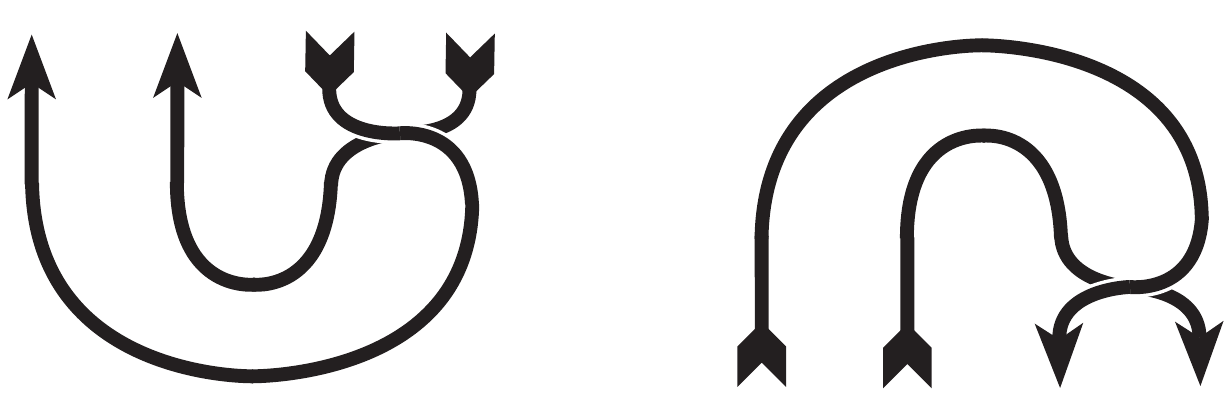,width=125pt}}\\
\hline \end{tabular}
\]

While the Frobenius multiplication is typically non-commutative (except in the degenerate case that
$\sigma_{A,A}=1_{A,A}$, which forces ${\bf C}$ to be trivial) the induced compact structure is always commutative:
\[
\begin{tabular}{c|c|}
\hline ${\rm D}({\bf C})$ & \raisebox{-0.64cm}{\epsfig{figure=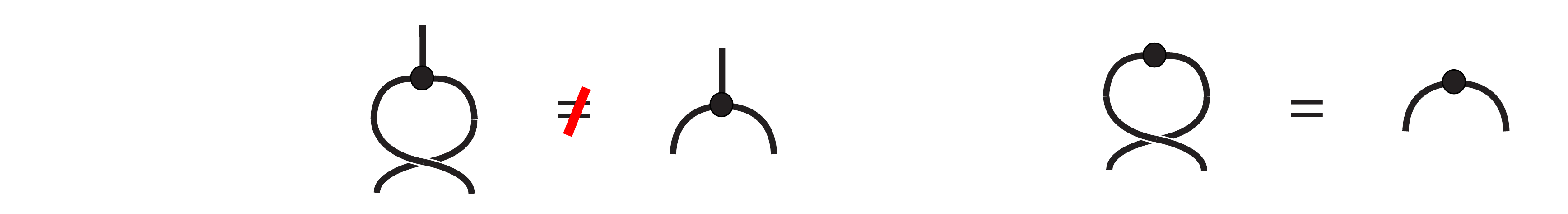,width=353pt}}\\
\hline
${\bf C}$ & \raisebox{-0.92cm}{\epsfig{figure=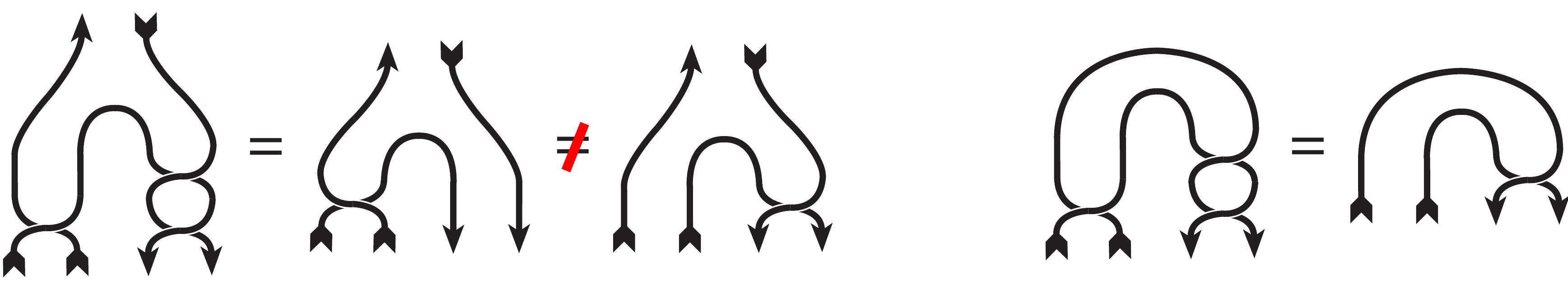,width=353pt}}\\
\hline \end{tabular}
\]

\begin{definition}[Selinger \cite{Selinger}]\label{def:CPMform}
A morphism  $f:A \to_{{\rm D}({\bf C})} B$  in ${\rm D}({\bf C})$ is \em completely positive \em if  its embedding in ${\bf C}$ is of the form:
\beq\label{CPMform}
f= (g\otimes \bar{g})\circ ( 1_A\otimes \eta_{C^*}  \otimes 1_{A^*})=\
\raisebox{-0.42cm}{\epsfig{figure=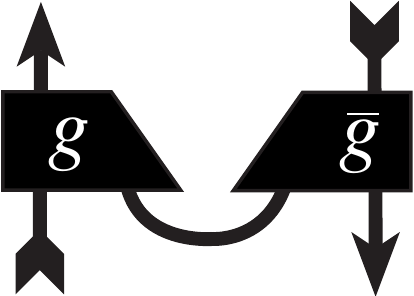,width=42pt}} :A \otimes A^* \to_{\bf C} B \otimes B^*,
\eeq
for some morphism $g: A \otimes C \to_{\bf C} B$.
It is \em normalized \em if we moreover have:
\beq
\epsilon_B\circ f=\ \raisebox{-0.42cm}{\epsfig{figure=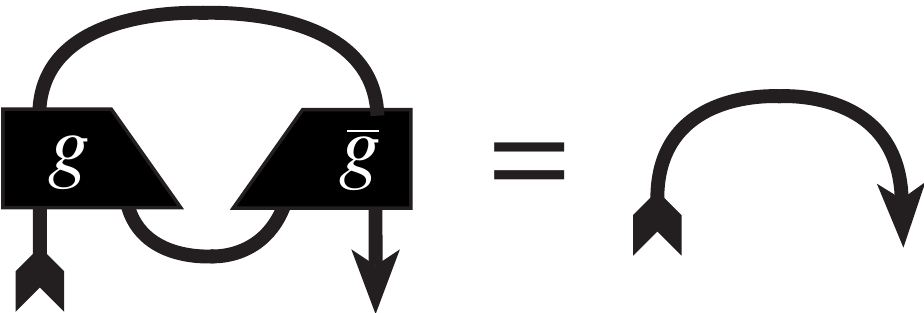,width=94pt}}= \epsilon_A\,.
\eeq
More specifically, a point $e:\II \to_{\rm D}({\bf C}) A$ in ${\rm D}({\bf C})$ is a \em mixed state  \em if  its embedding in ${\bf C}$  is of the form:
\beq \label{eq:mixedstate}
e=(g\otimes \bar{g})\circ\eta_{C^*}=\ \raisebox{-0.28cm}{\epsfig{figure=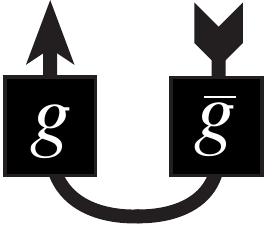,width=27pt}}
: \II \otimes \II^* \to_{\bf C} B \otimes B^*
\eeq
for some morphism $g:C \to_{\bf C} A$ in ${\bf C}$.  It is \em normalized \em if we moreover have:
\beq \label{eq:normalizedmixedstate}
\epsilon_A\circ e=  \raisebox{-0.25cm}{\epsfig{figure=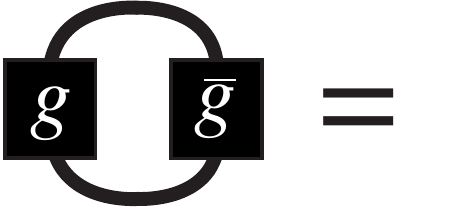,width=46pt}}\ \ \ \,.
\eeq
\end{definition}

\begin{example}
In ${\bf FdHilb}$  the concepts introduced in Definition \ref{def:CPMform} coincide with the usual ones; we explicitly establish this connection in the following section.
\end{example}

It is now easy to see that the failure of complete positivity in the case of the
${\cal F}$-comultiplication  (cf.~Section \ref{Sec:PresQonehalf})  is due to  the lack of symmetry between
the left and the right side of the picture:
\beq
\raisebox{-0.66cm}{\epsfig{figure=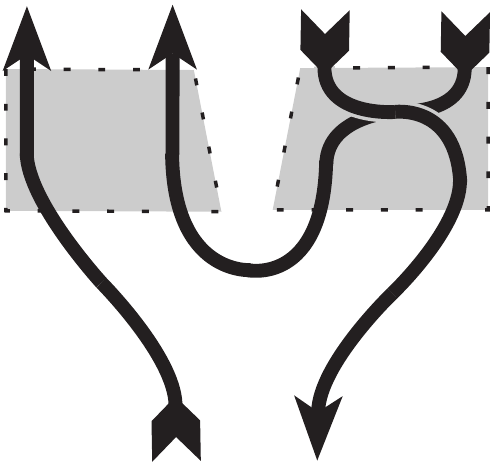,width=50pt}}
\eeq
This asymmetry is also what causes it to be non-commutative.

\begin{example}
Given a normalized mixed state $e_{A\ldots Z}: \II \to_{{\rm D}({\bf C})}  A\otimes\ldots\otimes Z$ in any such category ${\rm D}({\bf C})$, the specified Frobenius structure allows one build a Q${}_{1/2}$-calculus (provided the category has the appropriate inverses and square-roots) wherein the mixed state plays the role of the joint state.
\end{example}

\subsection{From operator presentation to ${\rm D}({\bf C})$-presentation}

At the convenience of the reader who is familiar with operator theory we now provide an explicit translation of typical operator theory concepts to the diagrammatic category ${\rm D}({\bf C})$.

By an \em operator \em we mean an endomorphisms in $\rho\in {\bf C}(A,A)$.  Such an operator $\rho$ is \em positive \em if it is of the form:
\[ \label{eq:positiveoperator}
\rho=g\circ g^\dagger=\ \raisebox{-0.60cm}{\epsfig{figure=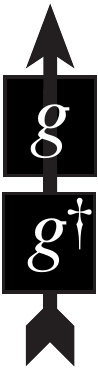,width=10pt}}\ .
\]

\begin{proposition}\label{lem:oppCPM}
For any object $A\in |{\bf C}|=|{\rm D}({\bf C})|$, operators ${\bf C}(A, A)$ are in bijective correspondence with morphisms ${\rm D}({\bf C})(\II, A)$ via the isomorphism
\beq
\xi_A:{\bf C}(A, A)\to{\rm D}({\bf C})(\II, A)::\rho=\
\raisebox{-0.36cm}{\epsfig{figure=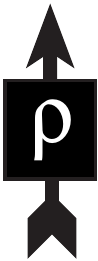,width=10pt}}\ \mapsto (\rho\otimes 1_{A^*})\circ\eta_{A^*}=\
\raisebox{-0.28cm}{\epsfig{figure=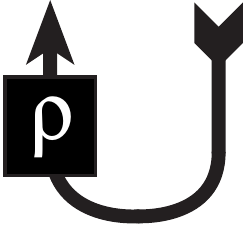,width=25pt}}\ .
\eeq
Along  this isomorphism, the positive operators in ${\bf C}(A, A)$ are in bijective correspondence with the mixed states in ${\rm D}({\bf C})(\II, A)$.
\end{proposition}
\bpf
That this map is a bijection follows easily from Definition \ref{defn:operatorcategory} of  ${\rm D}({\bf C})(\II, A)$ and the yanking equations (\ref{eq:yanking}), and that
positive operators are in correspondence with mixed quantum states  follows from the definitions of the latter, Eqs.~(\ref{eq:mixedstate}) and (\ref{eq:positiveoperator}), and from the definition of the conjugate, Eq.~(\ref{def:conjugate}), together with the yanking equations.
\endproof\newline

The following proposition expresses how operations on operators in ${\bf C}$ relate to operations on
the corresponding points in ${\rm D}({\bf C})$ along the isomorphism $\xi$, most notably:
that composition of operators in ${\bf C}$
\beq
- \circ -:{\bf C}(A,A)\times {\bf C}(A,A) \to {\bf C}(A,A):: \left(\raisebox{-0.36cm}{\epsfig{figure=ChoiJamolkovrho.pdf,width=10pt}}\,,
\raisebox{-0.36cm}{\epsfig{figure=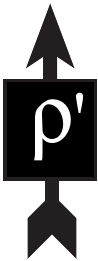,width=10pt}}\right)
\mapsto \raisebox{-0.59cm}{\epsfig{figure=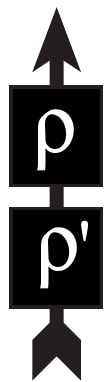,width=11.5pt}}\, ,
 \eeq
 corresponds to tensor product of the corresponding points in ${\rm D}({\bf C})$ composed with the non-commutative dagger Frobenius  multiplication ${\cal F}$,
 \beq
 {\cal F}\circ (- \otimes_{{\rm D}({\bf C})} -):{\rm D}({\bf C})(I,A)\times {\rm D}({\bf C})(I,A) \to {\rm D}({\bf C})(I,A):: \left(\raisebox{-0.28cm}{\epsfig{figure=ChoiJamolkov.pdf,width=25pt}},
 \raisebox{-0.28cm}{\epsfig{figure=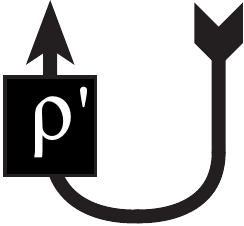,width=25pt}}\right) \mapsto \raisebox{-0.56cm}{\epsfig{figure=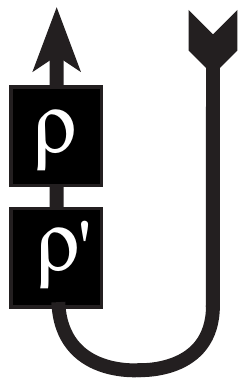,width=25pt}}.
 \eeq
We also show that the \em partial trace \em of operators in ${\bf C}$,
\beq
tr_B:{\bf C}(A\otimes B, A\otimes B)\to {\bf C}(A, A)::
\raisebox{-0.38cm}{\epsfig{figure=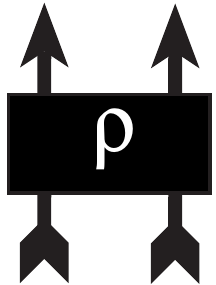,width=22pt}}\
\mapsto \
\raisebox{-0.38cm}{\epsfig{figure=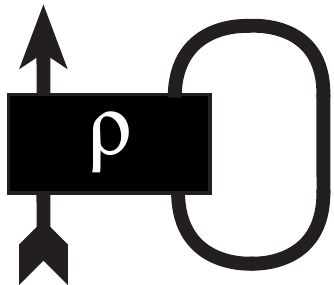,width=34pt}}\ ,
\eeq
corresponds in ${\rm D}({\bf C})$ to
\beq \label{eq:DCtrace}
tr^{\bf D}_B: {\rm D}({\bf C})(\II, A\otimes B)\to {\rm D}({\bf C})(\II, A)::
\ \raisebox{-0.22cm}{\epsfig{figure=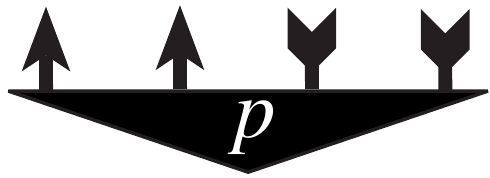,width=50pt}} \ \mapsto
\ \raisebox{-0.22cm}{\epsfig{figure=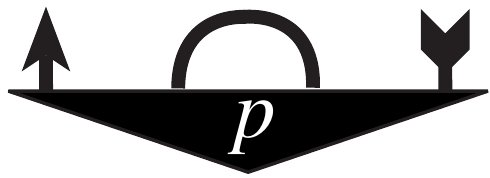,width=50pt}}.
\eeq

\begin{proposition}
(i) The following diagram commutes:
\beq
\begin{diagram}
\qquad{\bf C}(A, A) \times {\bf C}(A, A)
& \rTo^{\xi_{A}\times\xi_{A}} &
{\rm D}({\bf C})(\II, A)\times{\rm D}({\bf C})(\II, A)\\
\dTo^{- \circ -} & & \dTo_{{\cal F}\circ(-\otimes_{\rm D}-)}\\
{\bf C}(A,A)
& \rTo_{\xi_{A}} &
{\rm D}({\bf C})(\II, A)
\end{diagram}\qquad  ,
\eeq
and (ii) when setting $tr^{\bf D}_B$ as in Eq.~(\ref{eq:DCtrace}),
then the following diagram commutes:
\beq
\begin{diagram}
{\bf C}(A\otimes B,A\otimes B) & \rTo^{\xi_{A\otimes B}} & {\rm D}({\bf C})(\II, A\otimes B)\\
\dTo^{tr_B} & & \dTo_{tr^{\bf D}_B}\\
{\bf C}(A,A) & \rTo_{\xi_{A}} & {\rm D}({\bf C})(\II, A)
\end{diagram}\ .
\eeq
\end{proposition}
\bpf
We have:
\begin{enumerate}
  \item[(i)]
  \bit
\item ${\cal F}\circ\left(\xi_A\left(\raisebox{-0.36cm}{\epsfig{figure=ChoiJamolkovrho.pdf,width=10pt}}\right)\otimes_{\rm D}\xi_A\left(\raisebox{-0.36cm}{\epsfig{figure=ChoiJamolkovrhobis.pdf,width=10pt}}\right)\right)=
{\cal F}\circ\left(\raisebox{-0.28cm}{\epsfig{figure=ChoiJamolkov.pdf,width=25pt}}\otimes_{\rm D}\raisebox{-0.28cm}{\epsfig{figure=ChoiJamolkovbis.pdf,width=25pt}}\right)= \
\raisebox{-0.80cm}{\epsfig{figure=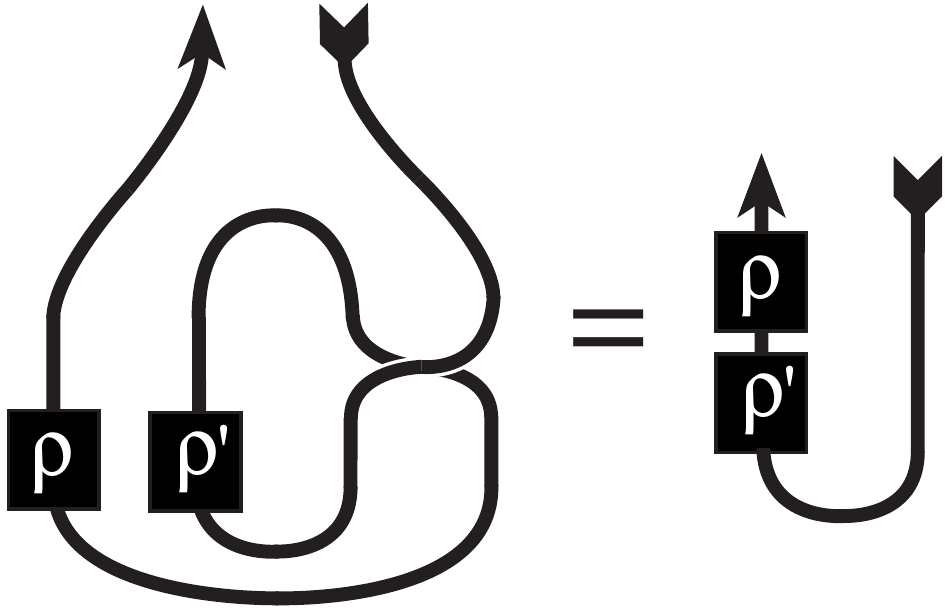,width=96pt}}$
\item $\xi_A\left( \raisebox{-0.36cm}{\epsfig{figure=ChoiJamolkovrho.pdf,width=10pt}} \circ \raisebox{-0.36cm}{\epsfig{figure=ChoiJamolkovrhobis.pdf,width=10pt}}\right)= \xi_A\left(  \raisebox{-0.59cm}{\epsfig{figure=FrobeniusProoftris.pdf,width=11.5pt}}\right)=\ \raisebox{-0.56cm}{\epsfig{figure=FrobeniusProofbis.pdf,width=25pt}} $
\eit
  \item[(ii)]
  \bit
\item $tr^{\bf D}_B\left(\xi_{A\otimes B}\left(\raisebox{-0.38cm}{\epsfig{figure=ChoiJamoltris.pdf,width=22pt}} \right)\right)=
tr^{\bf D}_B\left(\raisebox{-0.62cm}{\epsfig{figure=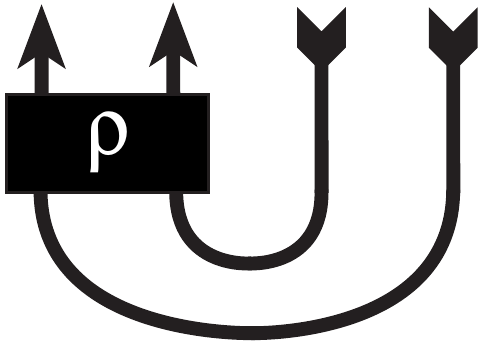,width=49pt}}\right) = \
\raisebox{-0.62cm}{\epsfig{figure=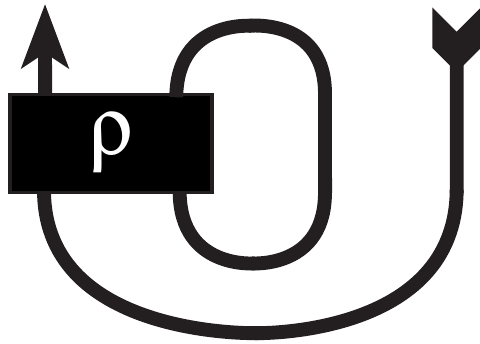,width=49pt}}$\ ;
\item $\xi_{B}\left(tr_B\left(\raisebox{-0.38cm}{\epsfig{figure=ChoiJamoltris.pdf,width=22pt}} \right)\right)
=\xi_{B}\left(\raisebox{-0.38cm}{\epsfig{figure=ptrace1bis.pdf,width=34pt}}\right)= \
\raisebox{-0.62cm}{\epsfig{figure=ptrace2bisbis.pdf,width=49pt}}$\ .
\eit
\end{enumerate}
\endproof

Hence the caps of the compact structure in ${\bf C}$ provides the partial trace, which in ${\rm D}({\bf C})$ becomes the counit of the Frobenius multiplication.


\begin{definition} \cite{Selinger}\label{def:SelingerCPM}
If ${\bf C}$ is any dCC then we define ${\rm CPM}({\bf C})$ to be the sub-dCC of ${\rm D}({\bf C})$ which has the same objects as ${\bf C}$ and which has completely positive maps as morphisms.
\end{definition}

The beauty of both ${\rm D}({\bf C})$ and ${\rm CPM}({\bf C})$ is that (density) operators become points rather than operations, and that completely positive maps, rather than being mappings from (density) operators to (density) operators, become morphisms.  Similarly, the Choi-Jamiolkowski isomorphism takes a particularly elegant form in ${\rm D}({\bf C})$ and ${\rm CPM}({\bf C})$, in that it becomes a bijective correspondence between elements and morphisms.


\begin{remark}
A similar presentation  of the \em internal endomorphism monoid \em in arbitrary  dCCs has already appeared in the literature  e.g.~\cite{Lauda,Jamie}, that is, a presentation as an object together with a non-commutative Frobenius structure which captures  composition of endomorphisms, namely:
\[
\left(  A^* \otimes A\ , \ {\cal G}:=\ \raisebox{-0.66cm}{\epsfig{figure=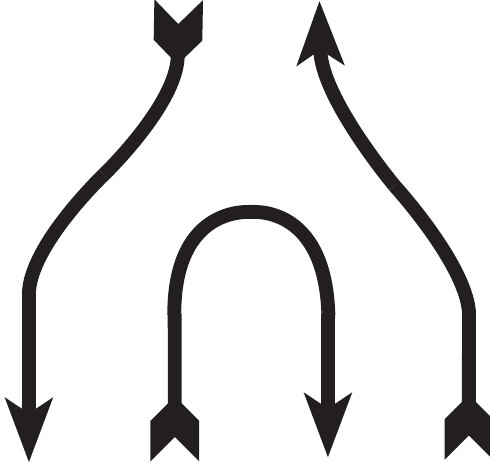,width=50pt}}\ ,\ \raisebox{-0.20cm}{\epsfig{figure=cup.pdf,width=24pt}} \right)\ ,
\]
where ${\cal G}$ is now easily seen to be a dagger Frobenius structure within ${\bf C}$ itself, that is, in particular, with respect to the $\otimes$-tensor.   While the $\otimes$-tensor and form of the Frobenius multiplication ${\cal G}$ are simpler to manipulate, the $\otimes_{\rm D}$-tensor  is essential for ${\rm D}({\bf C})$ (or ${\rm CPM}({\bf C})$) to be closed under tensoring  \cite{Selinger}, and the particular form of ${\cal F}$ is essential  for it to be an internal dagger Frobenius structure within ${\rm D}({\bf C})$.
\end{remark}

\section{Outlook}

There is much work to be done in developing the framework outlined in this paper.  For one, there are many concepts in classical Bayesian inference for which the quantum analogues are not yet known or are still poorly understood.  The development of the graphical approach described here can be informed by progress in this area and can contribute to it \footnote{It can also hope to provide some perspective on the wider body of work that seeks to interpret quantum theory as a theory of Bayesian inference, for instance, the program of `quantum Bayesianism' (See \cite{Qbism} and references therein)}.  Also, although we have confined our attention in this work to correlations between independent systems, the conditional density operator calculus can also be used to describe correlations across time for a single system.  This is a particularly interesting topic because, unlike the theory of classical Bayesian inference, there is a distinction between these kinds of correlations in quantum theory and such differences are important for understanding causality.  The distinction is captured in our graphical approach by the existence of two distinct compact structures, the interaction between which is described by means of dualizers~\cite{CPaqPer}.  Along these same lines, our approach should be useful for the project of finding a unified framework that incorporates retrodiction (inferences from a later time to an earlier time) as well as pre- and post-selection (inferences from information at both later and earlier times).  We would also like to provide an axiomatic characterization of the Q${}_{1/2}$-calculus that is analogous to our characterization of the classical Bayesian calculus, that is,  in terms of the motion of modifiers through the Frobenius structure, and to settle the question of whether one can prove in the Q${}_{1/2}$-calculus everything that can be proved in the conditional density operator calculus.  Finally, insofar as conditionning in Bayesian probability theory has strong connections with conditionning in a sequent calculus (which expresses the logic of provability), we expect that the present work will provide a stepping stone to a graphical representation of sequent calculi and to a better understanding of the interplay between probability theory and logic.

\section{Acknowledgement}

The authors thank Matt Leifer for discussions and for feedback on a draft of this article, in particular, the suggestion to explore the graphoid axioms.  RWS acknowledges support from the Perimeter Institute, which is funded by the Government of Canada through Industry Canada and by the Province of Ontario through the Ministry of Research and Innovation.  BC acknowledges support from EPSRC ARF EP/D072786/1, ONR Grant N00014-09-1-0248, EU FP6 STREP QICS, an FQXi Large Grant and Perimeter Institute who hosted him as a Long Term Visiting Scientist.


\end{document}